\documentclass[journal,a4paper]{IEEEtran}

\usepackage[]{graphicx}
\usepackage{url}
\usepackage{ifpdf}
\usepackage{hyperref}
\hypersetup{breaklinks,colorlinks,citecolor=blue}
\usepackage{cite}
\usepackage{amssymb}
\usepackage{amsmath}
\usepackage{mathtools}

\usepackage{subfigure}
\usepackage{color}

\newcommand{\analysissynthesis}[1]{\textcolor{black}{#1}}

\newcommand{\referee}[1]{\textcolor{black}{#1}}

\newcommand{\be}{\begin{equation}}
\newcommand{\ee}{\end{equation}}
\newcommand{\ba}{\begin{eqnarray}}
\newcommand{\ea}{\end{eqnarray}}
\newcommand{\bnum}{\begin{enumerate}}
\newcommand{\enum}{\end{enumerate}}

\newcommand{\eqn}[1]{(#1)}

\newcommand{\fig}[1]{Fig.~#1}

\newcommand{\sectn}[1]{Sec.~#1}

\newcommand{\etal}{\mbox{\it et al.}}
\newcommand{\eg}{\mbox{\it e.g.}}
\newcommand{\ie}{\mbox{\it i.e.}}

\DeclarePairedDelimiter{\diagfences}{(}{)}
\newcommand{\diag}{\operatorname{diag}\diagfences}
\newcommand{\spcend}{\ensuremath{\:}}
\newcommand{\img}{\ensuremath{{\rm i}}}
\newcommand{\cconj}{\ensuremath{\ast}} 
 
\newcommand{\reals}{\ensuremath{\mathbb{R}}}
\newcommand{\realsnn}{\ensuremath{\mathbb{R^+}}}

\newcommand{\integers}{\ensuremath{\mathbb{Z}}}
\newcommand{\naturals}{\ensuremath{\mathbb{N}}}
\newcommand{\complex}{\ensuremath{\mathbb{C}}}
\newcommand{\ltwo}{\ensuremath{\mathrm{L}^2}}
\newcommand{\sphere}{\ensuremath{{\mathbb{S}^2}}}
\newcommand{\sothree}{\ensuremath{{\mathrm{SO}(3)}}}

\newcommand{\vect}[1]{\ensuremath{\mbox{\boldmath ${#1}$}}}

\newcommand{\dx}{\ensuremath{\mathrm{\,d}}}
\newcommand{\dmu}[1]{\ensuremath{\dx \Omega(#1)}}

\newcommand{\deul}[1]{\ensuremath{\dx \varrho(#1)}}

\newcommand{\innerp}[2]{\ensuremath{\langle {#1},\: {#2} \rangle}}

\newcommand{\sa}{\ensuremath{\omega}}
\newcommand{\saa}{\ensuremath{\theta}}
\newcommand{\sab}{\ensuremath{\varphi}}
\newcommand{\sas}{\ensuremath{\saa, \sab}}
\newcommand{\eul}{\ensuremath{\mathbf{\rho}}}

\newcommand{\eula}{\ensuremath{\alpha}}
\newcommand{\eulb}{\ensuremath{\beta}}
\newcommand{\eulc}{\ensuremath{\gamma}}
\newcommand{\eulai}{\ensuremath{a}}
\newcommand{\eulbi}{\ensuremath{b}}
\newcommand{\eulci}{\ensuremath{g}}
\newcommand{\eulaiang}{\ensuremath{\eula_\eulai}}
\newcommand{\eulbiang}{\ensuremath{\eulb_\eulbi}}
\newcommand{\eulciang}{\ensuremath{\eulc_\eulci}}
\newcommand{\el}{\ensuremath{\ell}}
\newcommand{\m}{\ensuremath{m}}
\newcommand{\n}{\ensuremath{n}}

\newcommand{\spin}{\ensuremath{s}}

\newcommand{\elmax}{\ensuremath{{L}}}

\newcommand{\p}{\ensuremath{^\prime}}

\newcommand{\shfarg}[3]{\ensuremath{Y_{#1#2}({#3})}}
\newcommand{\shfargc}[3]{\ensuremath{Y_{#1#2}^\cconj({#3})}}

\newcommand{\shf}[2]{\ensuremath{Y_{#1#2}}}

\newcommand{\shc}[3]{\ensuremath{{#1}_{{#2}{#3}}}}

\newcommand{\sshc}[4]{\ensuremath{{}_{#4} {#1}_{{#2}{#3}}}}

\newcommand{\dmatbig}{\ensuremath{D}}
\newcommand{\Dlmn}{\ensuremath{ \dmatbig_{\m\n}^{\el} }}
\newcommand{\Dlmnc}{\ensuremath{ \dmatbig_{\m\n}^{\el\cconj} }}

\newcommand{\Dlmnp}{\ensuremath{ \dmatbig_{\m\n}^{\el}(\eul) }}
\newcommand{\Dlmnpc}{\ensuremath{ \dmatbig_{\m\n}^{\el\cconj}(\eul) }}

\newcommand{\dlmnhalfpim}{\ensuremath{ \Delta_{{\m\p}{\m}}^{\el} }}
\newcommand{\dlmnhalfpin}{\ensuremath{ \Delta_{{\m\p}{\n}}^{\el} }}

\newcommand{\wigc}[4]{\ensuremath{{#1}^{#2}_{{#3}{#4}}}}

\newcommand{\rotarg}[1]{\ensuremath{\mathcal{R_{#1}}}}

\newcommand{\f}{\ensuremath{f}}

\newcommand{\wav}{\ensuremath{\psi}}

\newcommand{\wscale}{\ensuremath{j}}

\newcommand{\Gmnm}{\ensuremath{G_{\m\n\m\p}}}

\newcommand{\Fmnm}{\ensuremath{\tilde{G}_{\m\n\m\p}}}
\newcommand{\Fmnmp}{\ensuremath{\tilde{G}_{\m\n\m\p{}\p}}}

\newcommand{\saai}{\ensuremath{t}}

\newcommand{\nsphere}{\ensuremath{{N_{\sind}}}}
\newcommand{\weight}{\ensuremath{w}}
\newcommand{\order}{\ensuremath{\mathcal{O}}}

\newcommand{\nmeas}{\ensuremath{M}}

\renewcommand{\eqn}[1]{Eq.~(#1)}

\renewcommand{\wav}{\ensuremath{\Psi}}

\newcommand{\bmtrx}[1]{\ensuremath{\boldsymbol{\mathsf{#1}}}}
\renewcommand{\nsphere}{\ensuremath{{N_\sphere}}}
\newcommand{\nsothree}{\ensuremath{{N_\sothree}}}

\begin{document}
\title{Sparse image reconstruction on the sphere:\\analysis \referee{and} synthesis}
\author{Christopher~G.~R.~Wallis, Yves Wiaux and Jason~D.~McEwen  \thanks{C.~G.~R.~Wallis and J.~D.~McEwen were supported by the Engineering and Physical
    Sciences Research Council (grant number
    EP/M011852/1).}
  \thanks{C.~G.~R.~Wallis and J.~D.~McEwen are with the Mullard Space Science Laboratory
    (MSSL), University College London (UCL), Surrey RH5 6NT, UK. Y. Wiaux is with the Institute of Sensors, Signals, and Systems, Heriot-Watt University, Edinburgh EH14 4AS, UK.
}  \thanks{E-mail: chris.wallis@ucl.ac.uk (C.~G.~R.~Wallis)}}

\markboth{IEEE TRANSACTIONS ON IMAGE PROCESSING ,~Vol.~--, No.~--, --}{Wallis \MakeLowercase{\textit{et al.}}: Sparse image reconstruction on the sphere}

\maketitle

\begin{abstract}
We develop techniques to solve ill-posed inverse problems on the sphere by sparse regularisation, exploiting sparsity in both axisymmetric and directional scale-discretised wavelet space.  Denoising, inpainting, and deconvolution problems, and combinations thereof, are considered as examples. Inverse problems are solved in both the analysis and synthesis settings, with a number of different sampling schemes.  The most effective approach is that with the most restricted solution-space, which depends on the interplay between the adopted sampling scheme, the selection of the analysis/synthesis problem, and any weighting of the $\ell_1$ norm appearing in the regularisation problem.  More efficient sampling schemes on the sphere improve reconstruction fidelity by restricting the solution-space and also by improving sparsity in wavelet space.  We apply the technique to denoise {\it Planck} \mbox{353 GHz} observations, improving the ability to extract the structure of Galactic dust emission, which is important for studying Galactic magnetism.
\end{abstract}


\IEEEpeerreviewmaketitle

\section{Introduction}

\IEEEPARstart{S}{pherical} images arise in many fields,  from cosmology (\eg\
\cite{Planck2015:Overview}) to biomedical imaging (\eg\ \cite{johansenberg:2009}),
where inverse problems are often encountered. Sparse priors have proved highly
effective in regularising Euclidean inverse problems, where sparsity is
imposed in a wavelet space or sparsifying dictionary.  In the
spherical setting, wavelet theory is only recently starting to approach
maturity, while a mature, general, and robust framework for sparse regularisation is
lacking.

Over the last couple of decades there have been many developments
regarding wavelet theory in spherical settings.  Many initial attempts to
extend wavelet transforms to the sphere differed primarily in the manner in
which dilations are defined on the sphere \cite{narcowich:1996, potts:1995,
freeden:1997a, torresani:1995, dahlke:1996, holschneider:1996, antoine:1999,
antoine:1998, wiaux:2005, sanz:2006, mcewen:2006:cswt2}.   These constructions
were essentially based on continuous methodologies, which, although
insightful, limited practical application to problems where the exact
synthesis of a function from its wavelet coefficients is not required.  A
number of early discrete constructions followed \cite{schroder:1995,
barreiro:2000, bogdanova:2004,
mcewen:2008:fsi, mcewen:szip}; however, many of these constructions do not
necessarily lead to stable bases \cite{sweldens:1997}.
More recently, a number of discrete wavelet frameworks have emerged that have
found considerable application, particularly in cosmology (\eg\
\cite{delabrouille:2009, bobin:2013, rogers:s2let_ilc_temp,
rogers:s2let_ilc_pol}), including needlets \cite{narcowich:2006, baldi:2009,
marinucci:2008}; scale-discretised wavelets \cite{wiaux:2007:sdw,
leistedt:s2let_axisym, mcewen:2013:waveletsxv, mcewen:s2let_spin,
mcewen:s2let_ridgelets, chan:s2let_curvelets, mcewen:s2let_localisation}; and
the isotropic undecimated and pyramidal wavelet transforms \cite{starck:2006}.
All three of these approaches have also been extended to analyse signals
defined on the three- dimensional ball formed by augmenting the sphere with
the radial line \cite{durastanti:2014, leistedt:flaglets,
mcewen:flaglets_sampta, leistedt:flaglets_spin, lanusse:2012}, such as the
galaxy distribution.

\analysissynthesis{Solving Euclidean inverse problems by imposing sparse regularising priors has
become increasingly popular in recent years.  This trend has been driven by
improving theoretical foundations for the recovery of sparse
signals, facilitated by the theory of compressive sensing
\cite{candes:2006a,candes:2006c,donoho:2006}, and empirical
results that have demonstrated the effectiveness of sparse priors for wide
classes of natural images.
Sparse reconstruction problems can be posed in either the synthesis or
analysis settings \cite{elad:2007}.  In the synthesis setting, the sparse (\eg\ wavelet)
coefficients of the signal are recovered, from which the signal is
synthesised. In the analysis setting, although sparsity is imposed in some sparsifying
(\eg\ wavelet) dictionary, the signal is recovered directly.  When the
dictionary considered is not an orthonormal basis but a redundant dictionary,
the synthesis and analysis approaches exhibit quite different properties since
the solution-space of the analysis problem is more restrictive than the
synthesis problem \cite{elad:2007,cleju:12,nam:2013}.  Empirical studies have shown
promising results for the analysis setting (\eg\ \cite{elad:2007,
carrillo:sara, carrillo:sara_algo}), which is hypothesised to be due to its
more restrictive solution-space.}

Some progress has been made towards solving sparse regularisation problems on
the sphere (\eg\ \cite{abrial:2007, rauhut:2011, mcewen:css2}).   Compressive
sensing for signals sparse in spherical harmonic space is considered in
\cite{rauhut:2011}, while inpainting problems are considered in
\cite{abrial:2007, mcewen:css2}, imposing sparsity in a redundant dictionary
\cite{abrial:2007} and the signal gradient \cite{mcewen:css2}.  

In this work we consider general linear inverse problems on the sphere,
including denoising, inpainting, and deconvolution problems, and combinations
thereof, and apply sparse regularising priors in scale-discretised wavelet
space, using both axisymmetric and directional wavelets \cite{wiaux:2007:sdw,
leistedt:s2let_axisym, mcewen:2013:waveletsxv, mcewen:s2let_spin}. \analysissynthesis{ Moreover, for the
first time we study in detail the properties and empirical performance of the
analysis and synthesis problems on the sphere.}  Furthermore, we investigate
the impact of sampling theorems and schemes on the sphere
\cite{driscoll:1994,mcewen:fssht,khalid:optimal_sampling} for sparse image
recovery, which have already been shown to play a significant role
\cite{mcewen:css2}. \analysissynthesis{ In particular, we study the impact of the efficiency of
sampling schemes in both the synthesis and analysis settings.}

While sparse regularisation is often very effective, we close this
introduction by cautioning against the blind application of sparse priors. For
example, for the cosmic microwave background (CMB), which is to very good
approximation a realisation of a Gaussian random field on the sphere, we
recall that inpainting by imposing sparsity in spherical harmonic space (via
the $\ell_1$ norm) has the undesirable property of breaking statistical
isotropy \cite{feeney:inpainting_isotropy}.  One must therefore take care in
applying priors appropriate for the problem at hand.

The remainder of the article is structured as follows. In
\sectn{\ref{sec:sampling_on_sphere_and_rotation_group}}  we review  harmonic
analysis on the sphere \sphere\ and rotation group \sothree\ and associated
sampling theorems and schemes. In \sectn{\ref{sec:sparce_reconstruction_on_the_sphere}} we
present the general framework to solve inverse problems on
the sphere using sparse reconstruction. In
\sectn{\ref{sec:numerical_experiments}} we study sparse image
reconstruction on the sphere through numerical experiments, comparing the analysis and synthesis
settings and evaluating the impact of the sampling scheme used. In \sectn{\ref{sec:planck_results}}
 we apply the methods proposed to denoise the 
{\it Planck} 353 GHz data. In \sectn{\ref{sec:conclusions}} we
conclude.


\section{Sampling on the Sphere and Rotation Group}\label{sec:sampling_on_sphere_and_rotation_group}

In this section we review the representation of signals on the sphere and the rotation group, in both the spatial and harmonic domains.
We consider discretised signals, sampled according to different sampling schemes and sampling theories, that differ in the number of samples required to capture all of the information content of signals.

\subsection{Signals on the Sphere}

We consider the space of square integrable functions defined on the sphere \sphere.
The canonical basis for the space of square integrable functions on
the sphere is given by the spherical harmonics $\shf{\el}{\m} \in
\ltwo(\sphere)$, with natural $\el\in\naturals$, integer
$\m\in\integers$ and $|\m|\leq\el$.  Due to the orthogonality and
completeness of the spherical harmonics, any square integrable
function on the sphere $x \in \ltwo(\sphere)$ may be represented by
its spherical harmonic expansion
\begin{equation}
\label{eqn:sht_inv}
x(\sa) = 
\sum_{\el=0}^\infty
\sum_{\m=-\el}^\el
\shc{x}{\el}{\m} \:
\shfarg{\el}{\m}{\sa}
\spcend ,
\end{equation}
where the spherical harmonic coefficients are given by the usual
projection onto each basis function: 
\begin{equation}
\shc{x}{\el}{\m}
= \innerp{x}{\shf{\el}{\m}}
= \int_\sphere \dmu{\sa} \: x(\sa) \: \shfargc{\el}{\m}{\sa}
\spcend,
\end{equation}
where $\dmu{\sa} = \sin \saa \dx \saa \dx \sab$ is the usual
invariant measure on the sphere and $\sa = (\sas)$ denote spherical
coordinates with colatitude $\saa \in [0,\pi]$ and longitude $\sab \in
[0,2\pi)$.  Complex conjugation is denoted by the superscript
${}^\cconj$. Throughout, we consider signals on the sphere band-limited at
$\elmax$, that is signals such that $\shc{x}{\el}{\m}=0$, $\forall
\el\geq\elmax$, in which case the summation over \el\ in
\eqn{\ref{eqn:sht_inv}} may be truncated to the first \elmax\ terms.

In the discrete setting we can write the forward and inverse spherical harmonic transforms
as linear operators, respectively:
\ba
\hat{\vect{x}} = \tilde{\bmtrx{Y}} \vect{x},\label{eq:discrete_spherical_harmonic_transform}\\
\vect{x} = \bmtrx{Y} \hat{\vect{x}},
\ea
where $\tilde{\bmtrx{Y}} \in \complex^{L^2 \times \nsphere}$ and $\bmtrx{Y}
\in \complex^{\nsphere \times L^2}$, with $\nsphere$ denoting  the number
of samples on the sphere required to capture all the information content of a
signal band limited at $L$.   We denote the concatenated vector of \nsphere\
spatial measurements by $\vect{x}\in\complex^{\nsphere}$ and the concatenated
vector of $L^2$ harmonic coefficients by $\vect{\hat{x}}\in\complex^{L^2}$.
Here and throughout we denote the forward harmonic transform with a tilde and
the inverse transform without.
Since sampling theorems on the sphere do not reach optimal dimensionality, as
discussed in more detail below, the operators $\tilde{\bmtrx{Y}}$ and
$\bmtrx{Y}$ are not necessarily inverses of one another, \eg\ $\bmtrx{Y}
\tilde{\bmtrx{Y}} \neq \bmtrx{I}$ \referee{(although we note $\tilde{\bmtrx{Y}}\bmtrx{Y}
 = \bmtrx{I}$, where $\bmtrx{I}$ is the identity).}

When considering images on the sphere the sampling theorem adopted can be of
great significance. A sampling  theorem allows one to transform from real
space to harmonic space and back, without loss of information, from a finite
number of samples \nsphere.  Sampling theorems on the sphere differ in the
number of samples \nsphere\ required.  No existing sampling theorem on the
sphere achieves the optimal number of samples  of $L^2$ suggested by the
harmonic dimensionality of a band-limited signal. The canonical Driscoll \&
Healy \cite{driscoll:1994} sampling theorem on the sphere (hereafter DH)
requires $\sim 4\elmax^2$ samples to capture the information content of a
signal band-limited at $L$.  Recently, McEwen and Wiaux \cite{mcewen:fssht}
(hereafter MW) developed a novel sampling theorem requiring $\sim 2\elmax^2$
samples only, thereby reducing the spherical Nyquist sampling rate by a
factor of two.  More recently, Khalid, Kennedy and McEwen
\cite{khalid:optimal_sampling} developed a new sampling scheme (hereafter KKM)
that achieves the optimal number of $L^2$ samples. However, this scheme does
\emph{not} lead to a sampling theorem with theoretically exact spherical harmonic
transforms; nevertheless, good numerical accuracy is achieved in practice.

Fast algorithms to compute spherical harmonic transforms, which avoid any
pre-computation\footnote{Precompute quickly becomes infeasible for high band-limits
 due to $\order(\elmax^3)$ storage requirements \cite{mcewen:fssht}.},
have been developed for the DH and MW sampling theorems, which scale as
$\order(\elmax^3)$ \cite{driscoll:1994,healy:2003,mcewen:fssht}.  The
complexity of the fast algorithm for the KKM sampling schemes scales as
$\order(\elmax^4)$, which can be reduced by appealing to algorithms to perform
fast matrix-vector multiplications, thereby reaching close to
$\order(\elmax^3)$ in practice \cite{khalid:optimal_sampling}.

Alternative sampling schemes also exist (\eg\
{\tt HEALPix} \cite{gorski:2005}, {\tt IGLOO} \cite{crittenden:1998}, {\tt
GLESP} \cite{doroshkevich:2005}), although these are typically oversampled and
the accuracy of numerical quadrature can in some cases be limited (\eg\ {\tt
HEALPix}).   \referee{Finite rate of innovation schemes to recover signals on the sphere comprised of a number of Dirac delta functions have also been developed \cite{deslauriers:2013,dokmanic:2016,sattar:2016}. However, such super-resolution approaches cannot be applied to recover general signals on the sphere.  In this article we focus on efficient sampling schemes for general band-limited signals}, which are also
highly accurate (with accuracy close to machine precision): namely, the KKM
\cite{khalid:optimal_sampling}, MW \cite{mcewen:fssht} and DH
\cite{driscoll:1994} schemes.

\subsection{Signals on the Rotation Group}

When considering directional wavelets it is necessary to be able to 
decompose and reconstruct square integrable signals defined on the rotation group \mbox{$\sothree$}, the space of three-dimensional rotations, where rotations are parameterised by
the Euler angles $\eul=(\phi,\theta,\psi)$, with
$\phi \in [0,2\pi)$, $\theta \in [0,\pi]$ and $\psi \in [0,2\pi)$. 
We adopt the $zyz$ Euler convention corresponding to the rotation of a
physical body in a \emph{fixed} coordinate system about the $z$, $y$
and $z$ axes by $\psi$, $\theta$ and $\phi$, respectively.

The Wigner $\dmatbig$-functions $\Dlmn \in \ltwo(\sothree)$, with natural $\el\in\naturals$
and integer $\m,\n\in\integers$, $|\m|,|\n|\leq\el$, are the matrix
elements of the irreducible unitary representation of the rotation
group \sothree\  \cite{varshalovich:1989}.  Consequently, the \Dlmnc\ also form an orthogonal
basis in $\ltwo(\sothree)$.\footnote{We adopt the conjugate
  \dmatbig-functions as basis elements since this convention
  simplifies connections to wavelet transforms on the sphere.}  
Due to the orthogonality and completeness of the Wigner
$\dmatbig$-functions, any square integrable function on the rotation
group $x \in \ltwo(\sothree)$ may be represented by its Wigner
expansion\referee{
\begin{equation}
  \label{eqn:wig_inverse}
   x(\eul) = 
   \sum_{\el=0}^{\infty} \frac{2\el+1}{8\pi^2} \sum_{\m=-\el}^{\el}
   \sum_{\n=-\el}^\el
   \wigc{x}{\el}{\m}{\n} \Dlmnpc
   \spcend,
\end{equation}}
where the Wigner coefficients are given by the projection onto each
basis function: 
\ba
\wigc{x}{\el}{\m}{\n} &=& \innerp{x}{\Dlmnc},\\
&=& \int_\sothree \deul{\eul} \: x(\eul) \: \Dlmnp\spcend,
\ea
where $\deul{\eul} = \sin\theta \dx\theta \dx\phi \dx\psi$ is the
usual invariant measure on the rotation group. Note that
$\innerp{\cdot}{\cdot}$ is used to denote inner products over both the
sphere and the rotation group (the case adopted can be inferred from
the context).
Throughout, we consider signals on the rotation group band-limited at
$\elmax$, that is signals such that $\wigc{x}{\el}{\m}{\n}=0$, $\forall
\el\geq\elmax$,  in which case the summation over \el\ in
\eqn{\ref{eqn:wig_inverse}} may be truncated to the first \elmax\ terms.  

In the discrete setting we can write the forward and inverse Wigner transforms as linear operators, respectively:
\ba
\hat{\vect{x}} &=& {\tilde{\bmtrx{D}}} \vect{x}\\
\vect{x} &=& \bmtrx{D} \hat{\vect{x}},\label{eq:decrete_inverse_wigner_transform}
\ea
where $\tilde{\bmtrx{D}} \in \complex^{L(4L^2-1)/3 \times \nsothree}$ and
$\bmtrx{D} \in \complex^{\nsothree \times L(4L^2-1)/3}$, with $\nsothree$ denoting
the number of samples on the rotation group required to capture all the
information content of a signal band-limited at $L$.  The harmonic dimensionality of a
band-limited signal on the rotation group reads $L(4L^2-1)/3$. We denote the
concatenated vector of \nsothree\ spatial measurements by
$\vect{x}\in\complex^{\nsothree}$ and the concatenated vector of harmonic
coefficients by $\vect{\hat{x}}\in\complex^{L(4L^2-1)/3}$. Again, we denote
the forward harmonic transform with a tilde, and the inverse transform without, 
and note that the operators $\tilde{\bmtrx{D}}$ and
$\bmtrx{D}$ are not necessarily inverses of one another.

For signals with limited directional sensitivity, it is convenient
to consider a directional band-limit $N$, such that
$\wigc{x}{\el}{\m}{\n}=0$, $\forall \n\geq \min(N, \el)$.  In settings where the adopted band-limit (\ie\ resolution) is not clear from the context, we adorn $\tilde{\bmtrx{D}}$ and ${\bmtrx{D}}$ with superscripts denoting the spherical and directional band-limits adopted, \eg\ $\tilde{\bmtrx{D}}^{L,N}$. 

Sampling theorems on the rotation group can be constructed from a
straightforward extension of sampling theorems defined on the sphere. Kostelec
\etal\ \cite{kostelec:2008} extend the DH sampling theorem, leading to a
sampling theorem on the rotation group requiring $\sim 8L^3$ samples. McEwen
\etal\ \cite{mcewen:so3} extend the MW sampling theorem, leading to sampling
theorem requiring $\sim 4L^3$ samples.  The KKM sampling scheme has not yet
been extended to the rotation group.\footnote{Note that sampling schemes on
the sphere and rotation group based on Gauss-Legendre quadrature have also
been developed (\eg\ \cite{mcewen:fssht,khalid:so3_gl}).}  No sampling theorem
on the rotation group reaches the optimal harmonic dimensionality of
$\sim4L^3/3$. We adopt the sampling theorem on the rotation group of McEwen
\etal\ \cite{mcewen:so3} in this work, where fast algorithms are developed to
compute forward and inverse Wigner transforms that scale as $\order(N
\elmax^3)$.

\section{Sparse Image Reconstruction on the Sphere}\label{sec:sparce_reconstruction_on_the_sphere}

We develop the proposed framework to solve inverse problems on the sphere in
this section. We begin by reviewing wavelet transforms on the sphere, before
presenting a discrete, operator  formulation that illuminates the adjoint
operators of the wavelet transform.  Sparse regularisation problems on the
sphere are then posed, in both analysis and synthesis settings, before the
properties of these problems are discussed, along with algorithmic details
for solving the problems, which require fast adjoint operators.

\subsection{Wavelet Analysis and Synthesis}

We adopt the scale-discretised wavelet transform on the sphere \cite{wiaux:2007:sdw,leistedt:s2let_axisym,mcewen:s2let_localisation,mcewen:s2let_spin}, which supports directional wavelets.
The wavelet transform is given by the directional convolution of each wavelet, $\Psi^j \in \ltwo(\sphere)$, with the signal of interest, $x \in \ltwo(\sphere)$:
\ba
w^{j}(\eul) &=&  \langle x , \mathcal{R}_{\eul} \Psi^j\rangle\\
&=& \!\!\!\!\int_\sphere\!\!\!\! \dmu{\sa^\prime}x(\sa^\prime)(\mathcal{R}_{\eul}\Psi^j)^\ast(\sa^\prime),
\ea
where wavelet coefficients $w^j \in \ltwo(\sothree)$ incorporate directional information and so are defined on the rotation group. $\mathcal{R}_{\eul}$ is the rotation operator that rotates by the Euler angles $\eul$.  Wavelets are considered for a range of scales $j$, which runs from $J_{\rm min}$ to $J_{\rm max}$. For further details on the wavelet construction and transform see, \eg, \cite{mcewen:s2let_localisation,mcewen:s2let_spin}. The lowest frequency
content of the signal of interest is extracted by the axisymmetric convolution of a scaling function, $\Upsilon \in \ltwo(\sphere)$, with the function of interest:
\ba
s(\sa) &=& \langle x , \mathcal{R}_{\sa} \Upsilon\rangle\\
&=& \!\!\!\!\int_\sphere\!\!\!\! \dmu{\sa^\prime}x(\sa^\prime)(\mathcal{R}_{\sa}\Upsilon)^\ast(\sa^\prime),
\ea
where scaling coefficients $s \in \ltwo(\sphere)$ are defined on the sphere since low-frequency directional structure is not typically of interest.  In harmonic space these directional and axisymmetric convolutions read, respectively,
\ba
\left({w}^{j}\right)^\ell_{m n} &=&  \frac{8\pi^2}{2\ell+1} {x}_{\ell m} {\Psi^j}^{*}_{\ell n},\label{eq:wavelet_trans_harm_psi}\\
s_{\ell m} &=&  \sqrt{ \frac{4\pi}{2\ell+1}} {x}_{\ell m} {\Upsilon}^{*}_{\ell 0},\label{eq:wavelet_trans_harm_scalar}
\ea
(see, \eg, \cite{mcewen:s2let_spin}), where $({w}^{j})^\ell_{m n} = \langle w^{j} , \Dlmnc \rangle$, $s_{\ell m} = \langle s , Y_{\ell m} \rangle$, \referee{$\Psi^j_{\ell n} = \langle \Psi^j, Y_{\ell n} \rangle$} and ${\Upsilon}_{\ell 0}\delta_{m0} = \langle \Upsilon , Y_{\ell m} \rangle$, \referee{where $\delta_{nm}$ is the Kronecker delta function}. 

The original image on the sphere can be reconstructed from its wavelet and scaling coefficients by
\ba
  x(\sa)
  &=& \int_\sphere \dmu{\sa\p} \:
  s(\sa\p) \: (\rotarg{\sa\p} \: \Upsilon)(\sa)\nonumber\\
  &+&
  \sum_{\wscale={J_{\rm min}}}^{J_{\rm max}} \int_\sothree \deul{\eul} \:
  w^{\wscale}(\eul) \: (\rotarg{\eul} \: \Psi^\wscale)(\sa)
  \spcend,
\ea
provided the wavelets and scaling function satisfy an admissibility criterion \cite{wiaux:2007:sdw,leistedt:s2let_axisym,mcewen:s2let_localisation,mcewen:s2let_spin}.
In harmonic space, reconstruction reads
\ba
  x_{\ell m} 
  &=& 
  \sqrt{\frac{4 \pi}{2 \el + 1}} \:  
  \shc{s}{\el}{\m} \:
  \sshc{\Upsilon}{\el}{0} \nonumber\\
  &+&
  \sum_{\wscale={J_{\rm min}}}^{J_{\rm max}} \sum_{n=-\ell}^{\ell}
  \wigc{\bigl(w^{{\wscale}}\bigr)}{\el}{\m}{\n} \:
  \wav_{\ell\n}^{\wscale}.\label{eq:wavelet_inv_trans_harm}
\ea
In practice, for directional wavelet transforms we consider wavelets with an
azimuthal band-limit $N$, \ie\ $\wav_{\ell\n}^{\wscale} =0$, \mbox{$\forall n
\geq \min(N,\el)$}, which implies the directional wavelet coefficients $w^j$ also
exhibit a directional band-limit $N$.  Furthermore, directional wavelets with
even or odd azimuthal symmetry are typically considered, in which case only $N$ (rather than $2N-1$) directions are required \cite{mcewen:s2let_spin,mcewen:s2let_localisation,wiaux:2007:sdw}.

\subsection{Discrete Wavelet Analysis and Synthesis}

We formulate discrete, operator representations of the forward and inverse wavelet transforms that permit a clear construction of the adjoint wavelet operators.  We consider the harmonic representation of the wavelet transform, which is inherently discretised, where we concatenate the harmonic coefficients into a single vector.  The wavelet transform can be represented by its action on harmonic coefficients, followed by inverse harmonic transforms.  A similar represententation is formulated for the inverse wavelet transform.  By formulating wavelet transforms as a concatenation of operators, it is straightforward to construct operators representing adjoint wavelet transforms,  which are required for solving sparse regularisation problems.

The harmonic expressions for the wavelet transform given by \eqn{\ref{eq:wavelet_trans_harm_psi}} and \eqn{\ref{eq:wavelet_trans_harm_scalar}} may be written in terms of linear operators:
\ba
\hat{\vect{w}}^{j} &=& \bmtrx{N}^j\bmtrx{W}^j\hat{\vect{x}},\\
\hat{\vect{s}} &=& \bmtrx{S}\hat{\vect{x}},
\ea
where $\hat{\vect{w}}^{j}$ denotes Wigner coefficients of the wavelet coefficients $w^j$ and $\hat{\vect{s}}$ denotes spherical harmonic coefficients of the scaling coefficients $s$.
The operators $\bmtrx{W}^j \in \complex^{N^j (L^j)^2 \times L^2}$ and $\bmtrx{S}
\in \complex^{L_{\rm s}^2 \times L^2}$ implement harmonic space multiplication by the wavelet
${\Psi^j}^{*}_{\ell n}$ and scaling function ${\Upsilon}^{*}_{\ell 0}$, respectively, as described by \eqn{\ref{eq:wavelet_trans_harm_psi}} and \eqn{\ref{eq:wavelet_trans_harm_scalar}}, where the $\el$ normalisation factor is not included in the former but is included in the latter ($L^j$ and $N^j$ are defined in detail below).  The normalisation for the wavelets, given by $8\pi^2/(2\el+1)$, is applied by the operator $\bmtrx{N}^j \in \reals^{N^j (L^j)^2 \times N^j (L^j)^2 }$.  We separate out the normalisation in this case as it does not apply in the reconstruction of the 
signal seen in \eqn{\ref{eq:wavelet_inv_trans_harm}}.

Harmonic space representations of wavelet and scaling coefficients are represented at the minimum resolution required to capture all signal content.  Thus, the band-limit for each wavelet scale $j$ is limited to $L^j$ and for the scaling function to $L_{\rm s}$.  Wavelet $\Psi^j$ has support in the range $[\lambda^{j-1},\lambda^{j+1}]$, where $\lambda \in \reals$ is a scaling parameter that defines the scale dependance of each wavelet (for a standard dyadic scaling $\lambda=2$), while the scaling function $\Upsilon$ has support in the range
$\ell < \lambda^{J_{\rm min}}$ (see \cite{mcewen:s2let_spin,mcewen:s2let_localisation,leistedt:s2let_axisym,wiaux:2007:sdw} for further details).  Consequently, $L^j=\lambda^{j+1}$ and $L_{\rm s}=\lambda^{J_{\rm min}}$. The azimuthal band limit of a wavelet scale is limited by the overall azimuthal band limit or the band limit of that scale, therefore $N^j = \min(N,L^j)$.

We collect the harmonic representation of all wavelet and scaling coefficients in a single vector:
\ba
  \vect{\hat{\alpha}} &=& \left[\hat{\vect{s}}^\dagger, (\hat{\vect{w}}^{J_{\rm min}}){}^\dagger\!, 
  (\hat{\vect{w}}^{{J_{\rm min}+1}}){}^\dagger, \ldots,
   (\hat{\vect{w}}^{{J_{\rm max}}}{})^\dagger\right]^{\dagger}\\
  &=&  
    [\bmtrx{S} ^\dagger, 
    (\bmtrx{N}^{J_{\rm min}}\bmtrx{W}^{J_{\rm min}})^\dagger,
    (\bmtrx{N}^{J_{\rm min}+1}\bmtrx{W}^{J_{\rm min}+1})^\dagger,\nonumber\\
    &&\qquad\qquad\dots,
    (\bmtrx{N}^{J_{\rm max}}\bmtrx{W}^{J_{\rm max}})^\dagger]^\dagger
  \hat{\vect{x}}\\
  &=& \bmtrx{N}\bmtrx{W} \vect{\hat{x}},
\ea
where $\cdot^{\dagger}$ denotes the Hermitian transpose or \linebreak adjoint,
\mbox{$\bmtrx{N} = \diag{\bmtrx{I}_{L_{\rm s}}, \bmtrx{N}^{J_{\rm min}}, \bmtrx{N}^{J_{\rm min}+1}, \ldots, \bmtrx{N}^{J_{\rm max}}}$}, and
\mbox{$\bmtrx{W} = \diag{\bmtrx{S}, \bmtrx{W}^{J_{\rm min}}, \bmtrx{W}^{J_{\rm min}+1}, \ldots, \bmtrx{W}^{J_{\rm max}}}$}.
The collection of scaling and wavelet coefficients can be calculated from their harmonic representations by a series of inverse spherical harmonic and Wigner transforms by
\ba
  \vect{{\alpha}} &=& \left[{\vect{s}}^\dagger, ({\vect{w}}^{J_{\rm min}}){}^\dagger\!, 
  ({\vect{w}}^{{J_{\rm min}+1}}){}^\dagger, \ldots,
   ({\vect{w}}^{{J_{\rm max}}}{})^\dagger\right]^{\dagger}\\
  &=& \bmtrx{H} \vect{\hat{\alpha}},
\ea
where
$\bmtrx{H} = \diag{\bmtrx{Y}, \bmtrx{D}^{J_{\rm min}}, \bmtrx{D}^{J_{\rm min}+1}, \ldots, \bmtrx{D}^{J_{\rm max}}}$.

The forward wavelet transform, denoted by the operator $\tilde{\bmtrx{\Psi}}$, can then be expressed by the concatenation of operators defined above, yielding
\begin{equation} \label{eqn:wavelet_forward_discrete}
  \vect{\alpha} = \tilde{\bmtrx{\Psi}} \vect{x}= \bmtrx{H}\bmtrx{N}\bmtrx{W}\tilde{\bmtrx{Y}} \vect{x} .
\end{equation}
In other words, the wavelet transform $\tilde{\bmtrx{\Psi}}$ is composed of a spherical harmonic transform $\tilde{\bmtrx{Y}}$, harmonic wavelet multiplication $\bmtrx{W}$, harmonic normalisation $\bmtrx{N}$, and inverse spherical harmonic and Wigner transforms $\bmtrx{H}$. 

The inverse wavelet transform, denoted by the operator ${\bmtrx{\Psi}}$ can be represented in a similar manner.  Firstly, we note the spherical harmonic and Wigner coefficients of the wavelet and scaling coefficients can be calculated by a series of forward harmonic transforms by
\begin{equation}
  \vect{\hat{\alpha}} = \tilde{\bmtrx{H}} \vect{\alpha} ,
\end{equation}
where
$\tilde{\bmtrx{H}} = \diag{\tilde{\bmtrx{Y}}, \tilde{\bmtrx{D}}^{J_{\rm min}}, \tilde{\bmtrx{D}}^{J_{\rm min}+1}, \ldots, \tilde{\bmtrx{D}}^{J_{\rm max}}}$.  From \eqn{\ref{eqn:wavelet_forward_discrete}}, the inverse wavelet transform reads
\begin{equation} \label{eqn:wavelet_inverse_discrete}
  \vect{x} = \bmtrx{\Psi} \vect{\alpha} 
  = \bmtrx{Y} (\bmtrx{NW})^{-1}\tilde{\bmtrx{H}}\vect{\alpha}
  = \bmtrx{Y} \bmtrx{W}^{\dagger}\tilde{\bmtrx{H}}\vect{\alpha},
\end{equation}
where the final equality follows by noting $(\bmtrx{NW})^{-1} = \bmtrx{W}^\dagger$, which can be inferred from \eqn{\ref{eq:wavelet_inv_trans_harm}}, which in turn follows by the wavelet admissibility criterion.  In other words, the inverse wavelet transform $\bmtrx{\Psi}$ is composed of forward spherical harmonic and Wigner transforms $\tilde{\bmtrx{H}}$, harmonic wavelet multiplication and summation $\bmtrx{W}^{\dagger}$, and an inverse spherical harmonic transform $\bmtrx{Y}$.

\subsection{Adjoints of Discrete Wavelet Analysis and Synthesis}

The operator descriptions of the forward and inverse wavelet transforms formulated in the previous subsection permit a straightforward construction of the adjoint operators, which are required to solve inverse problems on the sphere when imposing sparsity in wavelet space.  
From \eqn{\ref{eqn:wavelet_forward_discrete}}, the wavelet transform operator reads
\mbox{$\tilde{\bmtrx{\Psi}} = \bmtrx{H}\bmtrx{N}\bmtrx{W}\tilde{\bmtrx{Y}}$}, with corresponding adjoint:
\begin{equation} \label{eq:Psi_adjoint}
  \tilde{\bmtrx{\Psi}}^\dagger  = \tilde{\bmtrx{Y}}^\dagger \bmtrx{W}^\dagger \bmtrx{N} \bmtrx{H}^\dagger ,
\end{equation}
where we have noted that $\bmtrx{N}$ is self-adjoint, \ie\ $\bmtrx{N}^\dagger = \bmtrx{N}$.
From \eqn{\ref{eqn:wavelet_inverse_discrete}}, the inverse wavelet transform operator reads
$\bmtrx{\Psi} = \bmtrx{Y} \bmtrx{W}^{\dagger}\tilde{\bmtrx{H}}$, with corresponding adjoint:
\begin{equation} \label{eq:Psi_inv_adjoint}
  \bmtrx{\Psi}^\dagger =  \tilde{\bmtrx{H}}^\dagger \bmtrx{W} \bmtrx{Y}^\dagger .
\end{equation}

In the discrete setting, recall that the adjoint and inverse operators are not equivalent, \ie\ $\tilde{\bmtrx{Y}}^\dagger \neq {\bmtrx{Y}}$, $\tilde{\bmtrx{H}}^\dagger \neq {\bmtrx{H}}$, and \mbox{$\tilde{\bmtrx{\Psi}}^\dagger \neq {\bmtrx{\Psi}}$}.  Consequently, for practical application, fast algorithms must be developed to compute not only forward transforms and their inverses but also the adjoints of both the forward and inverse transforms.  This has been performed already \cite{mcewen:css2} for the spherical \referee{harmonic} transforms of the MW sampling theorem \cite{mcewen:fssht}, \ie\ to apply $\tilde{\bmtrx{Y}}^\dagger$ and ${\bmtrx{Y}}^\dagger$.  To apply directional wavelets, we also require fast algorithms to compute adjoint Wigner transforms.  In Appendix \ref{app:fast_so3_adjoints}, we develop fast algorithms to compute $\tilde{\bmtrx{D}}^\dagger$ and ${\bmtrx{D}}^\dagger$ for the Wigner transforms corresponding to the sampling theorem on the rotation group of \cite{mcewen:so3}.


\begin{figure}
\begin{center}
\begin{tabular}{c}

\includegraphics[width=0.8\linewidth,  trim=2.5cm 4cm 2cm 3.37cm, clip=true]{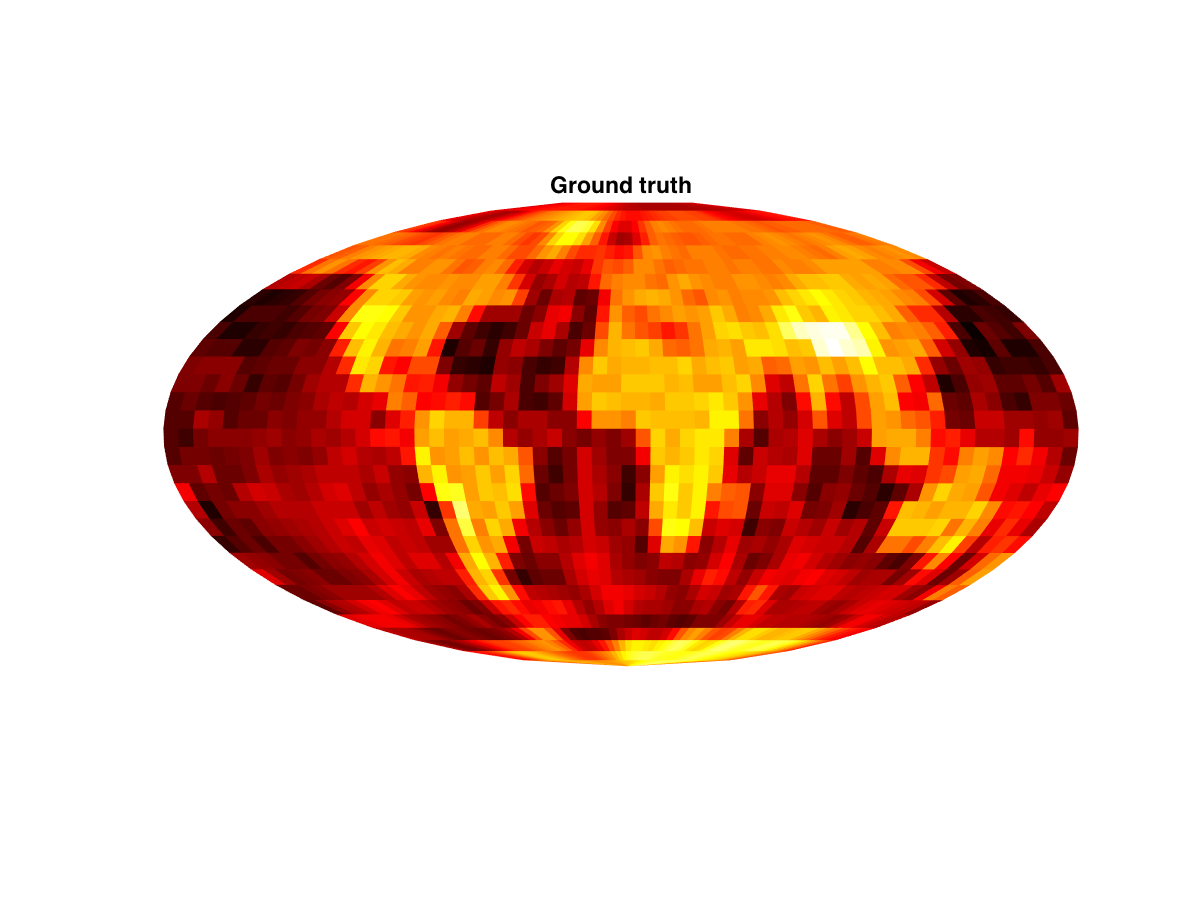}\\
\includegraphics[width=0.8\linewidth,  trim=2.5cm 4cm 2cm 3.37cm, clip=true]{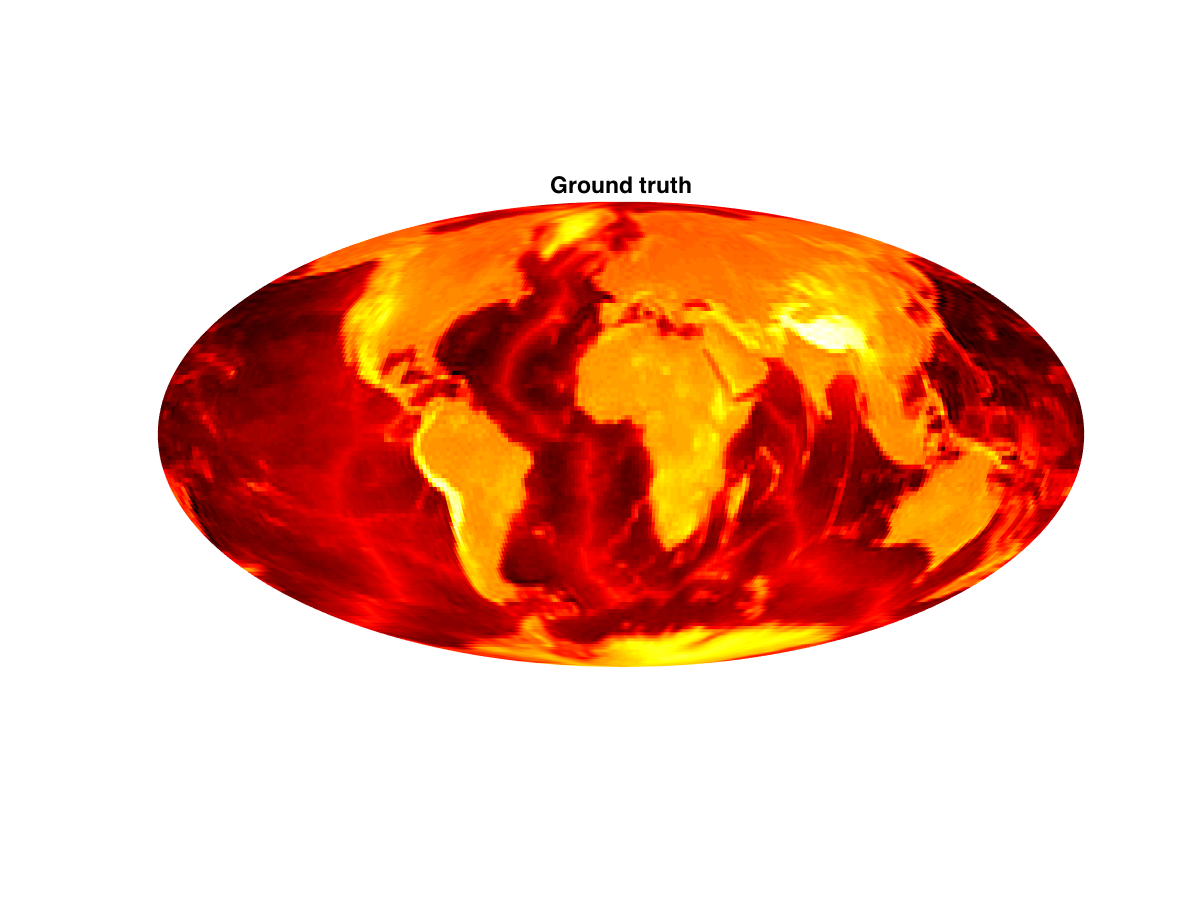}\\
\end{tabular}
\caption{Test images of Earth topographic data constructed to be band-limited at $L = 32$ (top) and $L=128$ (bottom).  These 
images constitute the ground truth in our numerical experiments. Here and subsequently data on the sphere are 
displayed using the Mollweide projection, with zero values shown in black, unit values shown in yellow, and the colour 
of intermediate values interpolated between these extremes.}
\label{fig:low_res_ground_truth}
\end{center}
\end{figure}

\subsection{Sparse Regularisation}\label{sec:sparce_reconstruction_on_the_sphere:spare_regularisation}
\label{sec:sparce_reconstruction_on_the_sphere:sparse_regularisation}

We consider linear, ill-posed inverse problems defined on the sphere,
including, for example, denoising, inpainting, and deconvolution problems. Consider $M$ measurements $\vect{y} \in \reals^M$ of the signal on the sphere \mbox{$\vect{x}\in \reals^{\nsphere}$}, acquired according to the measurement equation
\ba
\vect{y} = \bmtrx{\Phi}\vect{x} + \vect{n},\label{eq:measurements}
\ea
where $\bmtrx{\Phi} \in \reals^{M \times N_\sphere}$ is the measurement operator and \mbox{$\vect{n} \in \reals^{\nsphere}$} is measurement noise, assumed to be Gaussian, \ie\ $\vect{n} \sim \mathcal{N}(0,\sigma)$, where $\sigma = \| \hat{\vect{x}}\|_2 \times10^{(-{\rm SNR}/20)} $, \referee{where SNR is in dB}. For example, the measurement operator $\bmtrx{\Phi}$ may model the beam or point-spread function of a sensor (in a deconvolution problem) or a masking of the signal (in an inpainting problem).

We \referee{regularise} the \referee{spherical} ill-posed inverse problem of \eqn{\ref{eq:measurements}} by promoting sparsity in wavelet space by posing synthesis and analysis problems on the sphere.  The synthesis problem reads
\begin{equation}
\vect{\alpha}^\star = \arg \min_{\vect{\alpha}} \| \vect{\alpha}\|_1 ~~{\rm s.t.}~~\|\vect{y} - \bmtrx{\Phi}\bmtrx{\Psi}\vect{\alpha}\|_2 < \epsilon, \label{eq:prob_synthesis_real}
\end{equation}
where the signal is then recovered from its wavelet coefficients by $\vect{x}^\star = \bmtrx{\Psi} \vect{\alpha}^\star$. 
The analysis problems reads 
\begin{equation}
\vect{x}^\star = \arg \min_{\vect{x}} \| \tilde{\bmtrx{\Psi}}\vect{x}\|_1 ~~{\rm s.t.}~~\|\vect{y} - \bmtrx{\Phi}\vect{x}\|_2 < \epsilon, \label{eq:prob_analysis_real}
\end{equation}
where we recover the signal $\vect{x}^\star$ directly.\footnote{Our framework can be applied to spin $\spin \in \integers$ signals on the sphere (see \eg\ \cite{mcewen:fssht}) in a straightforward manner, noting that the spin wavelet transform of \cite{mcewen:s2let_spin} can be represented by the operators
\mbox{${}_\spin \tilde{\bmtrx{\Psi}} = {}_\spin\bmtrx{H}\bmtrx{N}\bmtrx{W}{}_\spin\tilde{\bmtrx{Y}}$} and $\bmtrx{\Psi} = {}_\spin\bmtrx{Y} \bmtrx{W}^{\dagger} {}_\spin\tilde{\bmtrx{H}}$, where 
the forward and inverse scalar spherical harmonic transforms are replaced by spin versions, \eg\ replacing $\bmtrx{Y}$ by ${}_\spin\bmtrx{Y}$. }

The $\ell_1$ norm appearing in the sparsity constraint must be defined appropriately for the spherical setting \cite{mcewen:css2}, as discussed in more detail below.  The square of the residual noise follows a scaled $\chi^2$
distribution with $M$ degrees of freedom, \ie\ $\| \vect{y} -
\bmtrx{\Psi} \vect{x}^\star\|_2^2 \sim \sigma^2 
\chi^2(\nmeas)$.  Consequently, we choose $\epsilon$
to correspond to a given percentile of the $\chi^2$ distribution \cite{mcewen:css2}.

\analysissynthesis{When solving the synthesis and analysis problems of \eqn{\ref{eq:prob_synthesis_real}} and \eqn{\ref{eq:prob_analysis_real}} we are free to choose different sampling schemes (\eg\ KKM, MW, DH sampling).  In Euclidean space, the analysis problem has shown promising results in empirical studies, which we recall is hypothesised to be due to the more restrictive solution-space of the analysis setting.  This relationship between the size of the solution-space and the analysis and synthesis settings does not in general carry over to the spherical setting since on the sphere sampling is not typically optimal.  Consequently, recovering the signal directly in the analysis setting does not necessarily lead to the most restrictive solution-space.  The most restrictive solution-space depends on the interplay between the adopted sampling scheme, the selection of the analysis/synthesis problem, and any weighting of the $\ell_1$ norm, which is made explicit in the following subsection.}

\subsection{Algorithmic Details}

The $\ell_1$ norm appearing in \eqn{\ref{eq:prob_synthesis_real}} and \eqn{\ref{eq:prob_analysis_real}} must be defined appropriately for the spherical setting, taking into account the sampling scheme adopted.  In \cite{mcewen:css2}, where the total variation (TV) norm is considered, the associated discrete TV norm is weighted by the exact quadrature weights of the sampling theorem adopted in order to approximate the continuous norm.  Through numerical experiments we have found the $\ell_1$ norm, and the solution of the sparse reconstruction problems, to be relatively insensitive to the exact form of weights: provided weights capture the area of each pixel  the underlying continuous norm is well approximated and it is not necessary to use exact quadrature weights.\footnote{\referee{Accounting for the area of each pixel in the definition of the $\ell_1$ norm is similar to the zeroth order $\ell_1$ norm approximation for functions defined on general manifolds considered in \cite{bronstein2016}.}} Consequently, for all sampling schemes we adopt the following weights for the wavelet and scaling coefficients, respectively, corresponding to scale $j$ and pixel $p$:
\ba
v^j_p &=& \frac{(\lambda^{j})^{\eta}}{\sum_{\ell m} \vert \Psi^j_{\ell m}\vert^2}\frac{4\pi^3\sin\theta_p}{n^j_\phi n^j_\theta n^j_\psi},\label{eq:weights_wavelets}\\
u_{p} &=& \frac{1}{\sum_{\ell m} \vert\Upsilon_{\ell m}\vert^2}\frac{2\pi^2\sin\theta_p}{n^j_\phi n^j_\theta},\label{eq:weights_scalar}
\ea
where $n^j_\phi$, $n^j_\theta$ and $n^j_\psi$ are the number of samples in the $\phi$, $\theta$ and $\psi$ directions, $\theta_p$ is the
$\theta$ coordinate of the sample $p$, and $\eta\in\realsnn$ is a decay parameter.

The weights approximate the area of each pixel, normalised by the energy of the wavelet and scaling function for the given scale $j$.  Furthermore, for the weighting of wavelet coefficients the additional factor $(\lambda^{j})^{\eta}$ is introduced.  The term $\lambda^{j}$ corresponds to the middle harmonic multipole $\ell$ to which the wavelet $\Psi^j$ is sensitive, while $\eta$ is introduced as a free parameter to incorporate prior knowledge of natural signals, \ie\ to control the wavelet decay imposed as a prior when solving the sparse regularisation problems.  Increasing $\eta$ promotes large scale features by increasing the weight applied to small wavelet scales, thereby increasing their penalty to the $\ell_1$ norm.  Moreover, for the synthesis setting, increasing $\eta$ reduces the effective size of the solution-space.

We use the \referee{Douglas}-Rachford (DR) \cite{combettes:2011} splitting algorithm to solve the sparse regularisation problems posed in \sectn{\ref{sec:sparce_reconstruction_on_the_sphere:sparse_regularisation}}. \referee{The Douglas-Rachford algorithm \cite{combettes:2011} is based on a splitting
approach that requires the computation of two proximity operators. In our case, one proximity operator is based
on the $\ell_1$ norm and the other on the data fidelity constraint.  The adaptation of the DR algorithm to the sphere is discussed further in \cite{mcewen:css2}.}
The DR algorithm requires the adjoint of the operators that appear in the problem specification, \eg\ the adjoint  sparsifying operators shown in \eqn{\ref{eq:Psi_adjoint}} and \eqn{\ref{eq:Psi_inv_adjoint}}.  In numerical experiments, if inverse operators are used in place of the adjoints, we have seen convergence failures. In \cite{mcewen:css2}, fast adjoints for the spherical harmonic transform corresponding to the MW sampling scheme \cite{mcewen:fssht} were dervied.  In an analogous manner, we derive in Appendix \ref{app:fast_so3_adjoints} fast adjoints for the Wigner transforms of \cite{mcewen:so3}. \referee{These efficient adjoints have a numeric complexity of order $NL^3$ compared to the naive adjoint operations which have complexity of order $N^2L^4$.}
The power method is used to calculate the norms of the operators (required when solving the optimisation problems).


\begin{figure*}
\begin{center}
\begin{tabular}{r c c c}

\includegraphics[width=0.2\linewidth,  trim=2.5cm 4cm 2cm 3.37cm, clip=true]{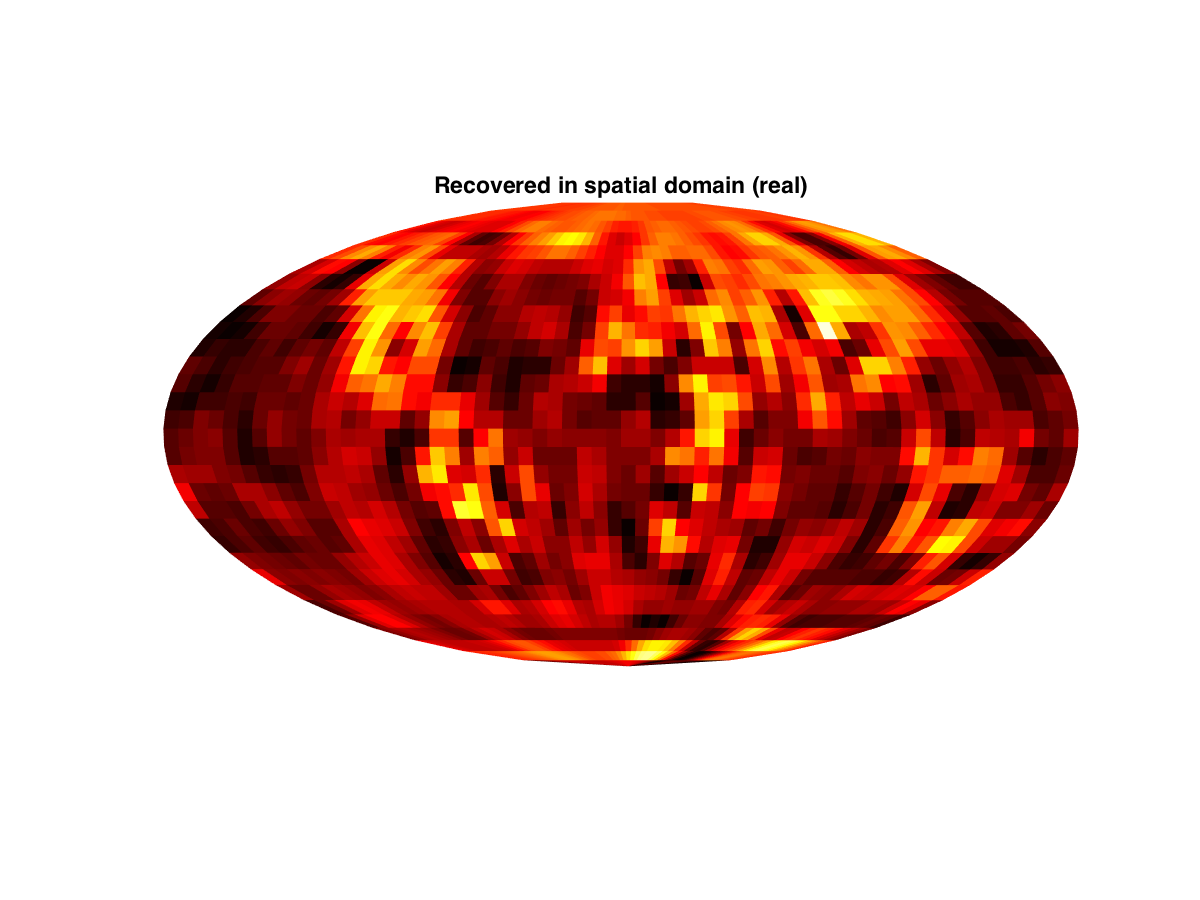}&
\includegraphics[width=0.2\linewidth,  trim=2.5cm 4cm 2cm 3.37cm, clip=true]{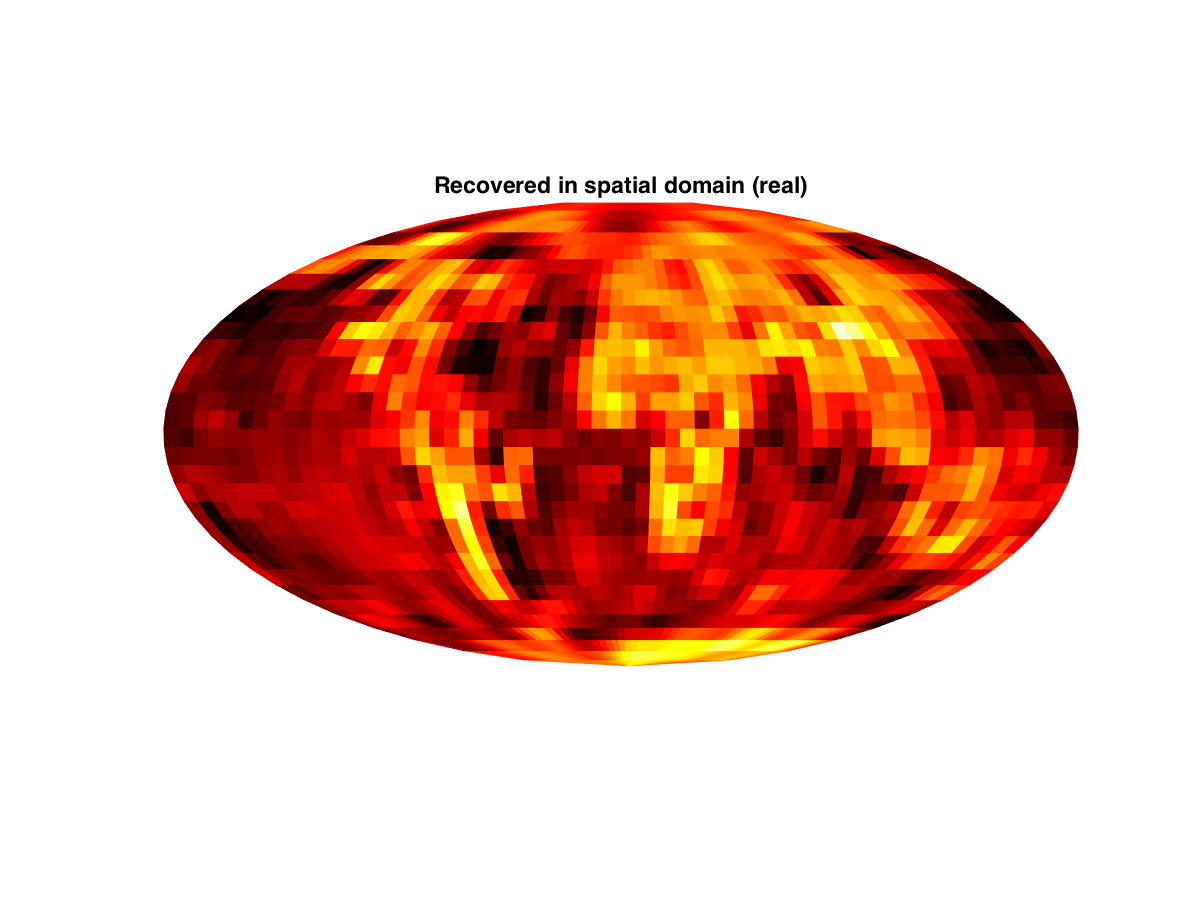}&
\includegraphics[width=0.2\linewidth,  trim=2.5cm 4cm 2cm 3.37cm, clip=true]{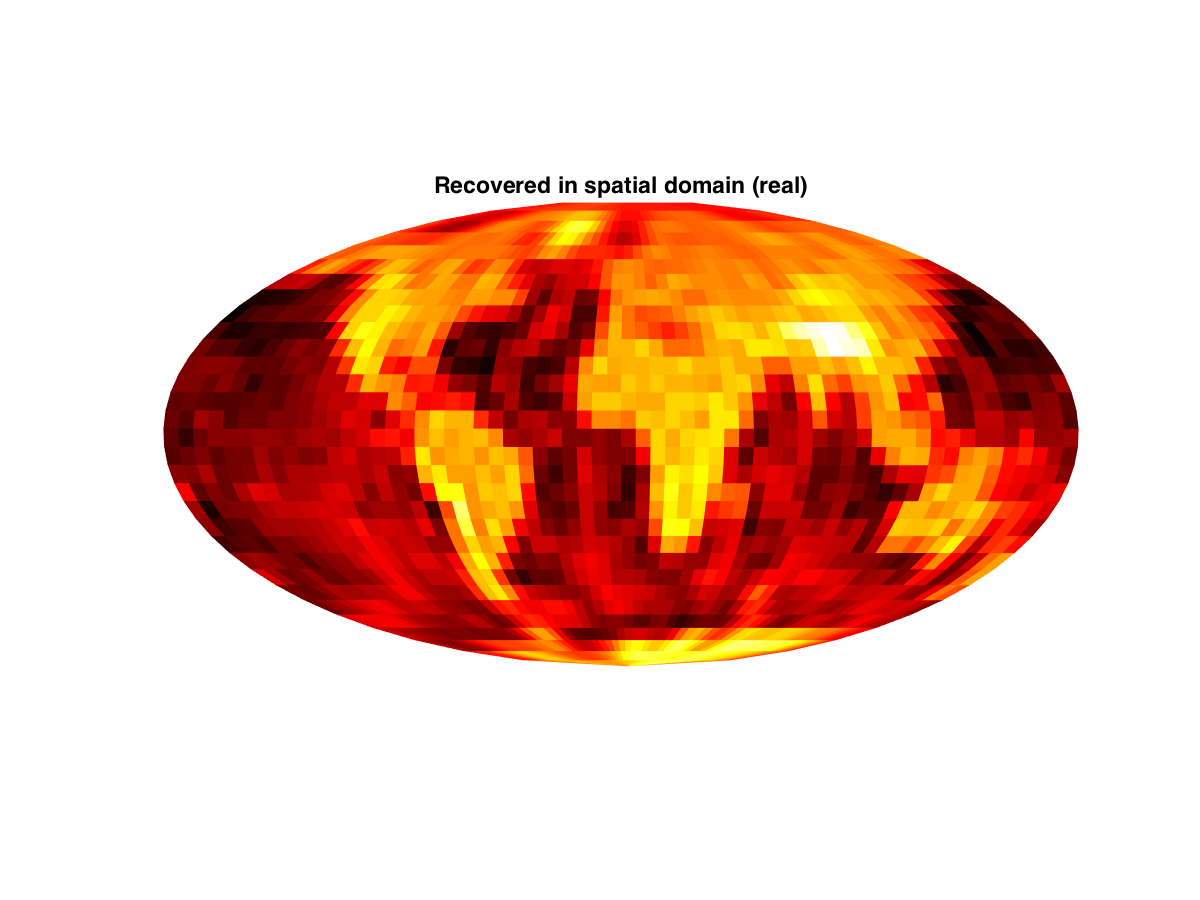}&\\
\shortstack{KKM Analysis, $M/L^2=0.3$ (15.8)}& \shortstack{$M/L^2=0.5$ (25.8)}& \shortstack{$M/L^2=1.0$ (64.3)}&\\

\includegraphics[width=0.2\linewidth,  trim=2.5cm 4cm 2cm 3.37cm, clip=true]{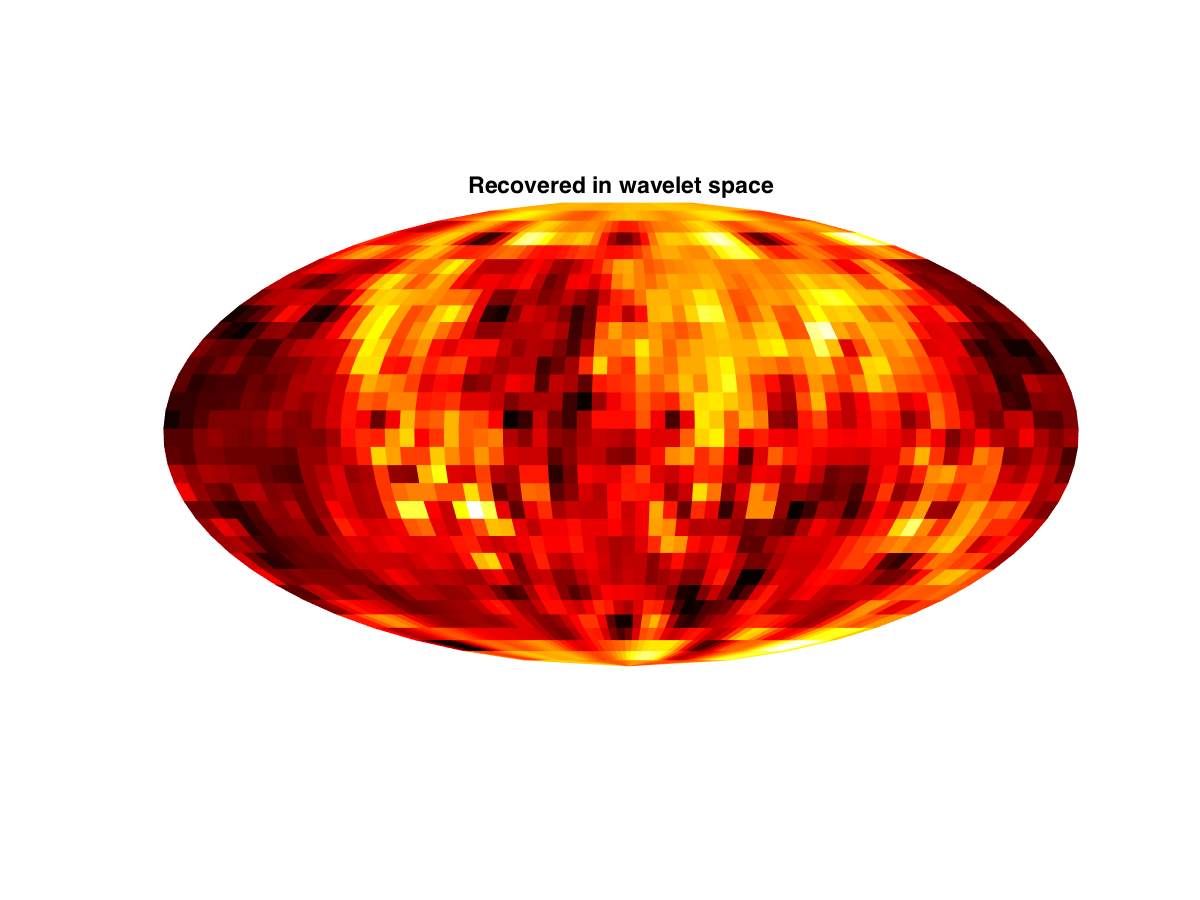}&
\includegraphics[width=0.2\linewidth,  trim=2.5cm 4cm 2cm 3.37cm, clip=true]{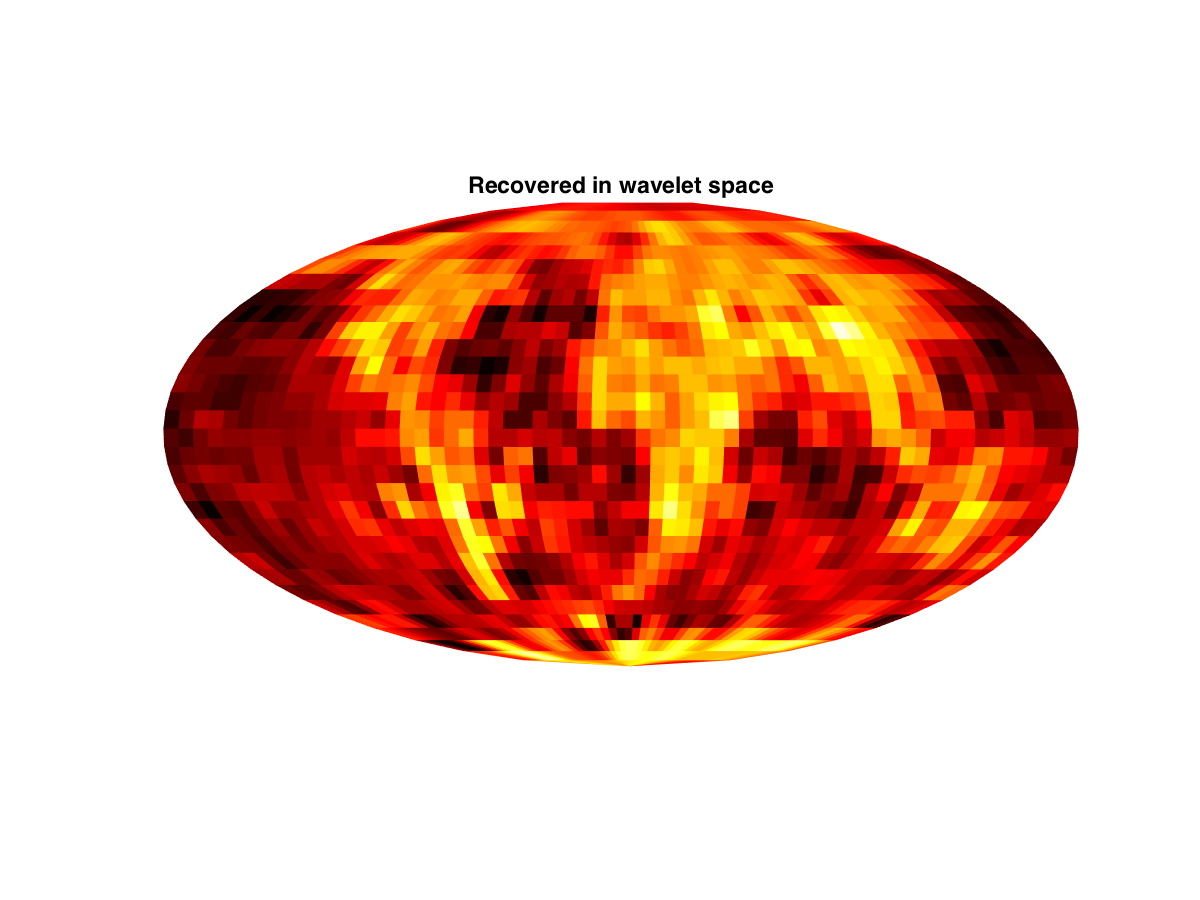}&
\includegraphics[width=0.2\linewidth,  trim=2.5cm 4cm 2cm 3.37cm, clip=true]{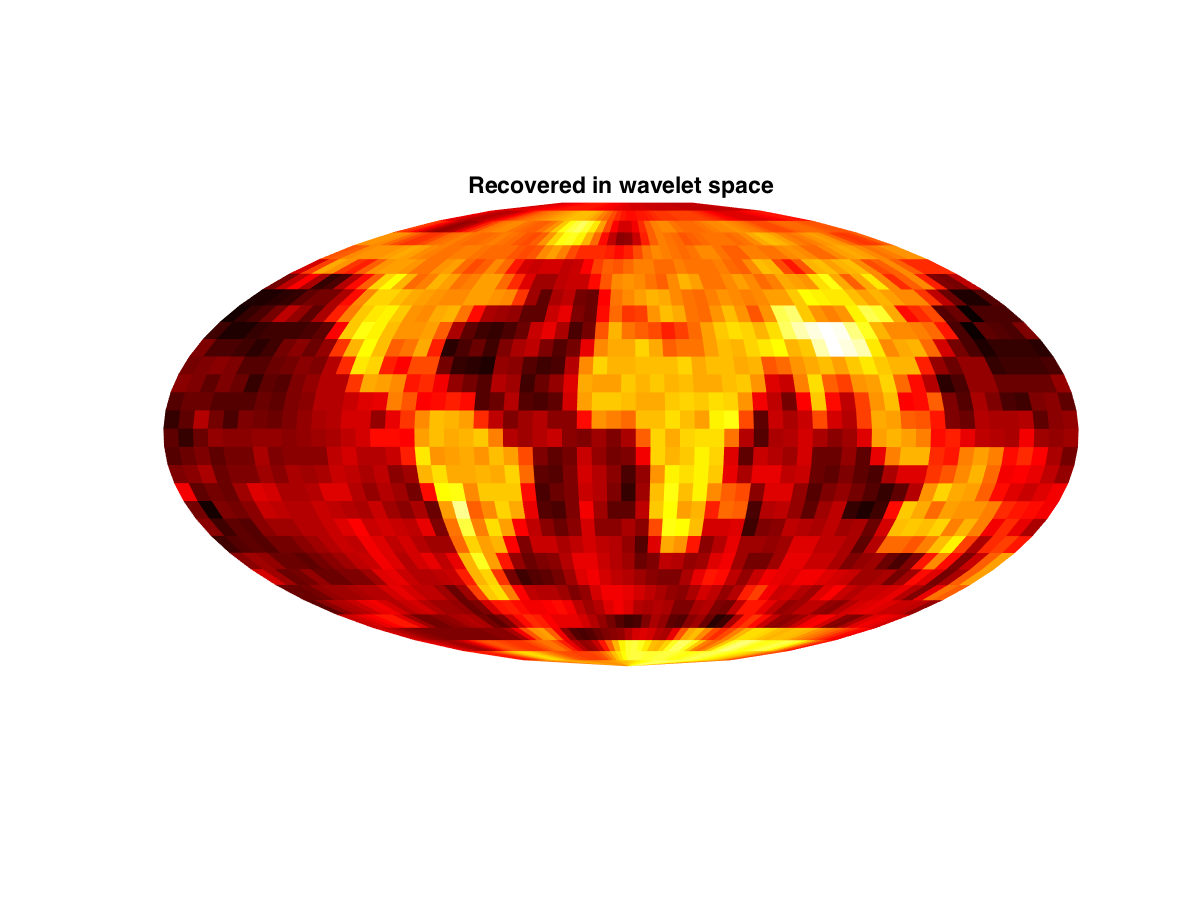}&\\
\shortstack{KKM Synthesis, $M/L^2=0.3$ (20.8)}& \shortstack{$M/L^2=0.5$ (28.5)}& \shortstack{$M/L^2=1.0$ (54.6)}&\\

\includegraphics[width=0.2\linewidth,  trim=2.5cm 4cm 2cm 3.37cm, clip=true]{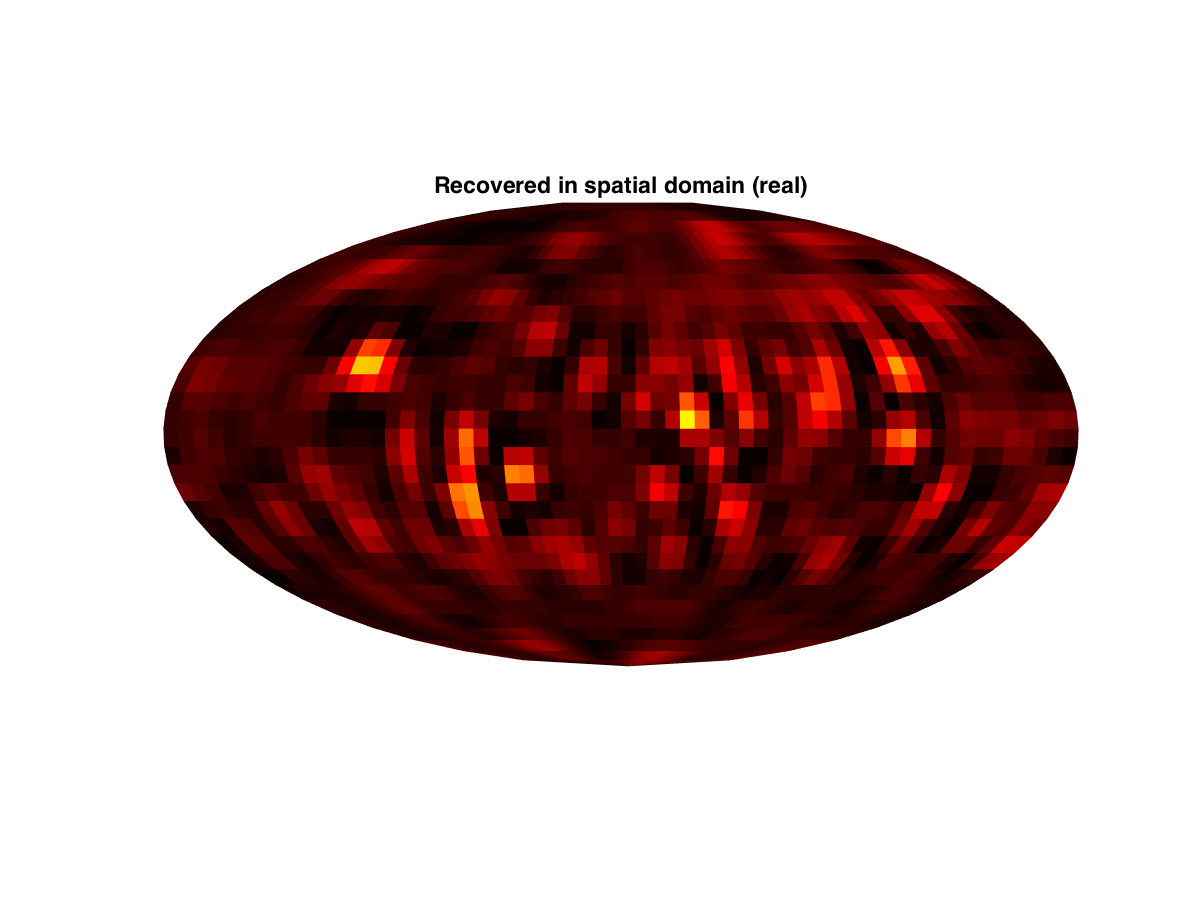}&
\includegraphics[width=0.2\linewidth,  trim=2.5cm 4cm 2cm 3.37cm, clip=true]{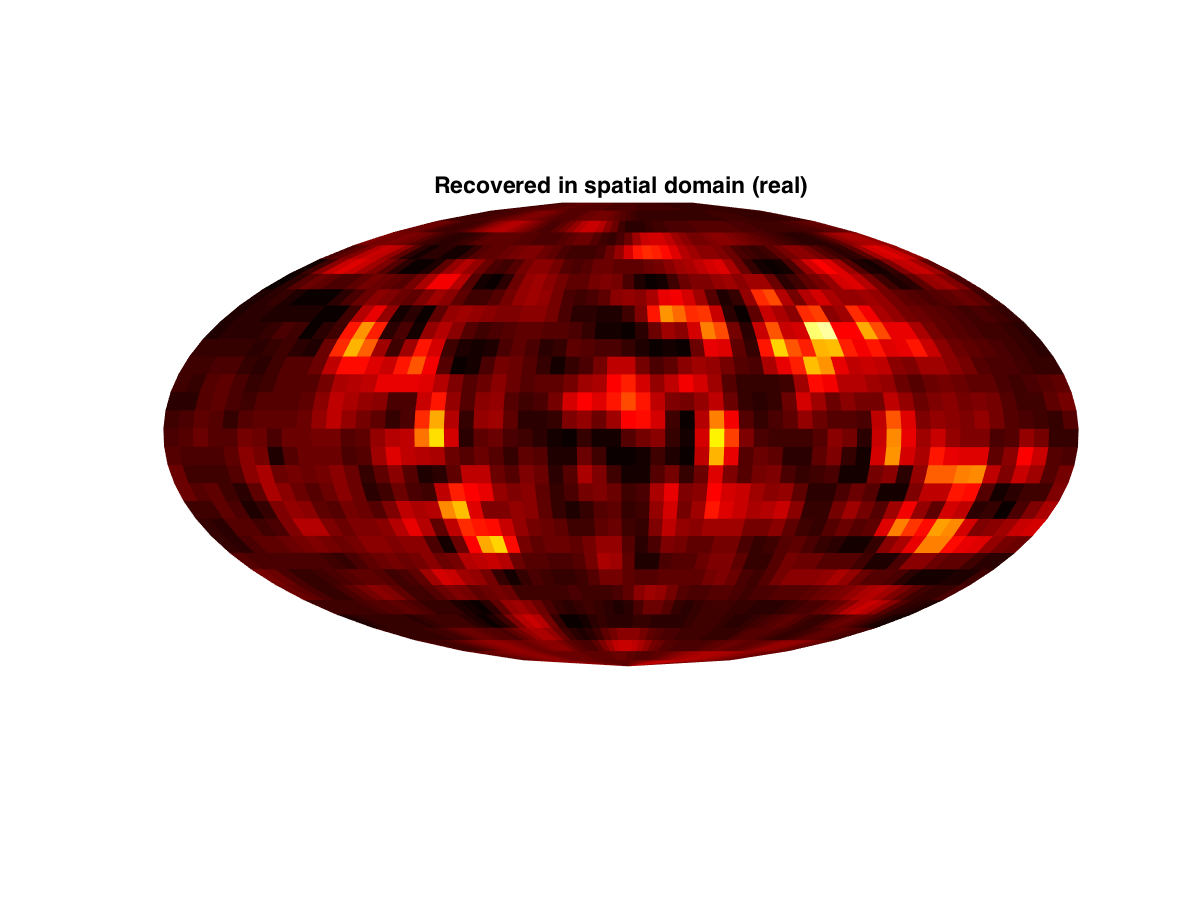}&
\includegraphics[width=0.2\linewidth,  trim=2.5cm 4cm 2cm 3.37cm, clip=true]{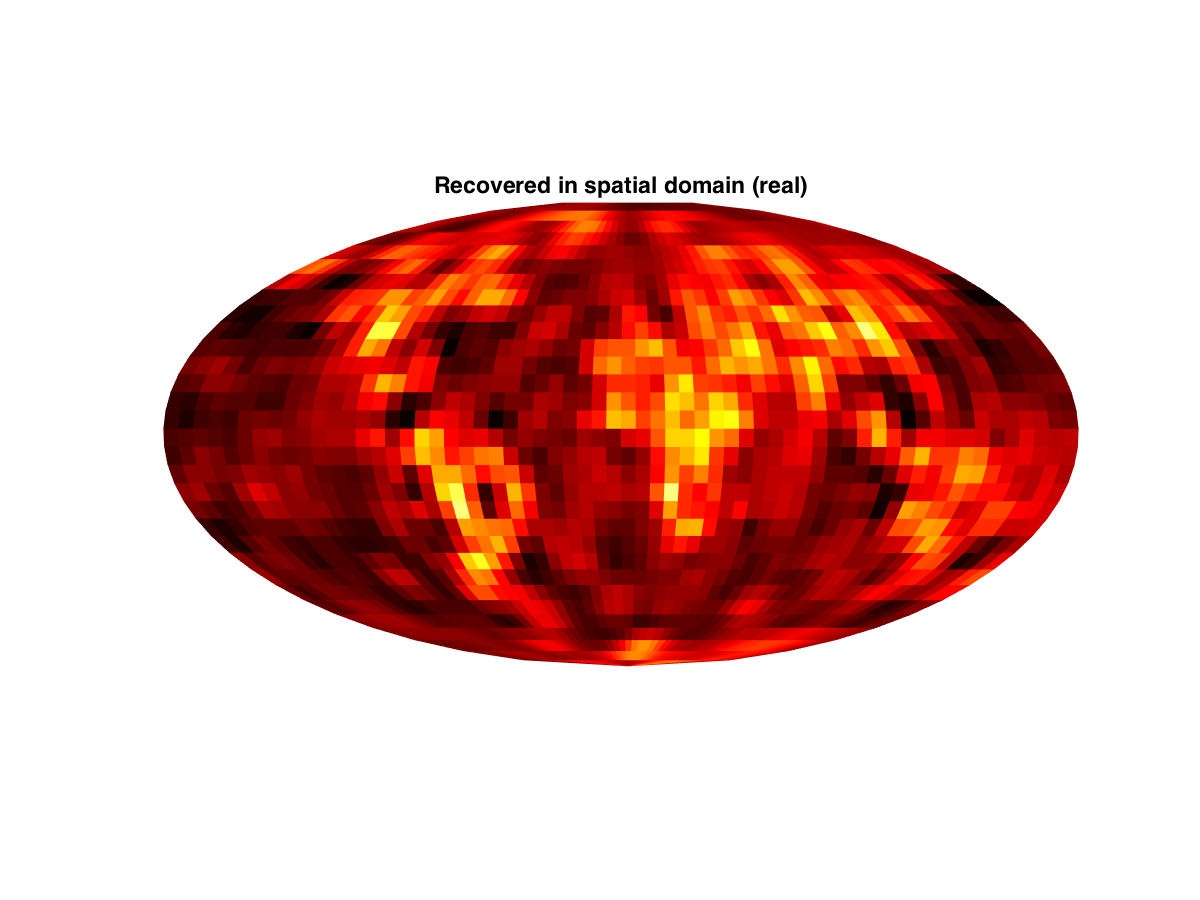}&
\includegraphics[width=0.2\linewidth,  trim=2.5cm 4cm 2cm 3.37cm, clip=true]{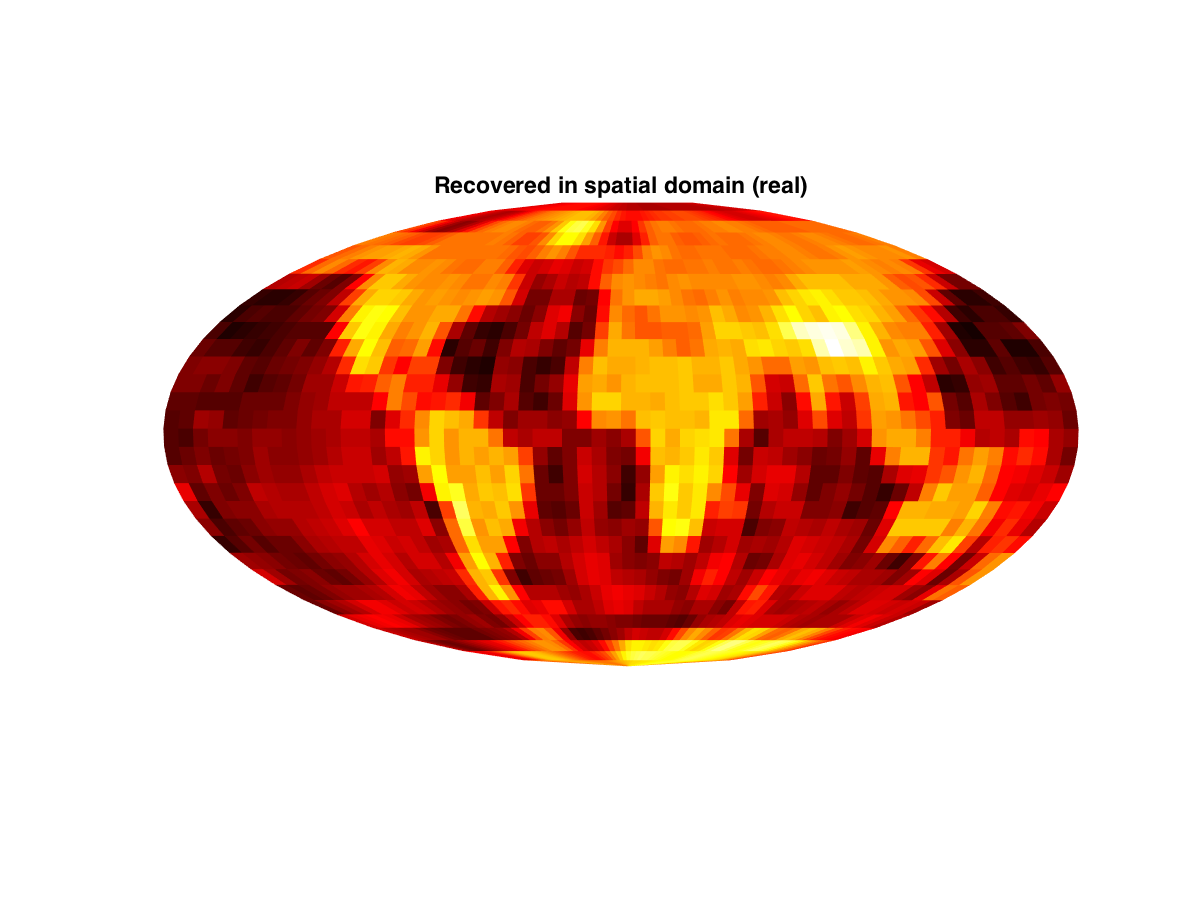}\\
MW Analysis, $M/L^2=0.3$ (5.2)& $M/L^2=0.5$ (8.9)& $M/L^2=1.0 (18.8)$& $M/L^2=1.9 (59.8)$\\

\includegraphics[width=0.2\linewidth,  trim=2.5cm 4cm 2cm 3.37cm, clip=true]{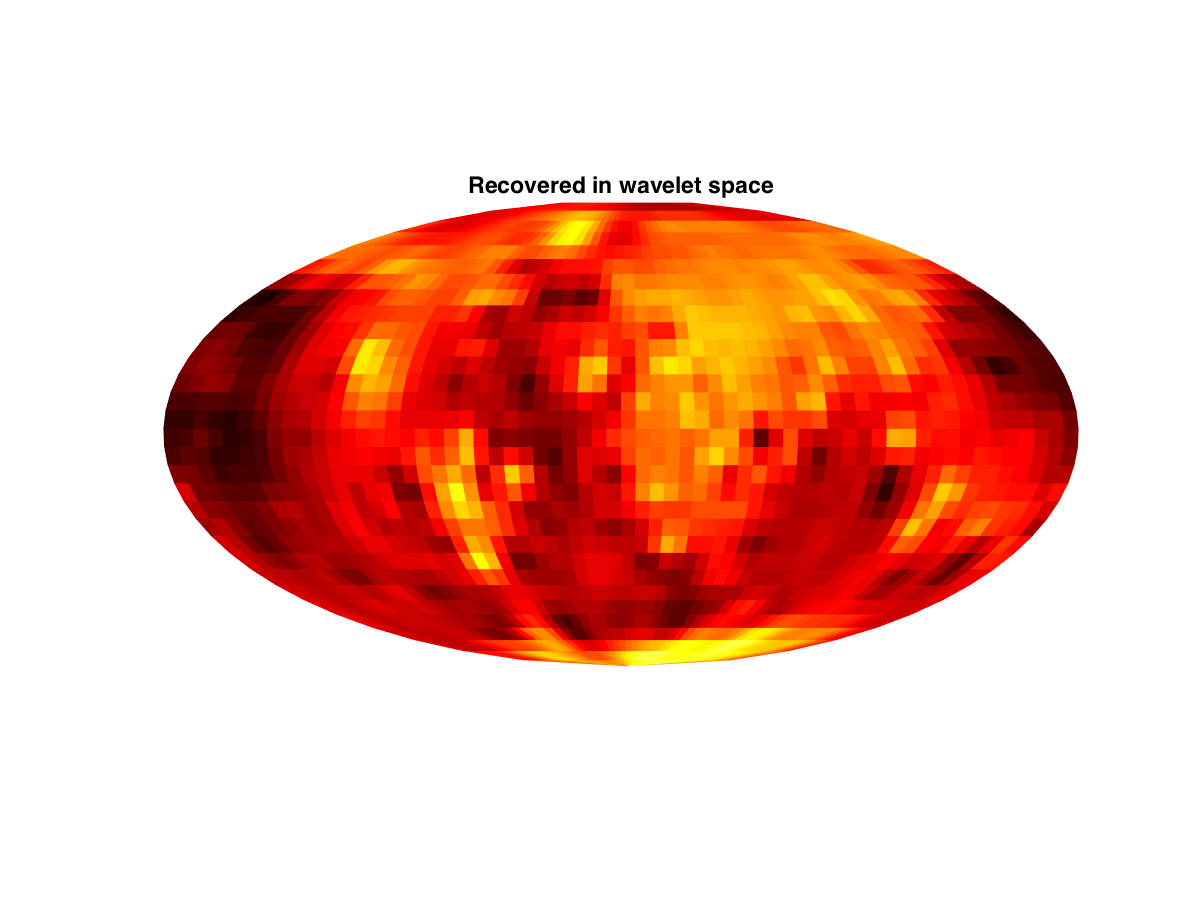}&
\includegraphics[width=0.2\linewidth,  trim=2.5cm 4cm 2cm 3.37cm, clip=true]{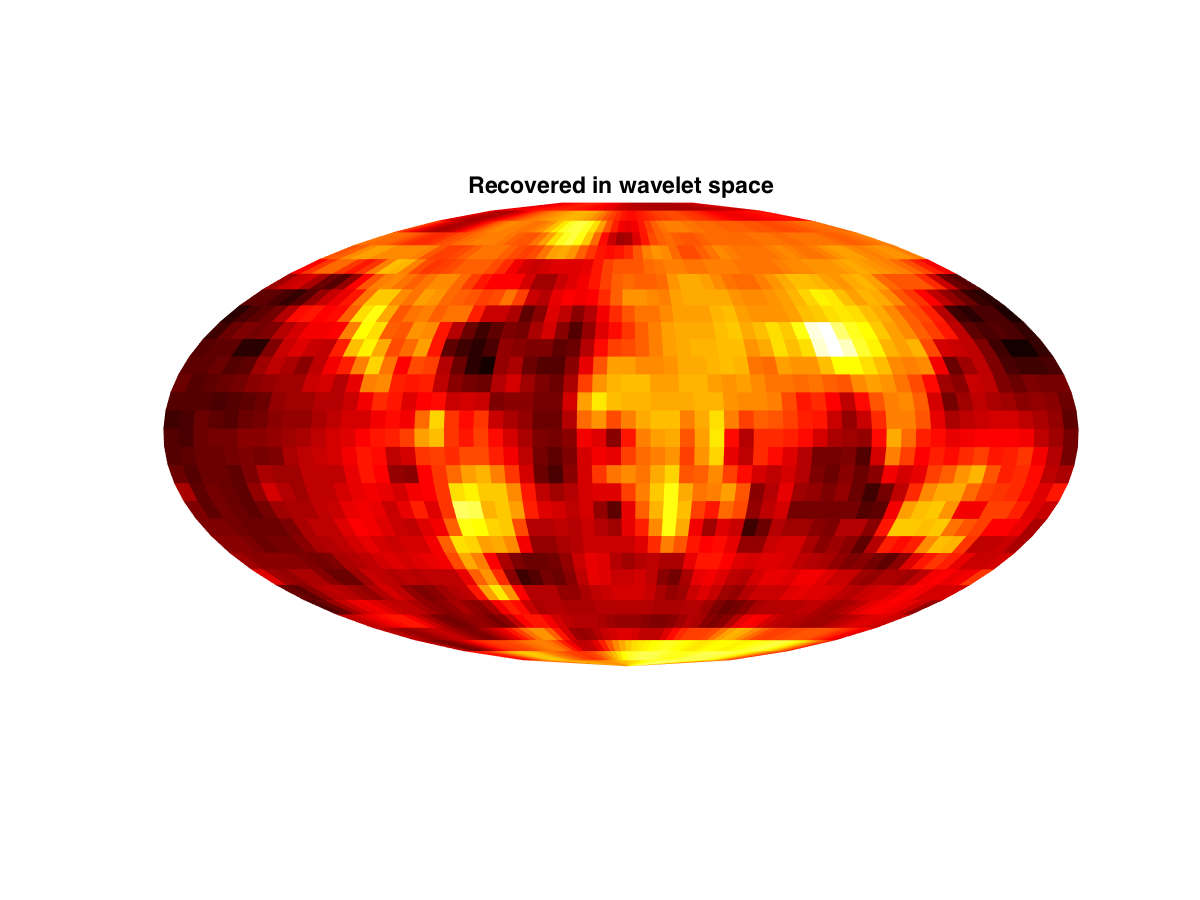}&
\includegraphics[width=0.2\linewidth,  trim=2.5cm 4cm 2cm 3.37cm, clip=true]{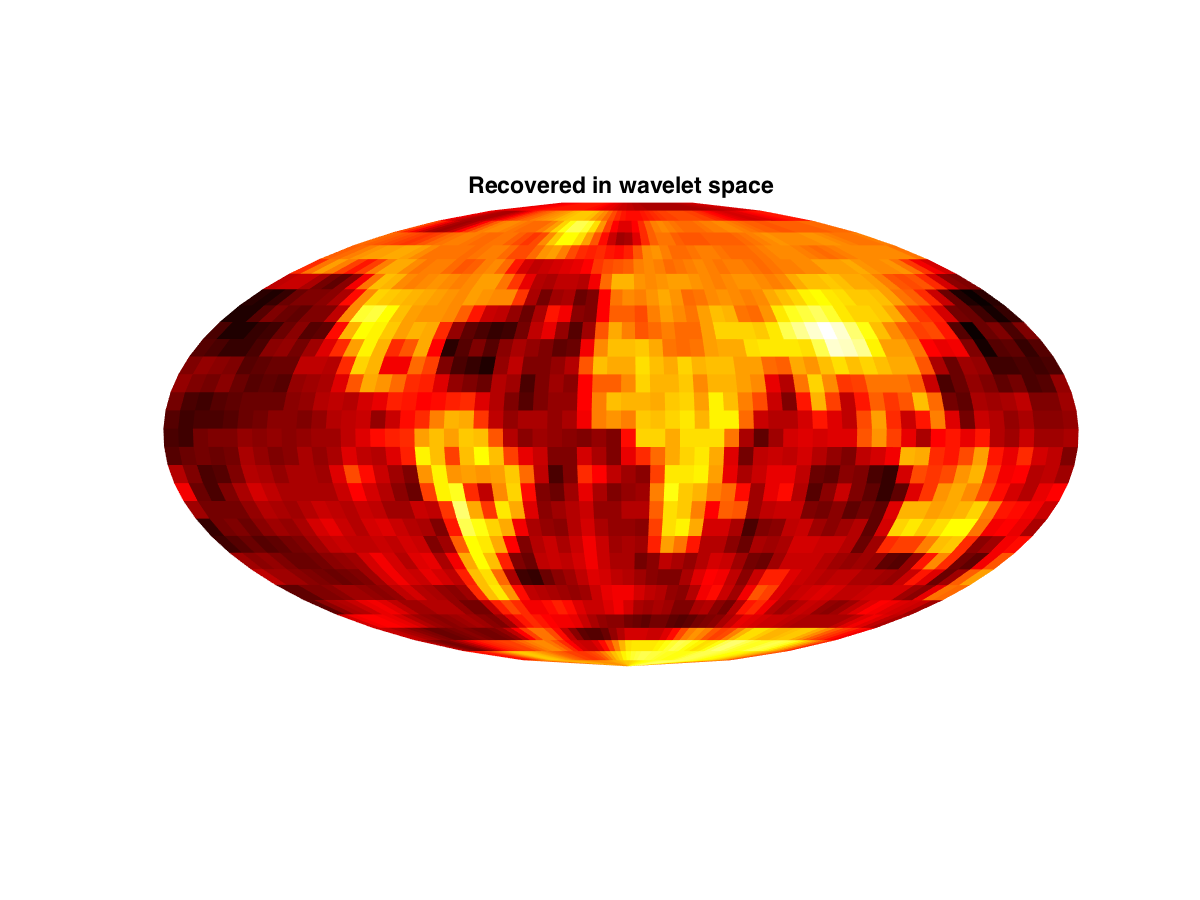}&
\includegraphics[width=0.2\linewidth,  trim=2.5cm 4cm 2cm 3.37cm, clip=true]{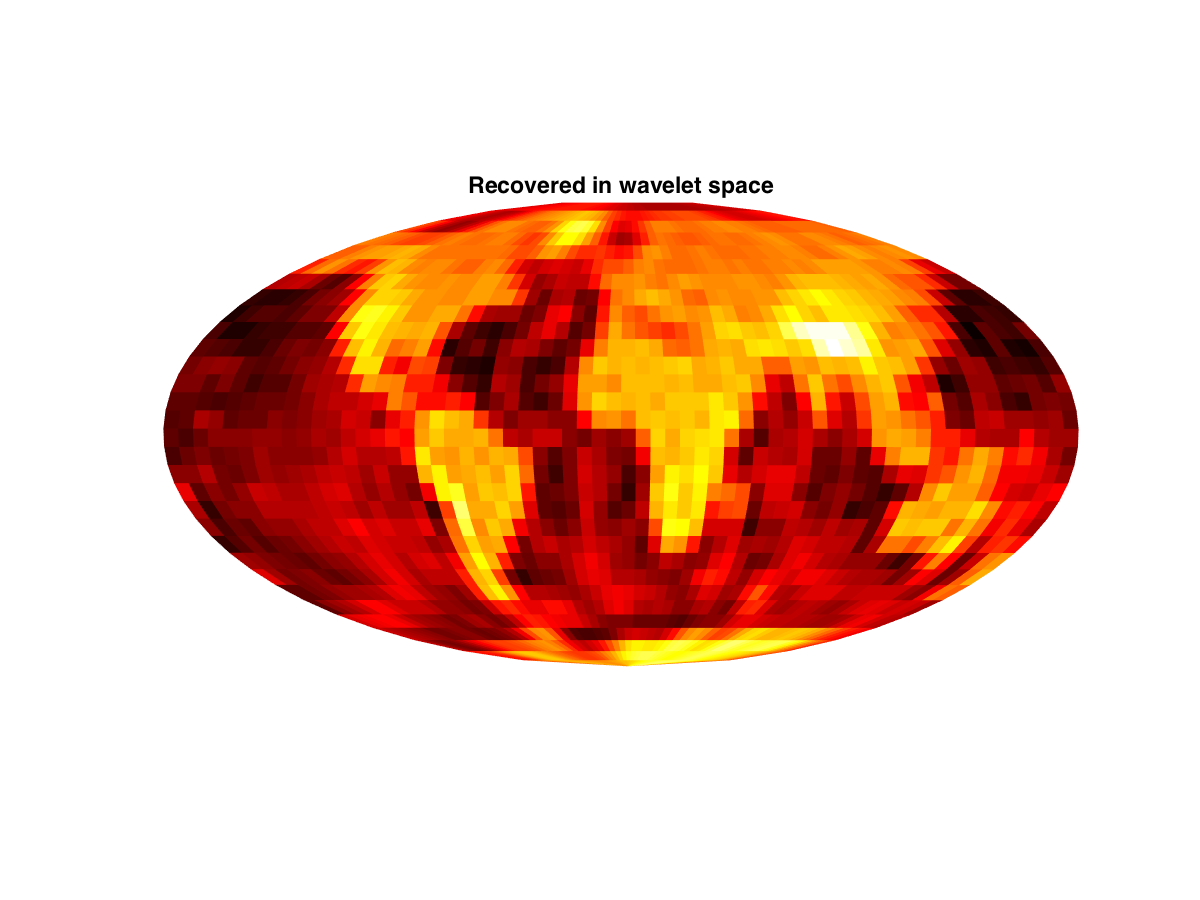}\\
MW Synthesis, $M/L^2=0.3$ (26.2)& $M/L^2=0.5$ (31.9)& $M/L^2=1.0$ (42.0)& $M/L^2=1.9$ (76.4)\\

\includegraphics[width=0.2\linewidth,  trim=2.5cm 4cm 2cm 3.37cm, clip=true]{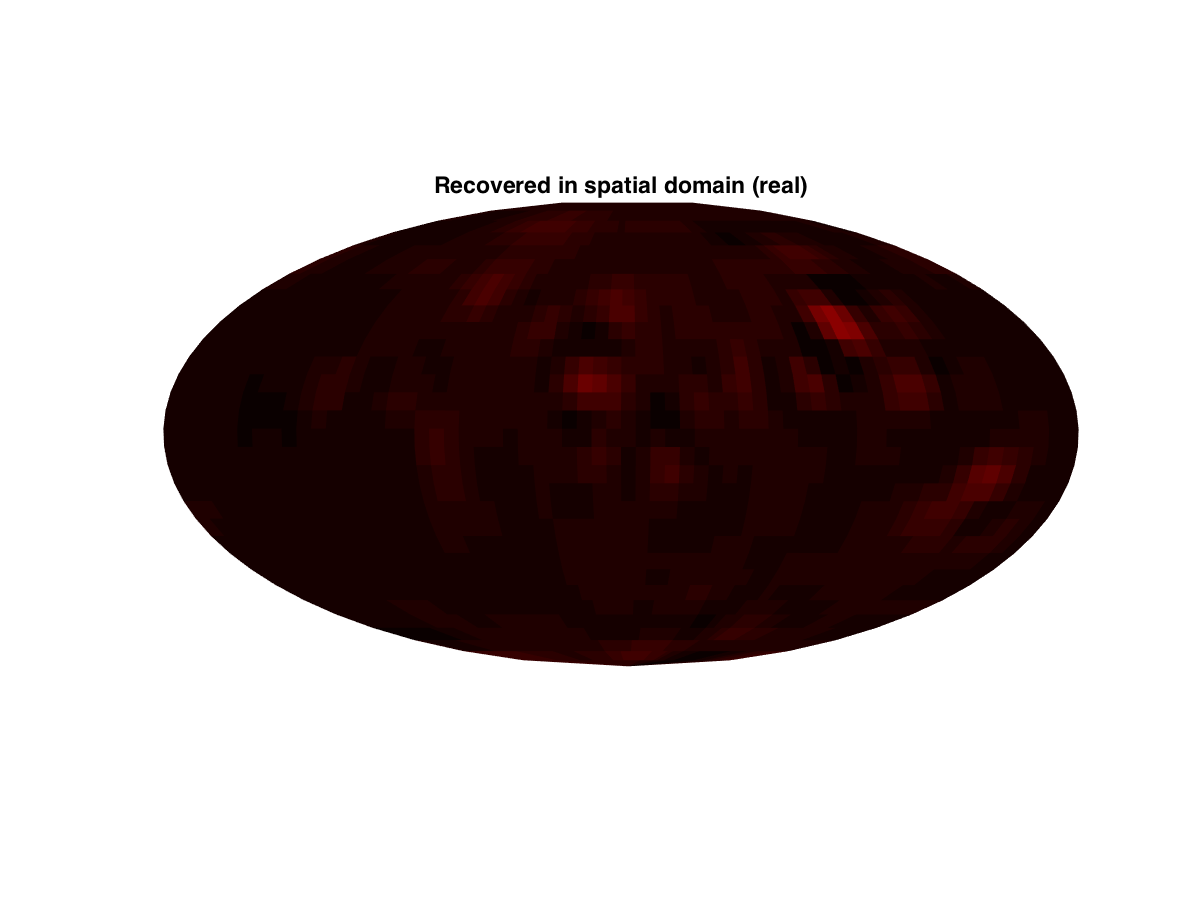}&
\includegraphics[width=0.2\linewidth,  trim=2.5cm 4cm 2cm 3.37cm, clip=true]{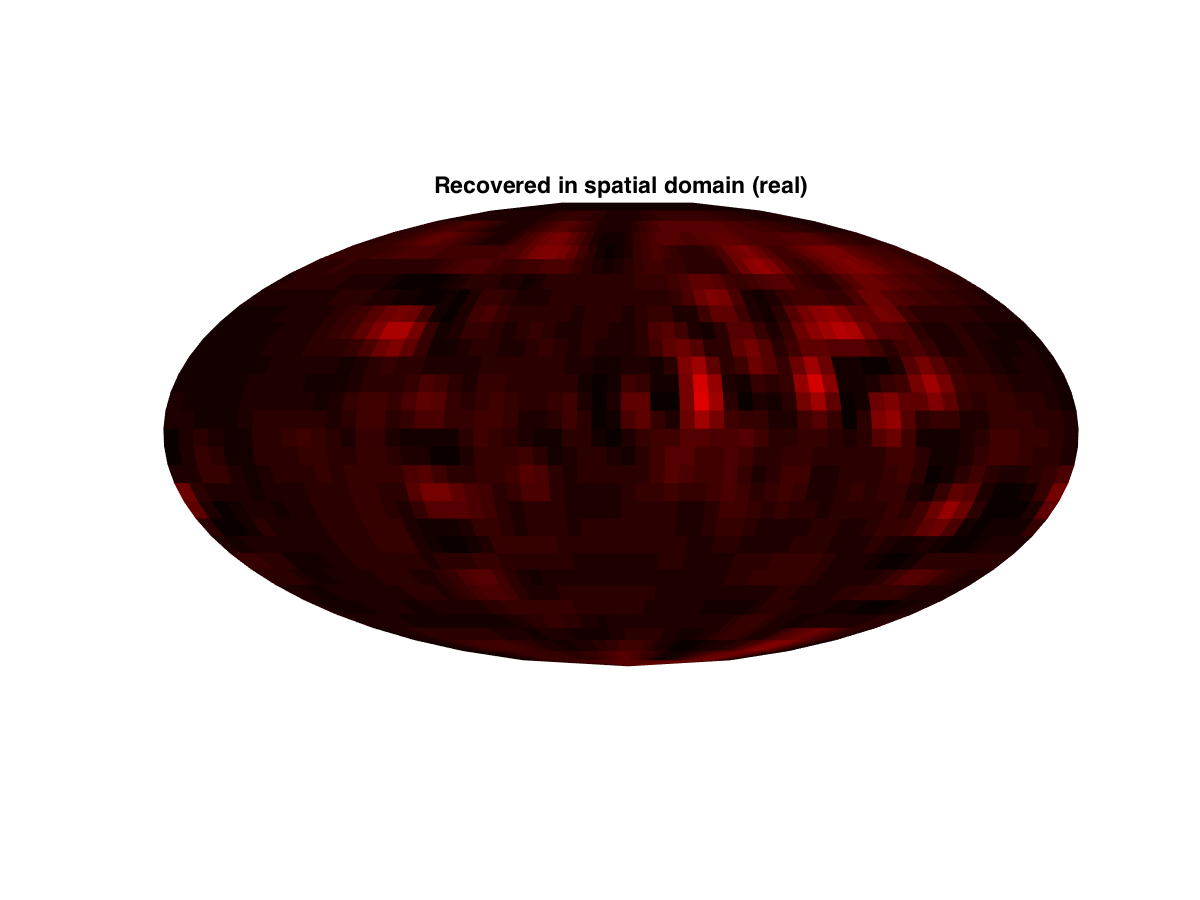}&
\includegraphics[width=0.2\linewidth,  trim=2.5cm 4cm 2cm 3.37cm, clip=true]{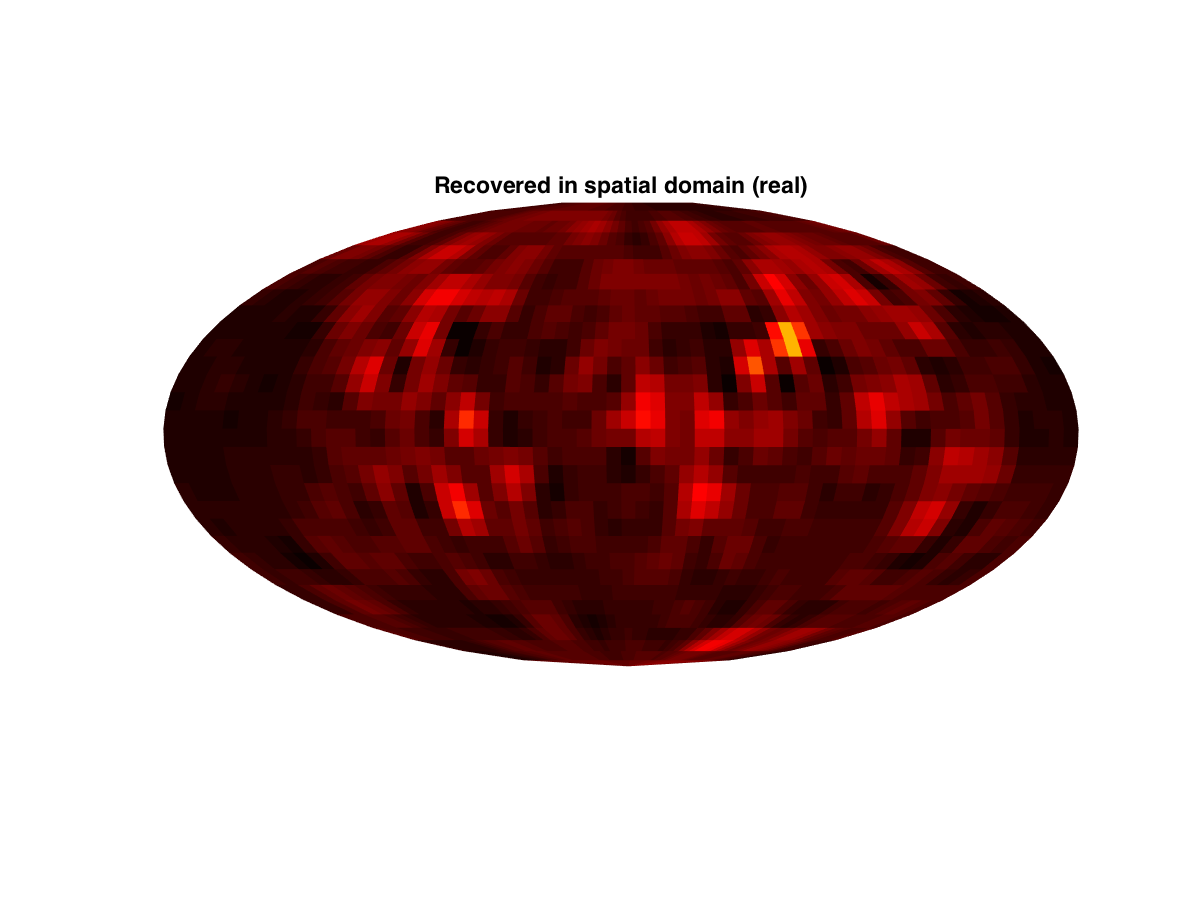}&
\includegraphics[width=0.2\linewidth,  trim=2.5cm 4cm 2cm 3.37cm, clip=true]{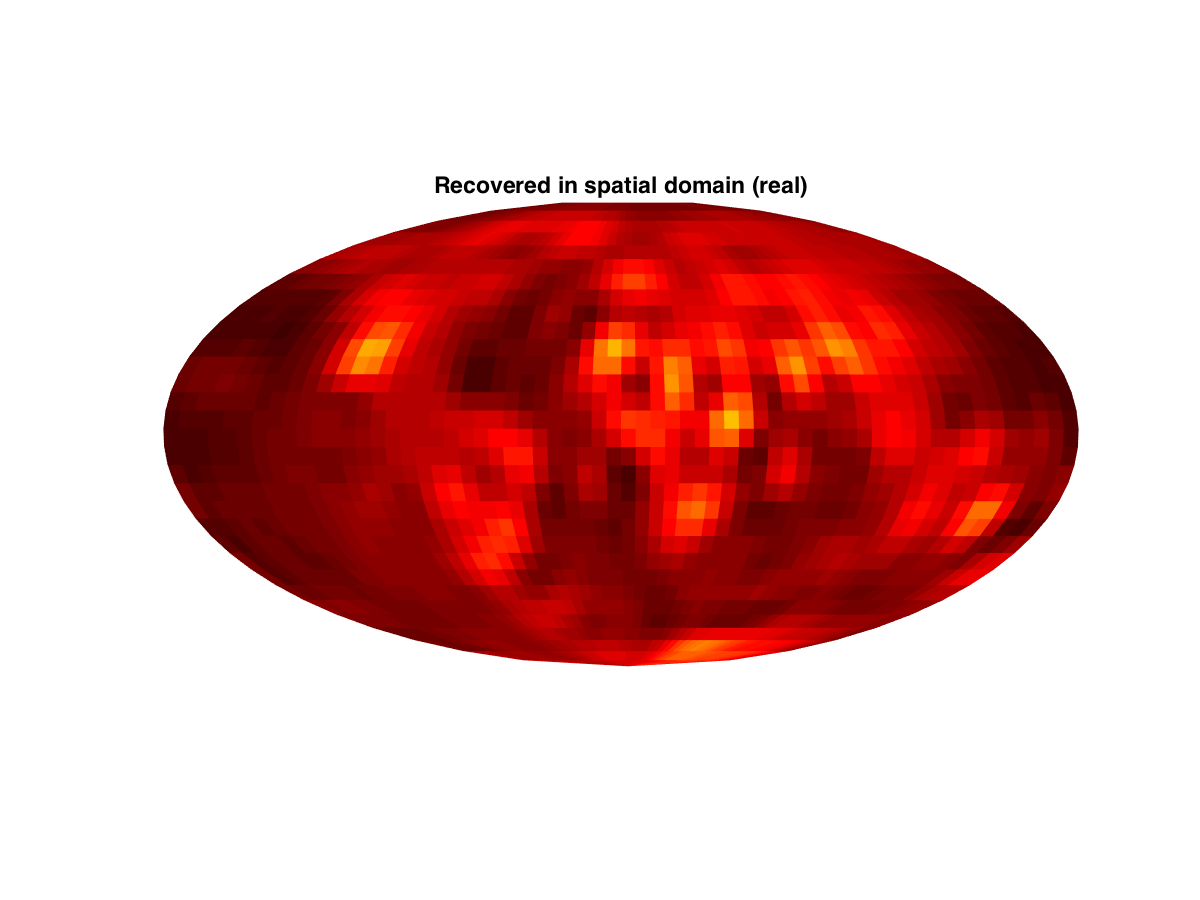}\\
DH Analysis, $M/L^2=0.3$ (1.7)& $M/L^2=0.5$ (3.16)& $M/L^2=1.0$ (6.6)& $M/L^2=1.9$ (16.0)\\

\includegraphics[width=0.2\linewidth,  trim=2.5cm 4cm 2cm 3.37cm, clip=true]{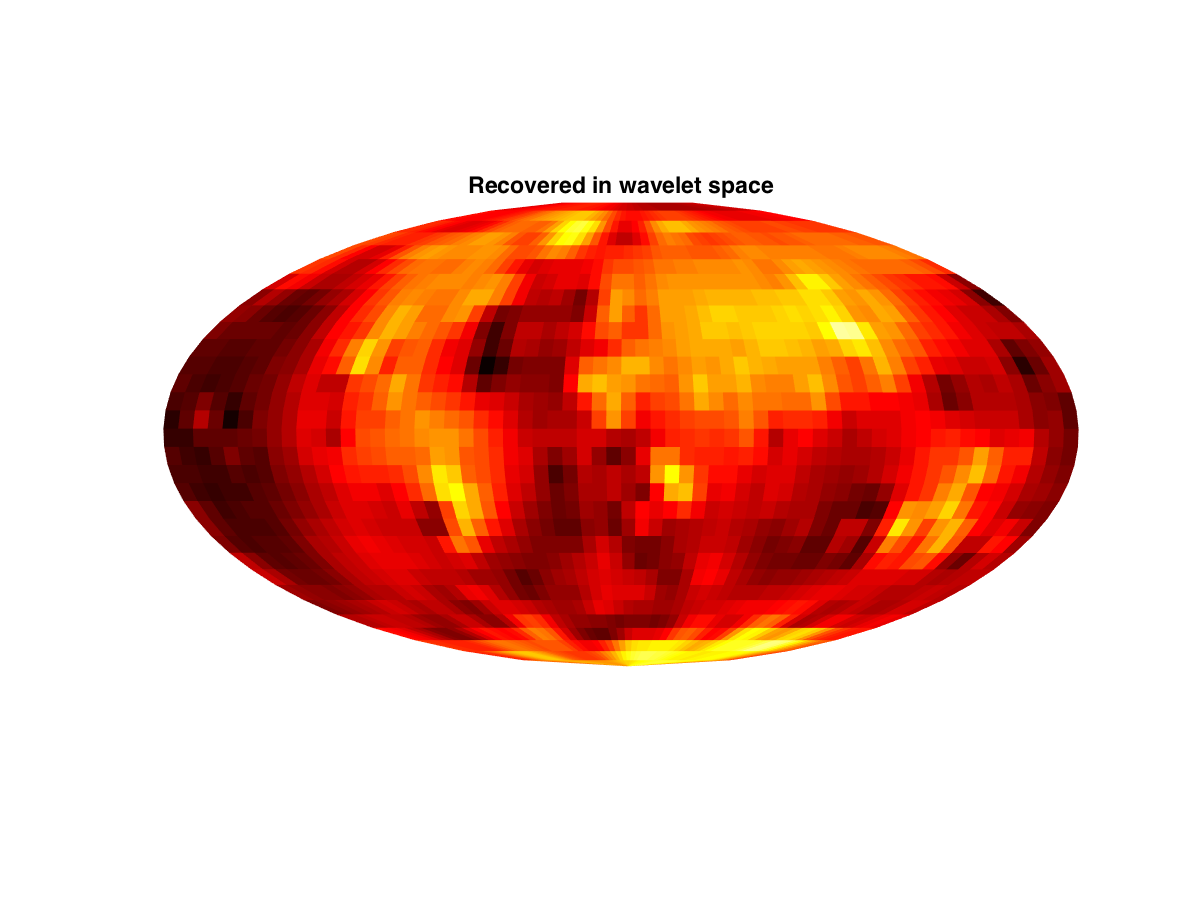}&
\includegraphics[width=0.2\linewidth,  trim=2.5cm 4cm 2cm 3.37cm, clip=true]{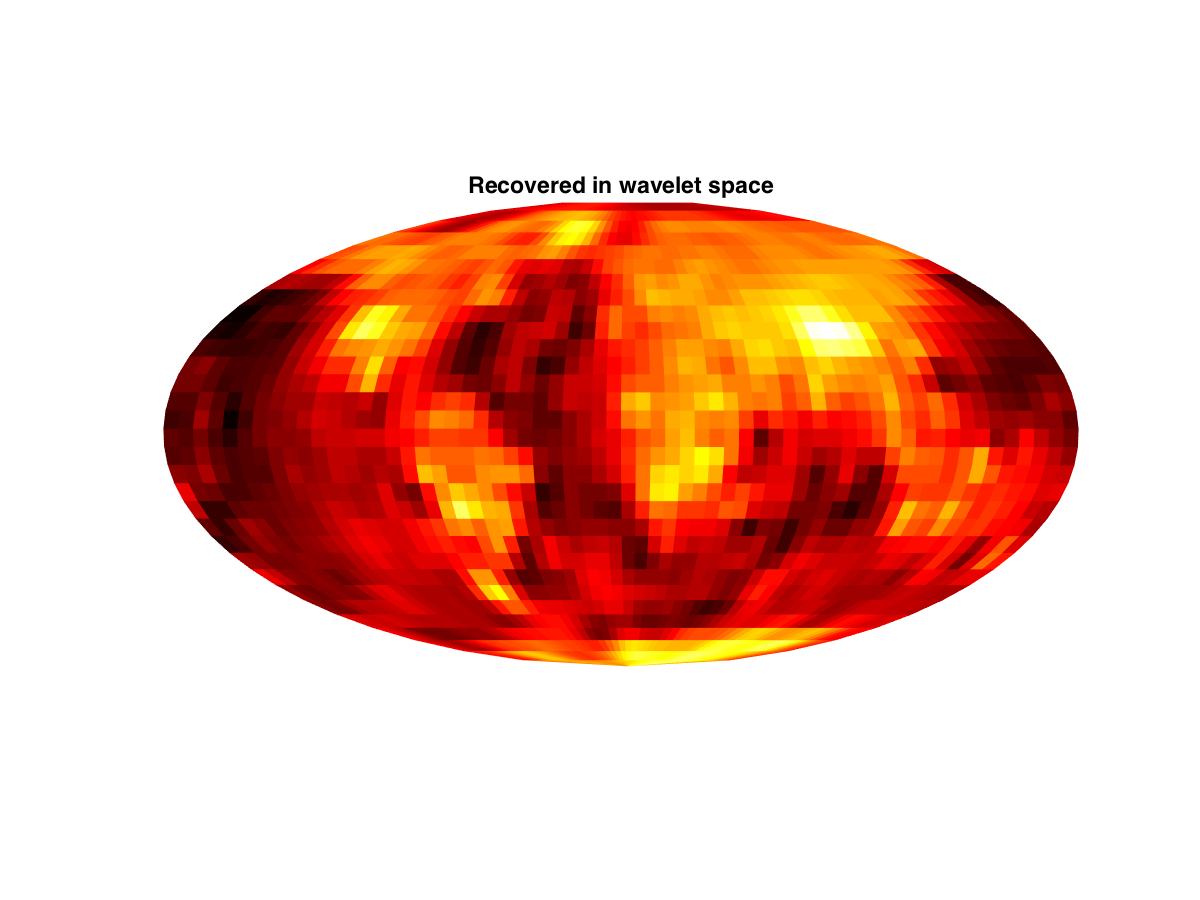}&
\includegraphics[width=0.2\linewidth,  trim=2.5cm 4cm 2cm 3.37cm, clip=true]{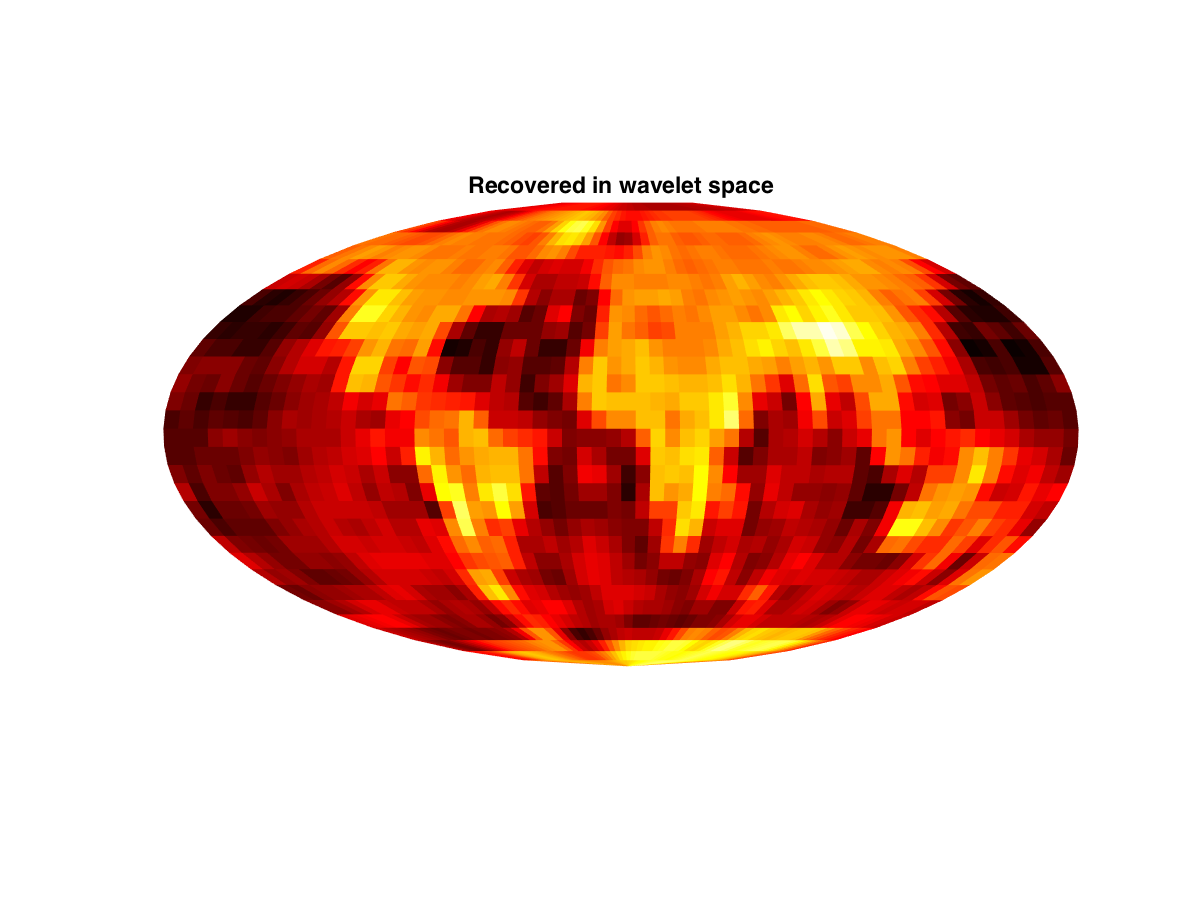}&
\includegraphics[width=0.2\linewidth,  trim=2.5cm 4cm 2cm 3.37cm, clip=true]{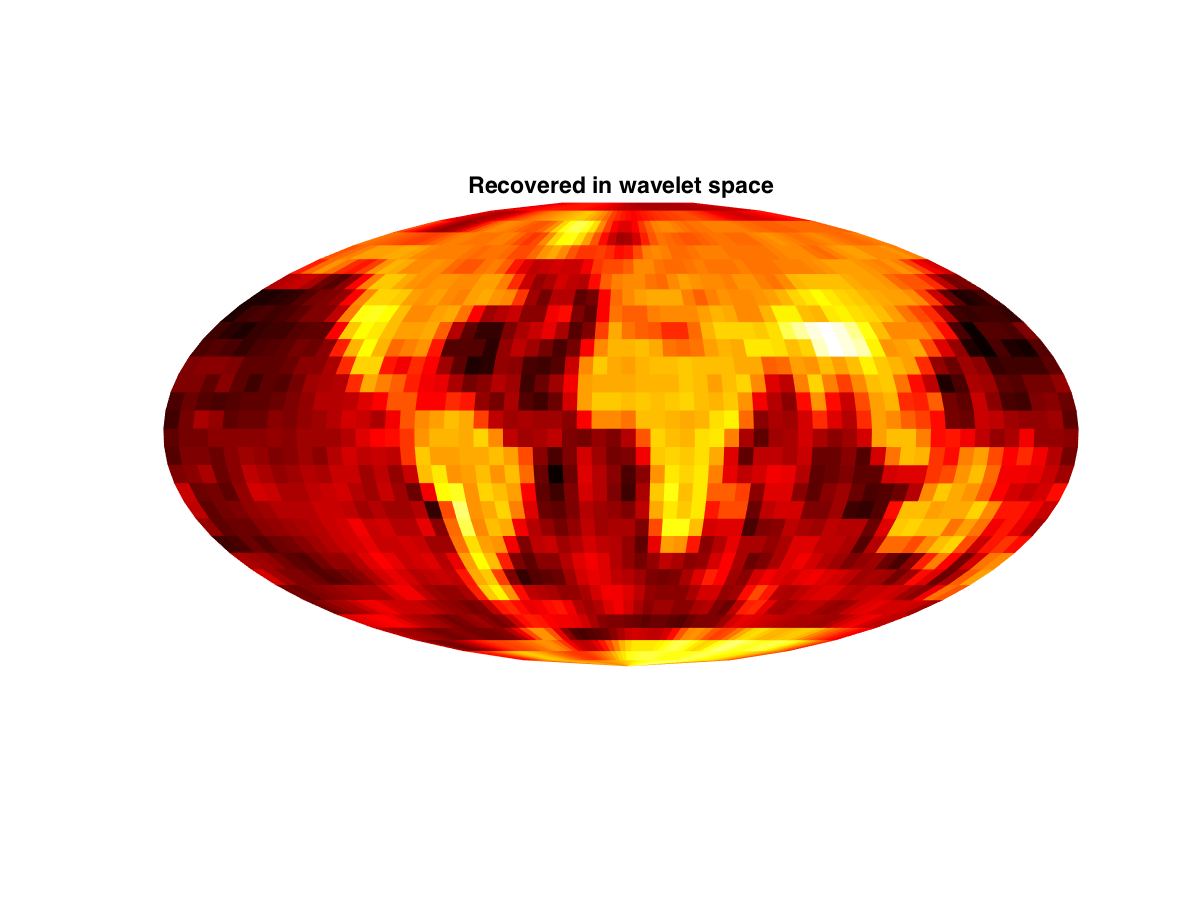}\\
DH Synthesis, $M/L^2=0.3$ (26.1)& $M/L^2=0.5$ (31.2)& $M/L^2=1.0$ (41.7)& $M/L^2=1.9$ (62.8)\\


\end{tabular}
\caption{Reconstructed images from the low resolution inpainting experiment described in 
 \sectn{\ref{sec:numerical_experiments:low_res}}. The first two rows show the reconstruction using KKM sampling
 when solving the analysis and synthesis problems, respectively. The third and fourth
 rows are solutions to the same problems with MW sampling, and fifth and sixth rows correspond to DH sampling. Each column corresponds to a different number of
 measurements. \referee{The SNR for each image is giving in brackets in dB.} The synthesis setting generally out-performs the analysis setting, while sampling schemes that require fewer samples 
 generally out perform those requiring more samples.}
\label{fig:low_res_reconstruction}
\end{center}
\end{figure*}

\section{Numerical Experiments}\label{sec:numerical_experiments}

We perform numerical experiments to both assess the effectiveness of imposing sparsity in
wavelet space and to test the impact of the sampling scheme used and whether or not the problem is solved in the analysis or synthesis setting.

We compare
the analysis and synthesis settings for the KKM, DH and MW sampling schemes. These experiments are performed at low resolution as
fast adjoint transforms for the DH and KKM sampling theorems are lacking. We chose to solve a noisy inpainting problem for these tests
(as an example of a common 
inverse problem). At high resolution we demonstrate image reconstruction with 
both axisymmetric and directional wavelet sparsity priors using MW sampling (for which we have constructed fast adjoint operators). We test the method at high resolution
on inpainting and deconvolution problems, and a combined inpainting and deconvolution problem, all in the presence of noise.

We generate low and high resolution test images from Earth topography data. The original Earth topography data are taken from the 
Earth Gravitational Model (EGM2008) publicly released by the U.S.\ National Geospatial-Intelligence Agency 
(NGA) EGM Development Team.\footnote{\referee{These data were downloaded from \url{http://geoweb.princeton.edu/people/simons/DOTM/Earth.mat} and extracted using the tools available 
at \url{http://geoweb.princeton.edu/people/simons/software.html}.}}  The ground truth for the low and high resolution experiments
performed in the remainder of this section are shown in \fig{\ref{fig:low_res_ground_truth}}.

Much of the work in this section takes advantage of publicly available codes: we use {\tt SSHT}\footnote{\url{http://www.spinsht.org} or \url{http://astro-informatics.github.io/ssht} (v1.0b1)} \cite{mcewen:fssht} and {\tt NSHT}\footnote{\url{http://www.zubairkhalid.org/nsht.html} (0.9b)} \cite{khalid:optimal_sampling} to compute spherical harmonic transforms; {\tt SO3}\footnote{\url{http://www.sothree.org} or \url{http://astro-informatics.github.io/so3} (v1.2b1)} \cite{mcewen:so3} to compute harmonic transforms on the rotation group; {\tt S2LET}\footnote{\url{http://www.s2let.org} or \url{http://astro-informatics.github.io/s2let} (v2.2b2)} \cite{leistedt:s2let_axisym,mcewen:s2let_spin} to compute wavelet transforms on the sphere; and {\tt SOPT}\footnote{\url{http://basp-group.github.io/sopt/} (v2.0)} \cite{carrillo:sara_algo} to solve inverse problems.
\begin{figure}
\begin{center}
\begin{tabular}{c}
\includegraphics[width=.9\linewidth,  trim=1cm 0cm 1cm 1.25cm, clip=true]{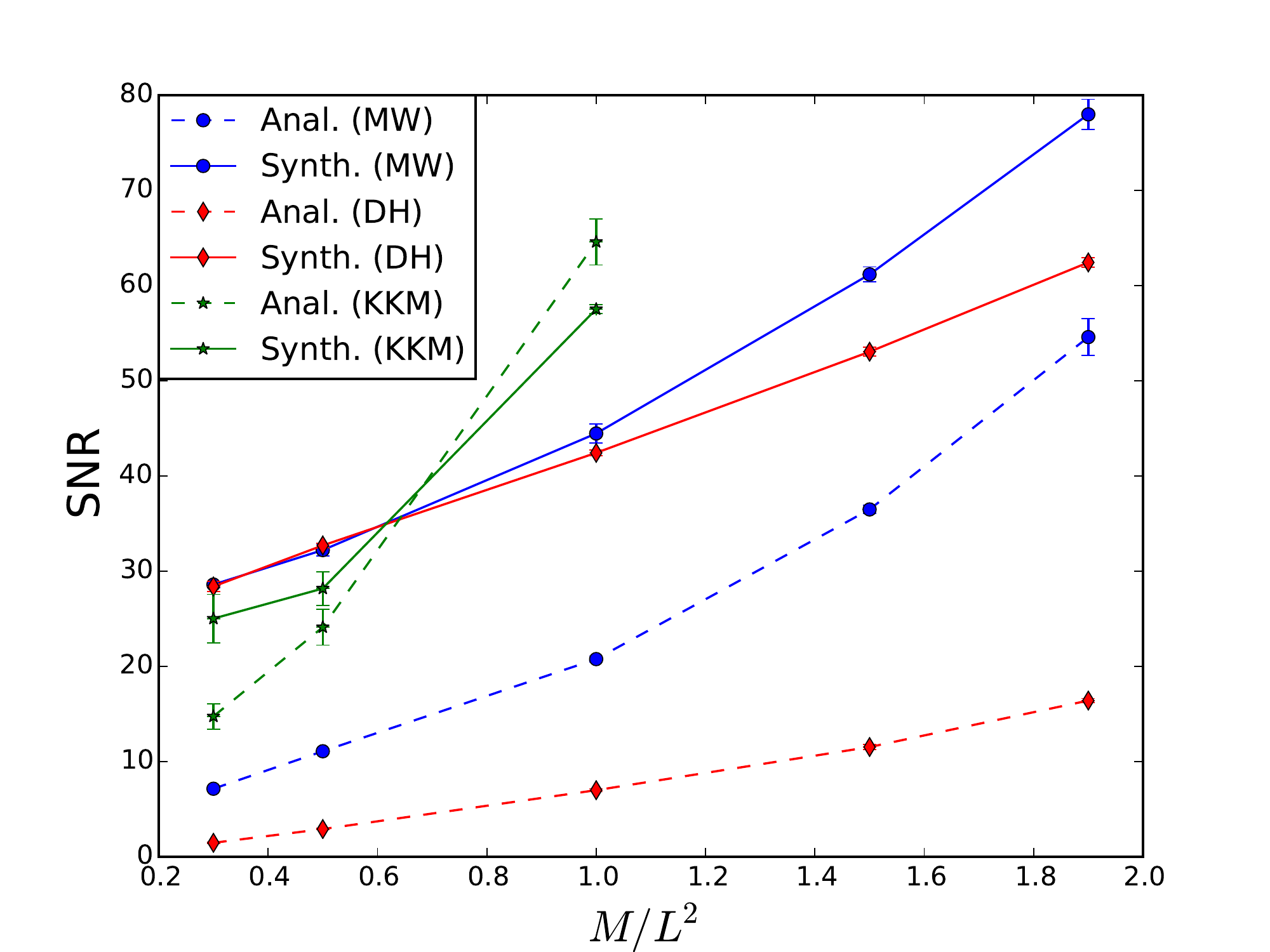}\\
\end{tabular}
\caption{Reconstruction performance averaged over 10 realisations for the DH (red diamonds), MW (blue circles) 
and KKM (green stars) sampling schemes, when solving 
the low resolution inpainting problem in the analysis setting (dashed lines) and the synthesis setting (solid lines). The KKM and MW 
sampling schemes provide 
enhancements in reconstruction quality when compared to the DH sampling scheme. The synthesis setting is shown to be generally superior to the analysis setting.}
\label{fig:low_res_snr}
\end{center}
\end{figure}

\referee{The main code used to perform these experiments was not optimised and significant parts
are run in serial in the high level language \textsc{MATLAB}. The runtimes for this
naive implementation were as follows: the low resolution experiments described in Section \ref{sec:numerical_experiments:low_res} 
take approximately 30 seconds for the MW sampling, 1 minute for the DH sampling, and 1 hour for the KKM sampling. The high resolution experiments performed with the MW sampling
described in Section \ref{sec:numerical_experiments:high_res} take around 10 minutes. All experiments were performed on a MacBook Pro (early 2015), with a 2.9 GHz Intel Core i5 processor and 16 GB of RAM.}

\subsection{Low Resolution Axisymmetric Experiments}\label{sec:numerical_experiments:low_res}

We first solve a simple inpainting problem at low resolution ($L=32$) using axisymmetric wavelets ($N=1$). 
 In this case the measurement operator in \eqn{\ref{eq:measurements}} is,
\ba
\bmtrx{\Phi} = \bmtrx{\Phi}_{\rm IP} \in \reals^{M \times N_\sphere}, \label{eq:measument_inpainting}
\ea
and represents a uniformly random masking of the spherical image, with one non-zero, unit value on each row specifying the 
location of the measured datum. The adjoint of the operator can be calculated trivially.
Measurements are taken according to \eqn{\ref{eq:measurements}} 
with noise included corresponding to a \referee{signal}-to-noise-ratio (SNR) of 46 dB, where ${\rm SNR}
 = 20 \log(\| \hat{\vect{x}} \| _2/ \| \hat{\vect{x}}^* - \hat{\vect{x}} \|_2)$, defined in harmonic space to avoid differences due to the number of 
 samples of each sampling scheme. 
We vary the number of 
measurements taken \referee{as} $M=N_{\rm m}L^2$, where $N_{\rm m}=[0.3,0.5,1.0,1.5,1.9]$. We run these experiments for each of the
 sampling schemes we consider, specifically the KKM, MW and DH sampling schemes. 
 
\analysissynthesis{Results can be seen in \fig{\ref{fig:low_res_reconstruction}}. The improvement given by the lower number of samples of the KKM or MW sampling 
schemes can be seen clearly. There is also typically
an improvement when solving \eqn{\ref{eq:prob_synthesis_real}}, the synthesis problem, as opposed to \eqn{\ref{eq:prob_analysis_real}}, the analysis problem.
We set $\eta=2.5$ (as we do for the remainder of the article unless otherwise stated), since this was shown to be the 
value resulting in the reconstructions of the highest SNR, \referee{although it should be noted that the resulting SNR was
very similar for $2.5<\eta<4.5$ when an emperical investigation was conducted.} In \fig{\ref{fig:low_res_snr}} we show 
the average SNR for 10 reconstructions, which supports the findings inferred from \fig{\ref{fig:low_res_reconstruction}}.}
\referee{These experiments were repeated on another data set from a different domain (natural light probe images) with similar results.}


\begin{figure*}
\begin{center}
\begin{tabular}{c c c}

{\bf Acquired Data} & {\bf Recovered Image Axisymmetric}& {\bf Recovered Image Directional}\\

\includegraphics[width=0.3\linewidth,  trim=2.5cm 4cm 2cm 3.37cm, clip=true]{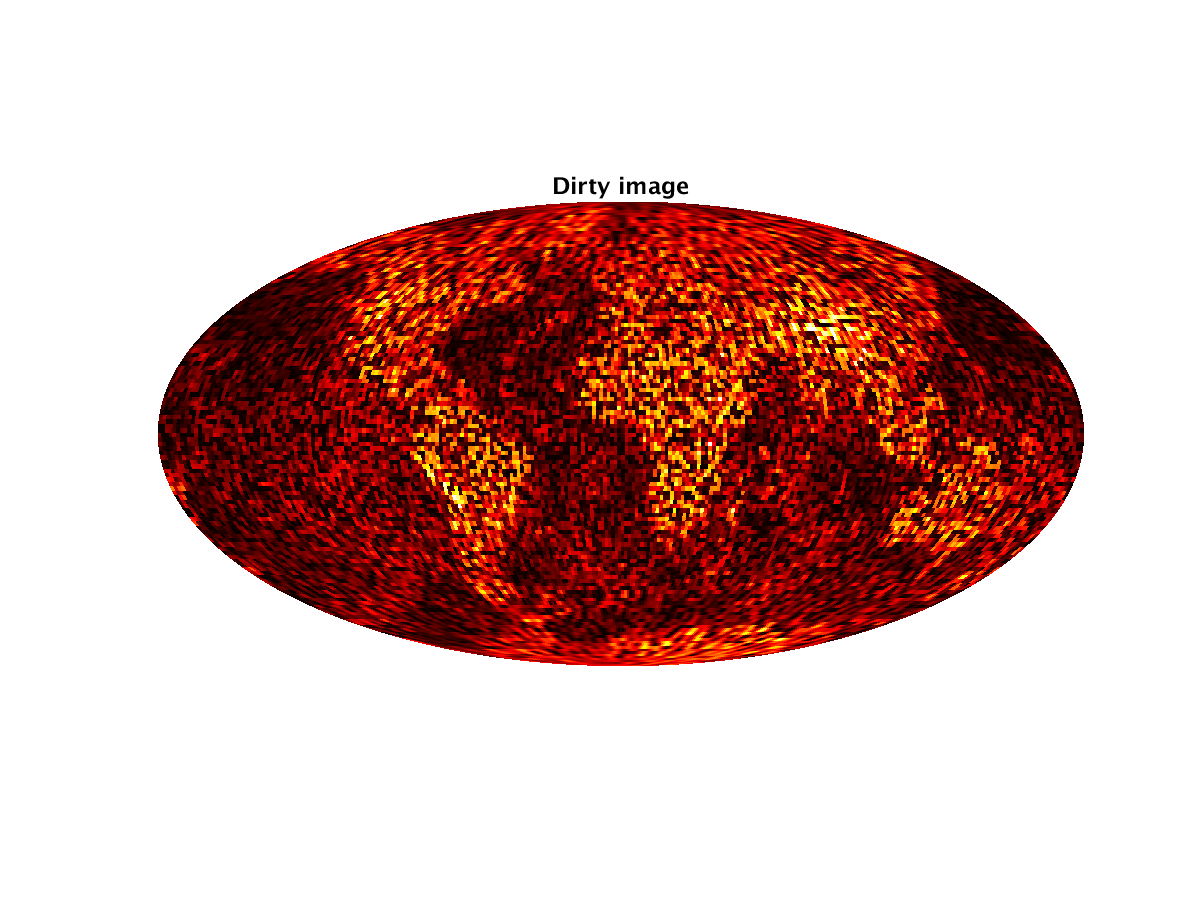}&
\includegraphics[width=0.3\linewidth,  trim=2.5cm 4cm 2cm 3.37cm, clip=true]{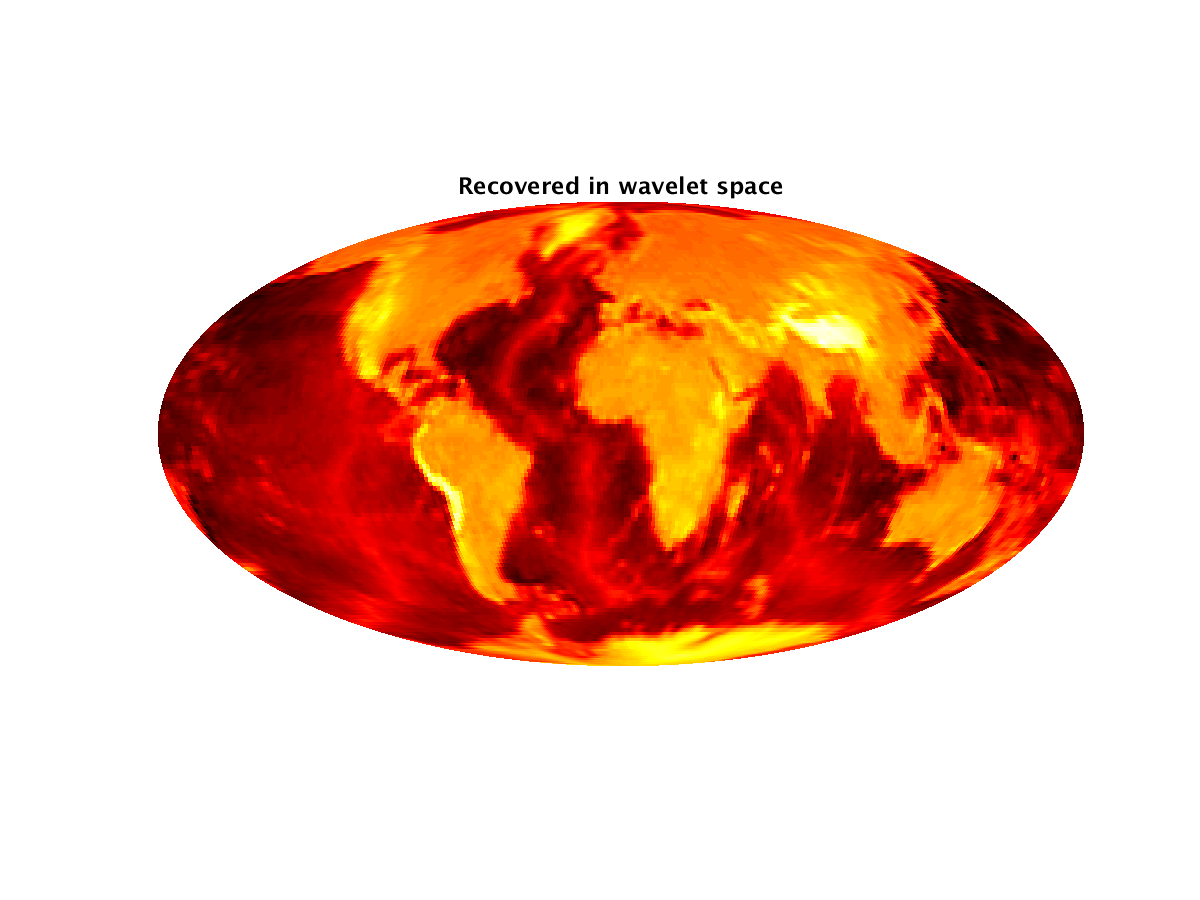}&
\includegraphics[width=0.3\linewidth,  trim=2.5cm 4cm 2cm 3.37cm, clip=true]{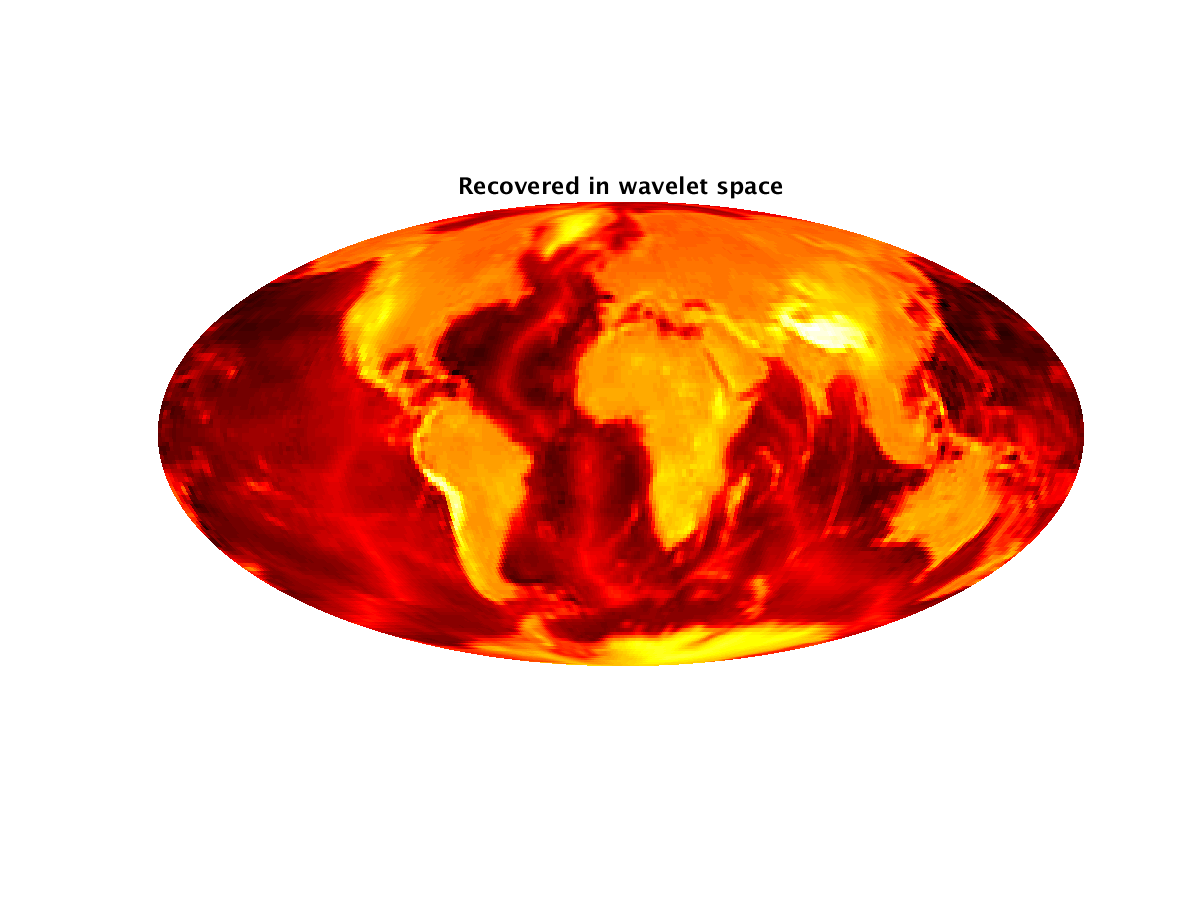}\\
\multicolumn{3}{c}{(a) Inpainting}\\

\includegraphics[width=0.3\linewidth,  trim=2.5cm 4cm 2cm 3.37cm, clip=true]{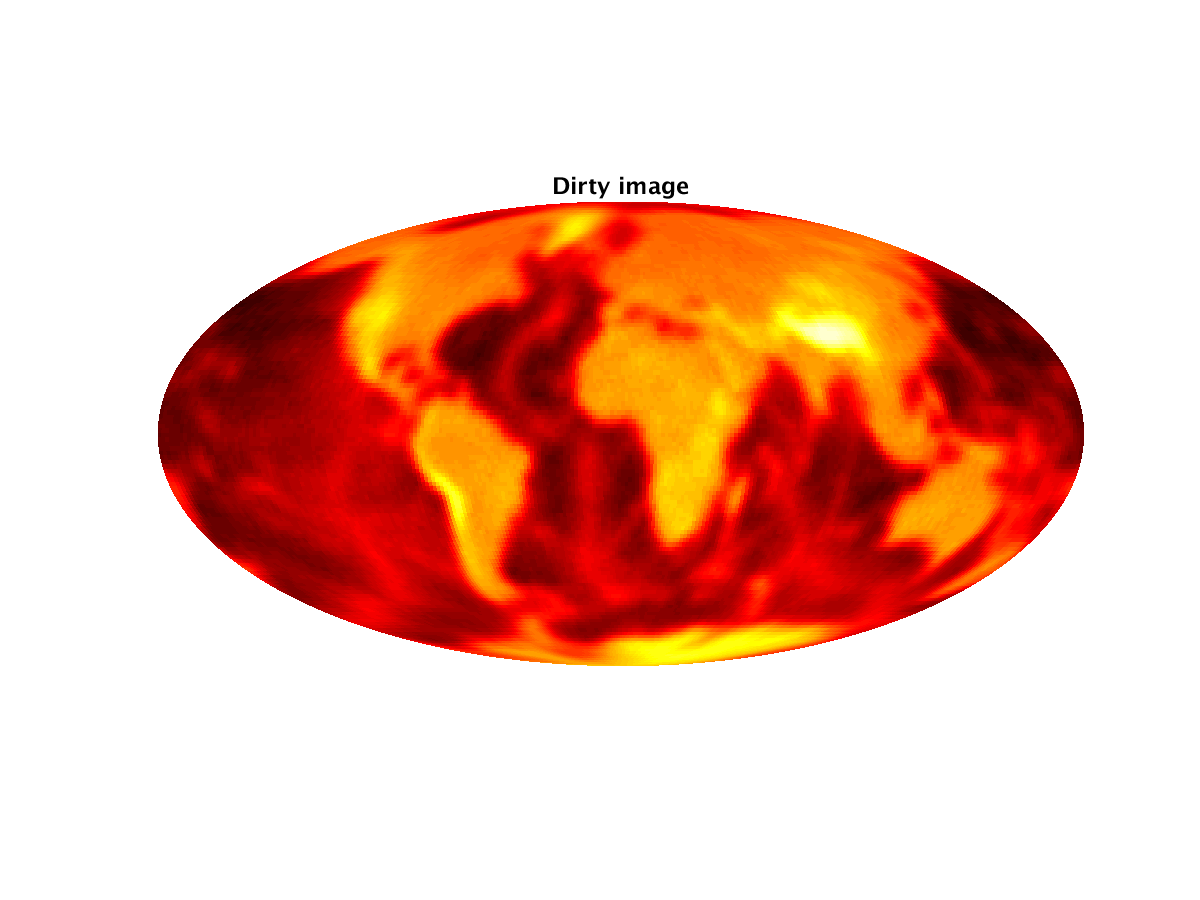}&
\includegraphics[width=0.3\linewidth,  trim=2.5cm 4cm 2cm 3.37cm, clip=true]{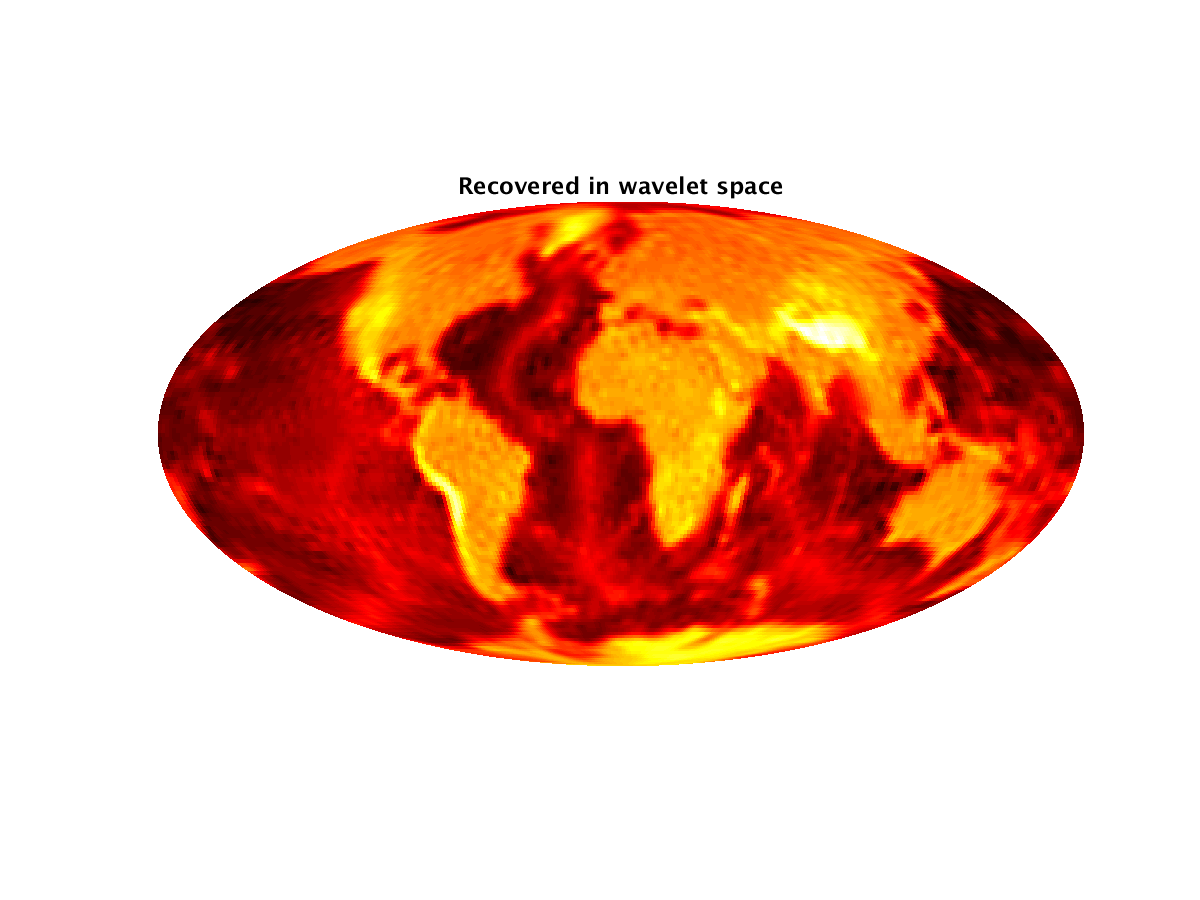}&
\includegraphics[width=0.3\linewidth,  trim=2.5cm 4cm 2cm 3.37cm, clip=true]{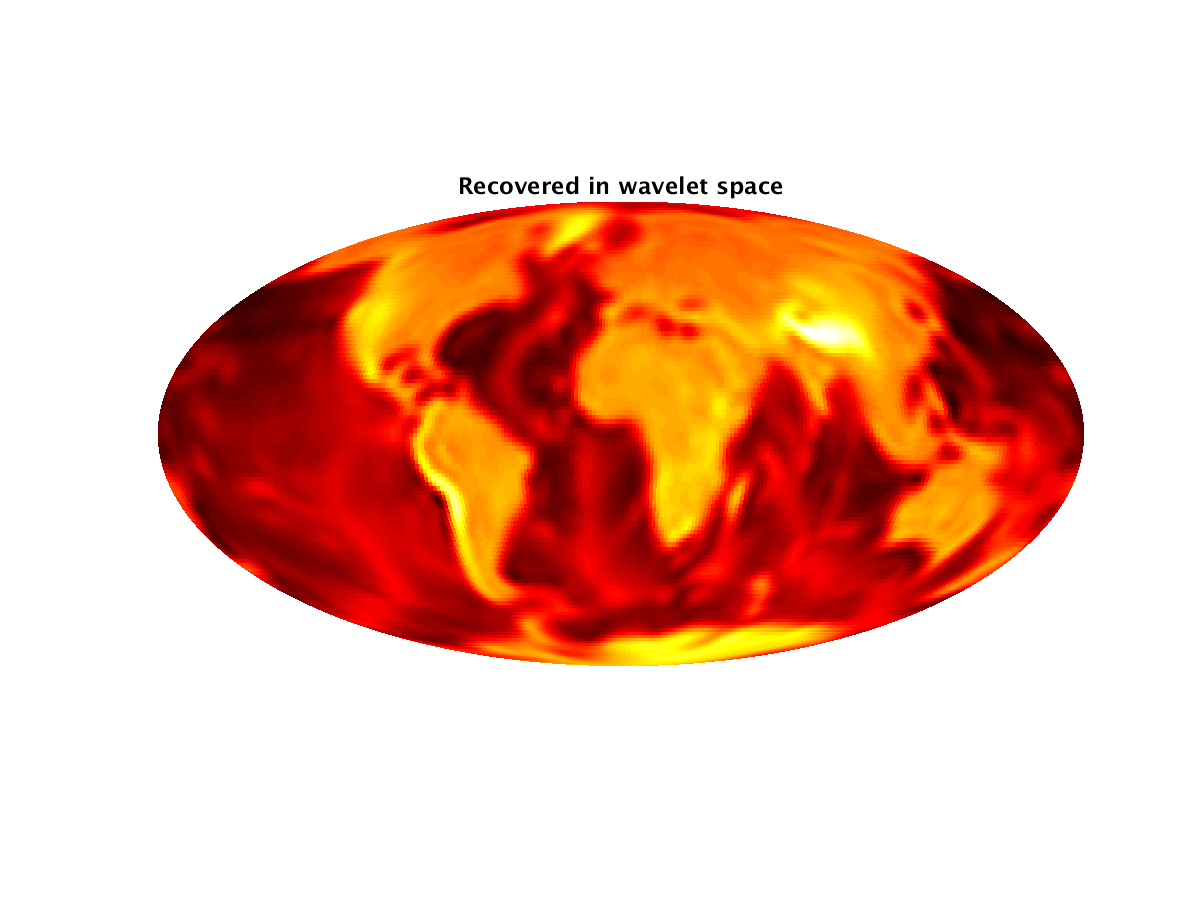}\\
\multicolumn{3}{c}{(b) Deconvolution} \\

\includegraphics[width=0.3\linewidth,  trim=2.5cm 4cm 2cm 3.37cm, clip=true]{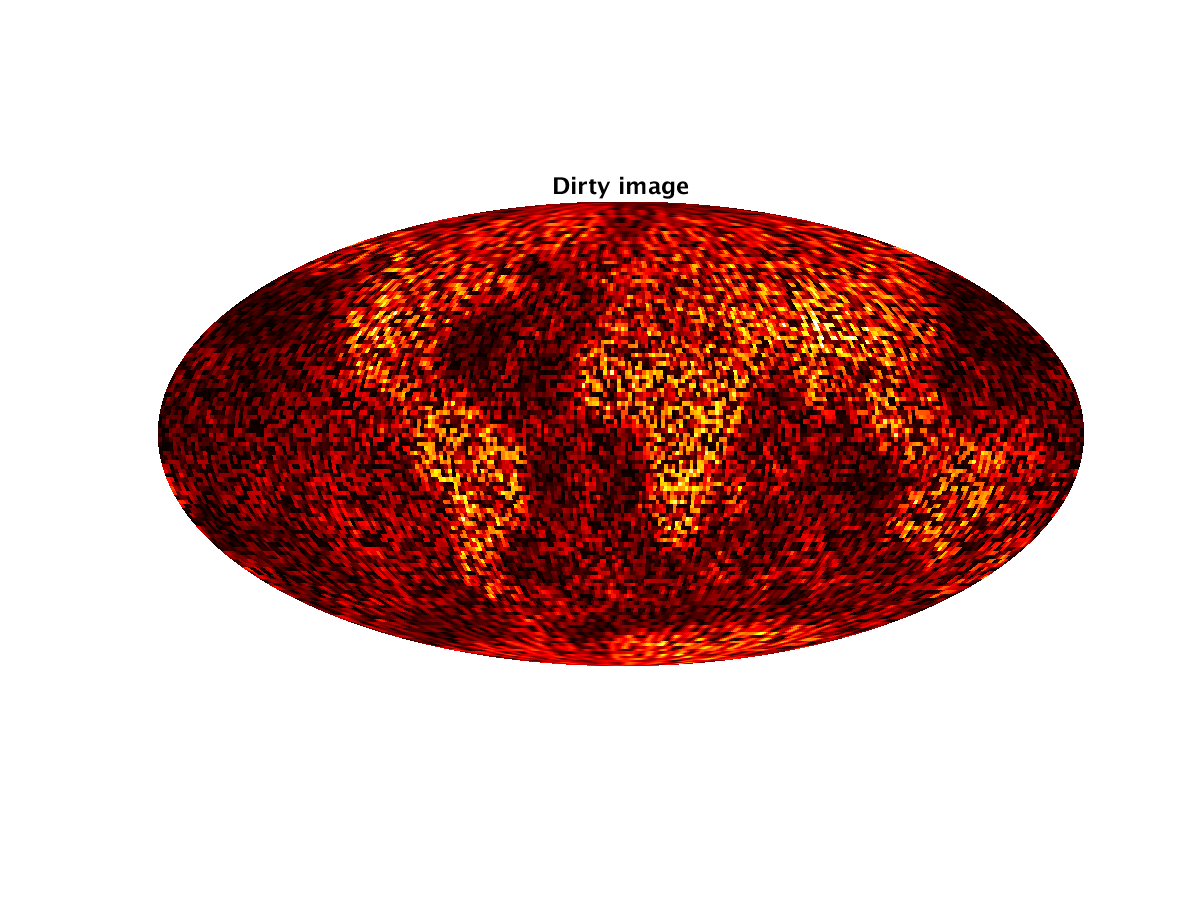}&
\includegraphics[width=0.3\linewidth,  trim=2.5cm 4cm 2cm 3.37cm, clip=true]{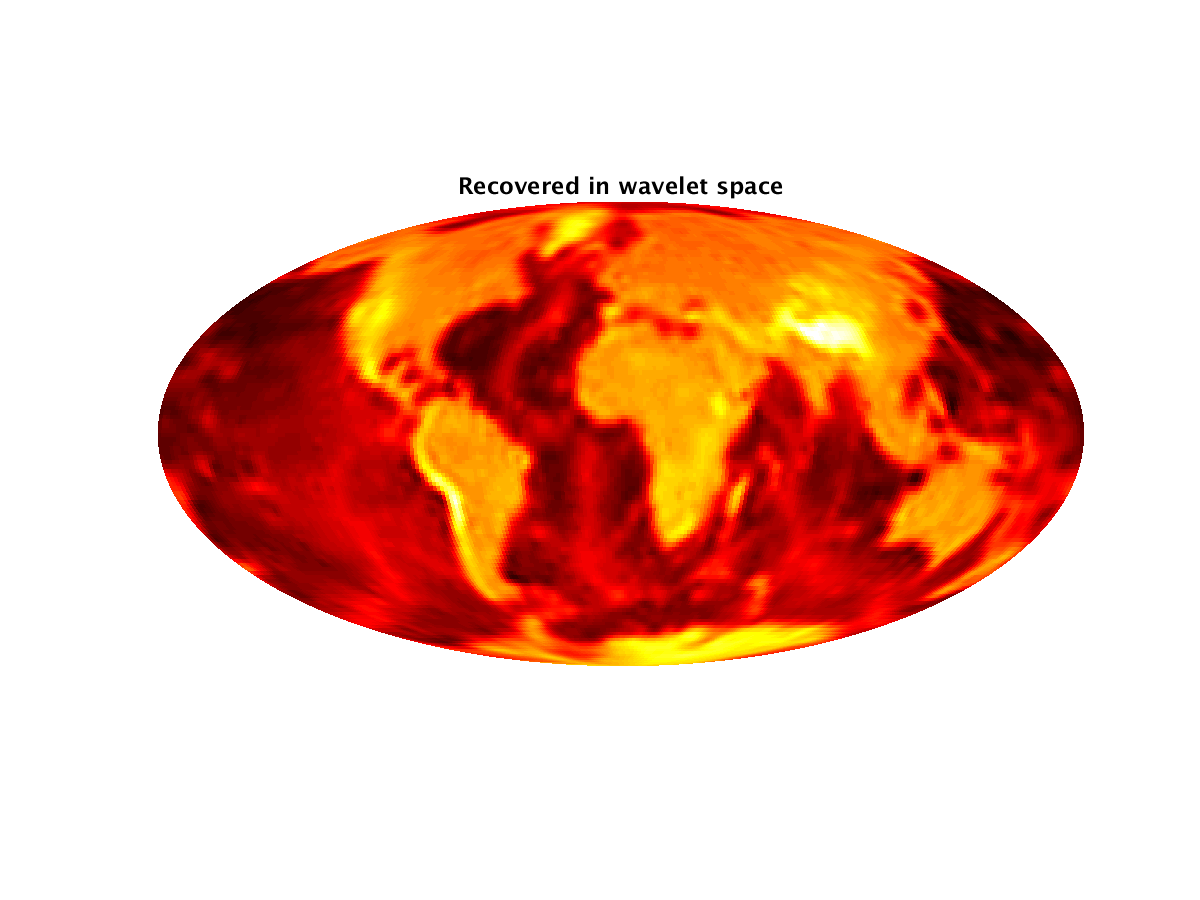}&
\includegraphics[width=0.3\linewidth,  trim=2.5cm 4cm 2cm 3.37cm, clip=true]{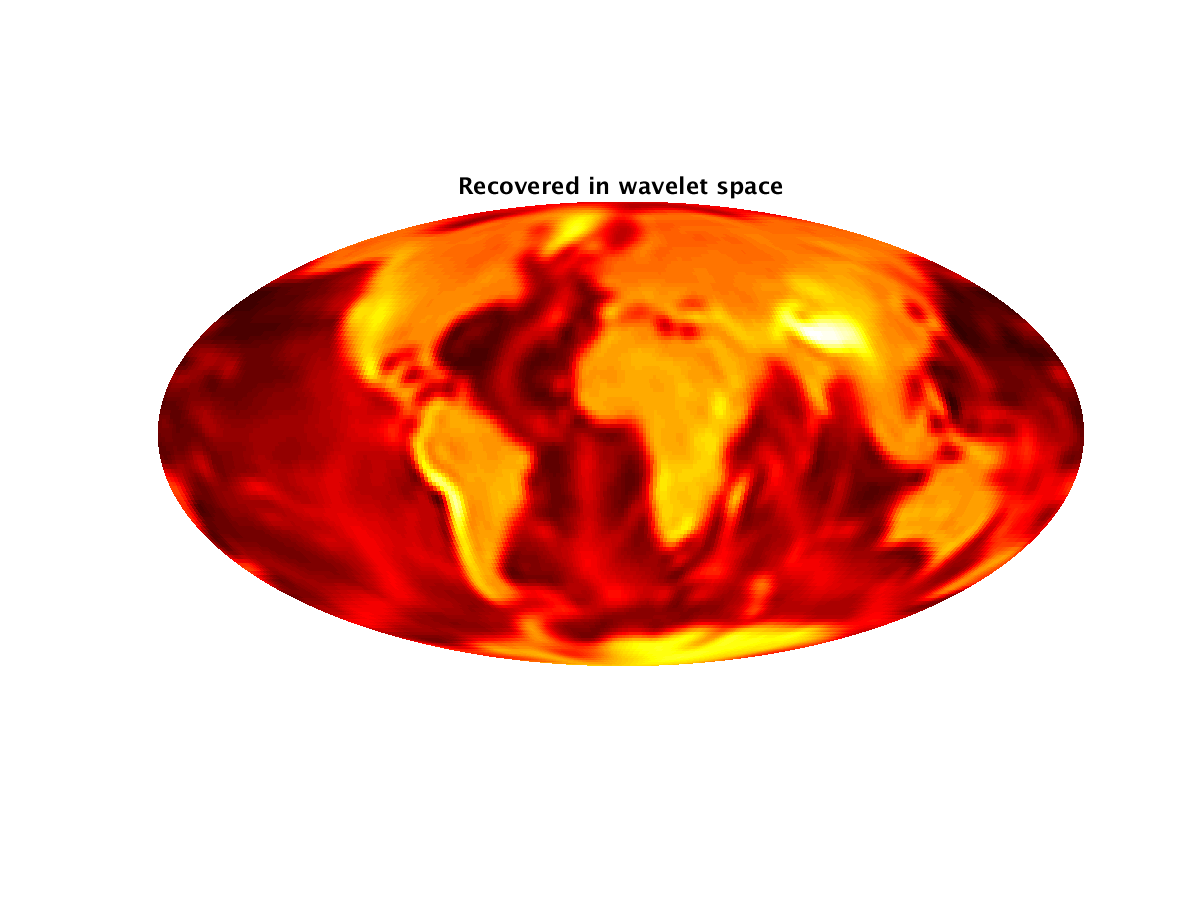}\\
\multicolumn{3}{c}{(c) Inpainting and deconvolution}\\

\end{tabular}
\caption{Reconstructed images from the high resolution experiments described in \sectn{\ref{sec:numerical_experiments:high_res}}. All problems are solved in the synthesis setting using MW sampling.
 Each solution is presented next to a band limited representation of the measured data. The SNR of the reconstructions are
 (a) 57.1 dB, (b) 51.8 dB and (c) 50.8 dB for the axisymmetric wavelets and (a) 58.4 dB, (b) 48.5 dB and (c) 48.5 dB for the directional wavelets. For the inpainting problem  directional wavelets yield superior performance, while for the deconvolution and joint inpainting and deconvolution problems the SNR recovered with axisymmetric wavelets is superior, albeit visual artefacts are mitigated when using directional wavelets. }
\label{fig:high_res_reconstruction}\label{fig:high_res_dir_reconstruction}
\end{center}
\end{figure*}

\subsection{High Resolution Experiments}\label{sec:numerical_experiments:high_res}

We run three example high resolution ($L=128$) experiments on the high resolution Earth map shown in \fig{\ref{fig:low_res_ground_truth}}.
We solve all the problems with the MW sampling in the synthesis setting with $\eta=2.5$.
We consider the MW sampling as there are currently no fast adjoint algorithms for the other sampling theorems and for the synthesis setting as it was shown to be superior in \sectn{\ref{sec:numerical_experiments:low_res}}

We consider both axisymmetric wavelets ($N=1$) and directional wavelets with $N=4$ which leads to wavelets with 
odd azimuthal symmetry.
Firstly, we solve a simple inpainting problem with a measurement operator given by 
\eqn{\ref{eq:measument_inpainting}}, with $N_m=1.0$. Secondly, we consider a deconvolution problem, where the measurement operator is
\ba
\bmtrx{\Phi} &=& \bmtrx{\Phi}_{\rm CV} \in \complex^{N_\sphere \times N_\sphere}\label{eq:measument_convolve}\\
&=& \bmtrx{Y}\bmtrx{G}\tilde{\bmtrx{Y}},
\ea
where $\bmtrx{G} \in \reals^{L^2 \times L^2}$ is a diagonal matrix whose elements are,
\ba
\bmtrx{G}_{\ell m \ell^\prime m^\prime} = e^{-\ell^2 \sigma^2}\delta_{\ell \ell^\prime}\delta_{m m^\prime},
\ea
where $\sigma = \pi/L$. Thirdly we solve the combined problem,
\ba
\bmtrx{\Phi} = \bmtrx{\Phi}_{\rm IP}\bmtrx{\Phi}_{\rm CV} \label{eq:measument_both}.
\ea
We consider measurement noise with SNR of 46 dB.

We show the results of these experiments and a band limited version of the measured data in \fig{\ref{fig:high_res_reconstruction}}. The SNR of the recovered images are encouragingly high, showing a
good similarity with the ground truth. The deconvolution problem shows minor visual artefacts in the axisymmetric  wavelet case. The visual artefacts are reduced in the directional setting
for the two problems involving deconvolution, however SNRs are slightly lower. The inpainting problem visually shows a marked 
improvement in the directional case over the axisymmetric case, as also illustrated by the improved SNR.


\section{Denoising {\it Planck} 353 GHz Data}\label{sec:planck_results}

The {\it Planck} satellite observed the entire sky at a range of microwave frequencies \cite{Planck2015:Overview}, yielding high resolution maps of the polarised Galactic dust emission from its high frequency polarised channel centred on 353 GHz and total intensity maps at even higher frequencies from other channels \cite{Planck:2016:asky_dust}. One of the many uses of these maps is the study of the Galactic magnetic field, where it is important to have high SNR maps of the clumps of dust in the Galaxy. It is common practise to smooth the data with a Gaussian kernel in order to suppress the high frequency noise \cite{Planck:2016:orientation_mag_feild}. This smoothing has the undesirable effect of not denoising large scales and, perhaps more damaging, removing important structure on small scales. Here we examine the use of sparsity in wavelet space as a prior to denoise the {\it Planck} 353GHz total intensity map.

\begin{figure}
\begin{center}
\includegraphics[width=1\linewidth,  trim=0cm 2cm 1cm 5.2cm, clip=true]{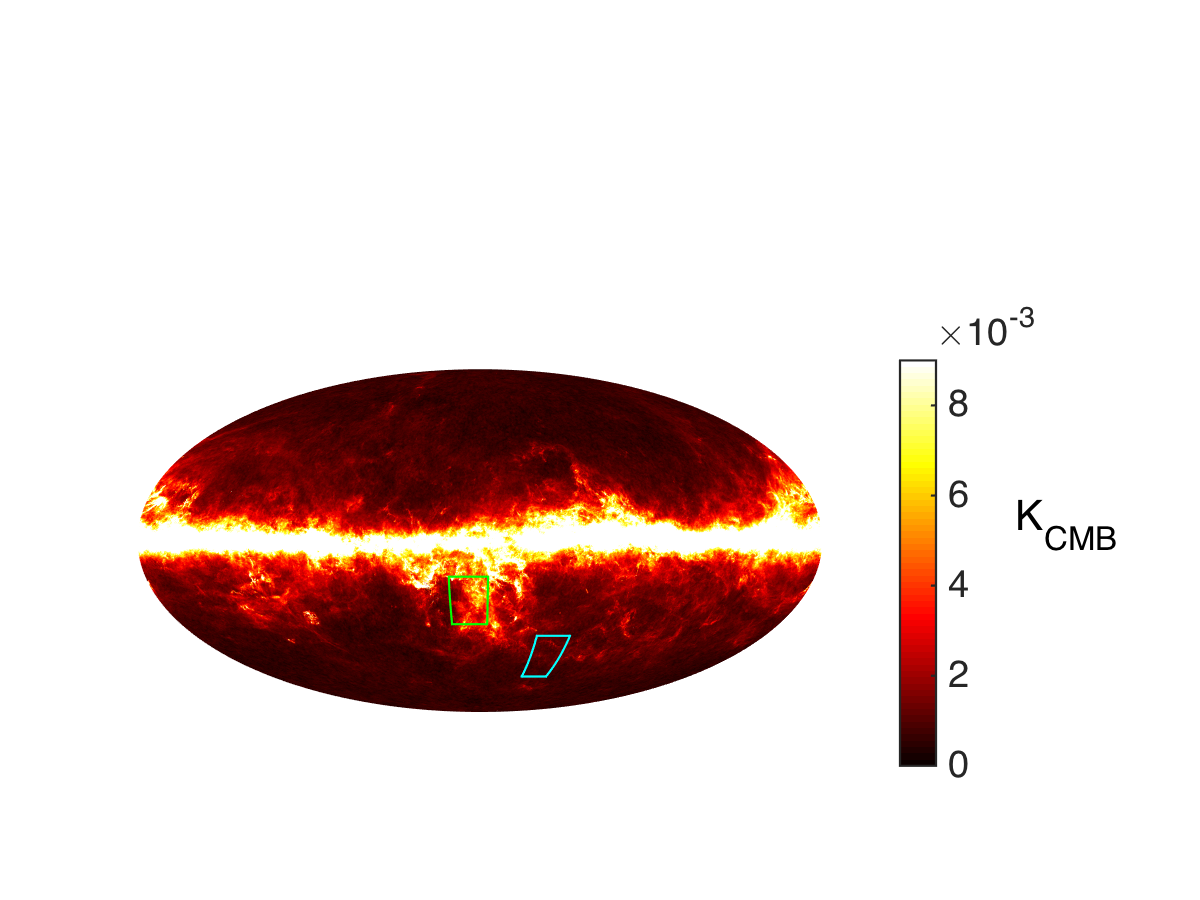}
\caption{Acquired data of the 353 GHz {\it Planck} total intensity map. A full description of how this data was aquired can be found in \sectn{\ref{sec:planck_results}}. \referee{The regions highlighted are shown in more 
detail in Figure \ref{fig:planck_denoising}.}}
\label{fig:planck_original_data}
\end{center}
\end{figure}

The {\it Planck} 353 GHz data is available to download\footnote{\referee{These data can be found at \url{http://pla.esac.esa.int/pla/\#maps}. We use the 353 GHz {full mission} data.}} in HEALPix\footnote{\url{http://healpix.sourceforge.net/}} format \cite{gorski:2005}. We use the HEALPix software to compute the spherical harmonic coefficients of this spherical image. We then band limit these to $L=2048$ and use {\tt SSHT} \cite{mcewen:fssht} to obtain a MW sampled image of the sphere. This is taken as our input data to then be denoised and can be seen in \fig{\ref{fig:planck_original_data}}. An estimate of the noise level is made by downloading the noise maps from the same archive, performing the same operation, and averaging the noise over all of the sky. This leads to an initial estimate of $\epsilon$, which is subsequently optimised through experimentation.

\begin{figure*}
\begin{center}
\begin{tabular}{c c c}

{\bf ~~Acquired Data} & {\bf ~~Recovered Image} & {\bf ~~Smoothed Image}\\
\includegraphics[width=0.298\linewidth,  trim=0.5cm 0.5cm 4.6cm 2cm, clip=true]{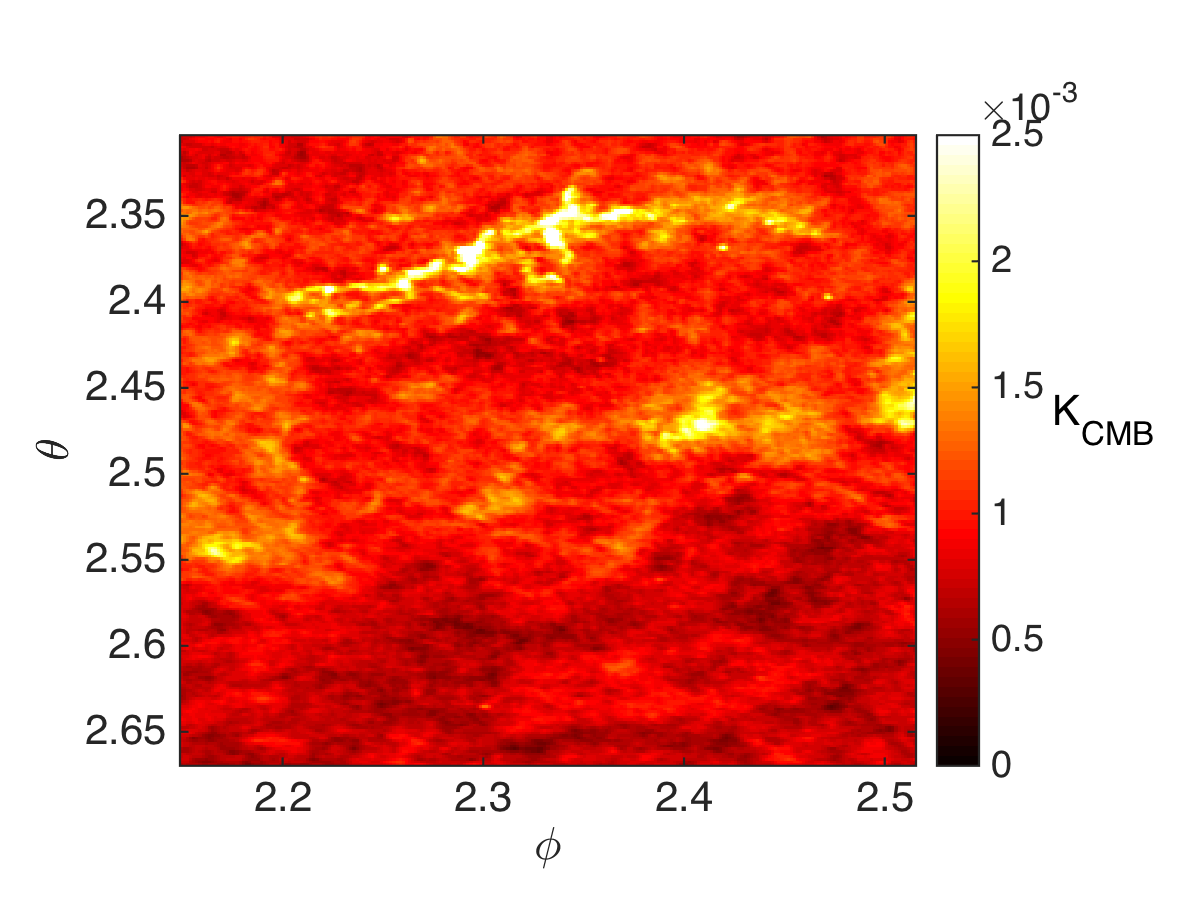}&
\includegraphics[width=0.249\linewidth,  trim=3cm 0.5cm 4.6cm 2cm, clip=true]{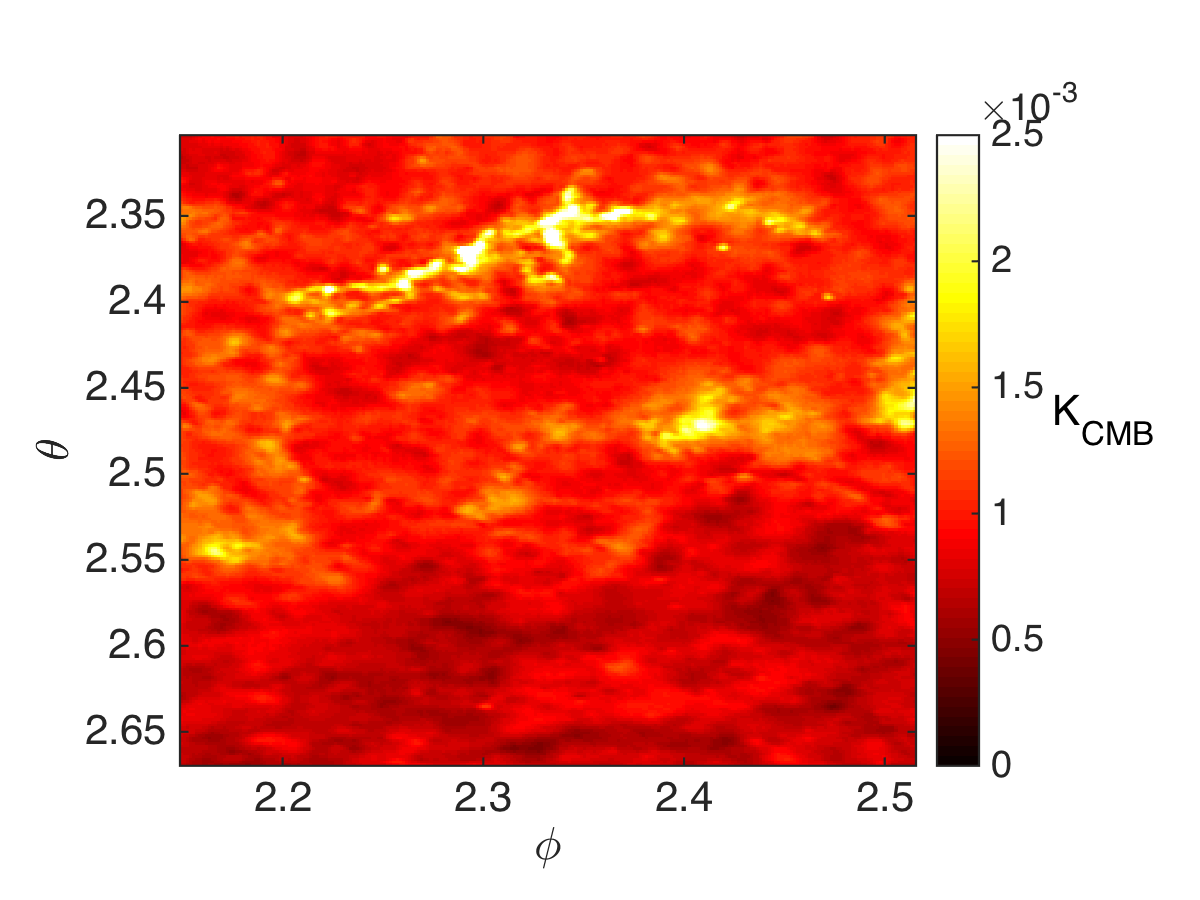}&
\includegraphics[width=0.33\linewidth,  trim=3cm 0.5cm 0.5cm 1.4cm, clip=true]{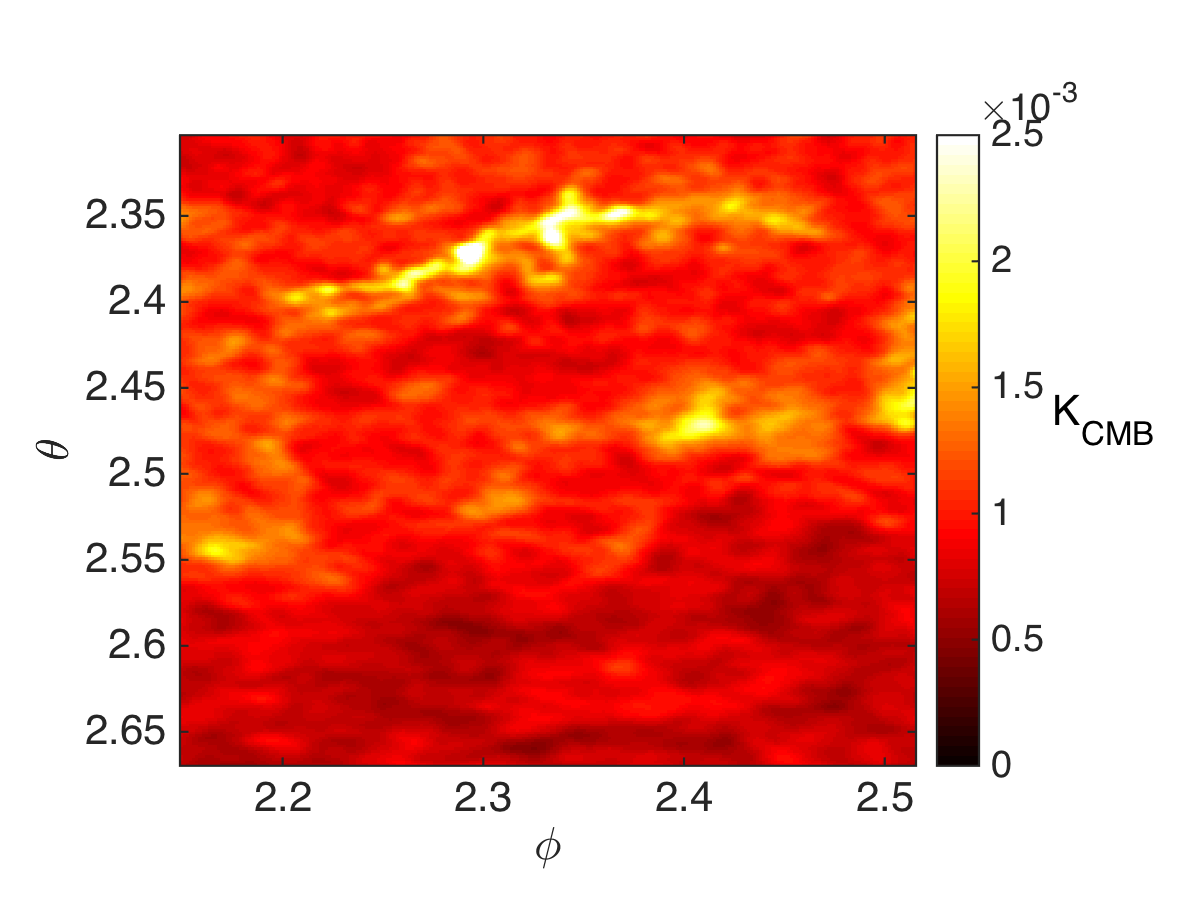}\\

\includegraphics[width=0.298\linewidth,  trim=0.5cm 0.5cm 4.6cm 2cm, clip=true]{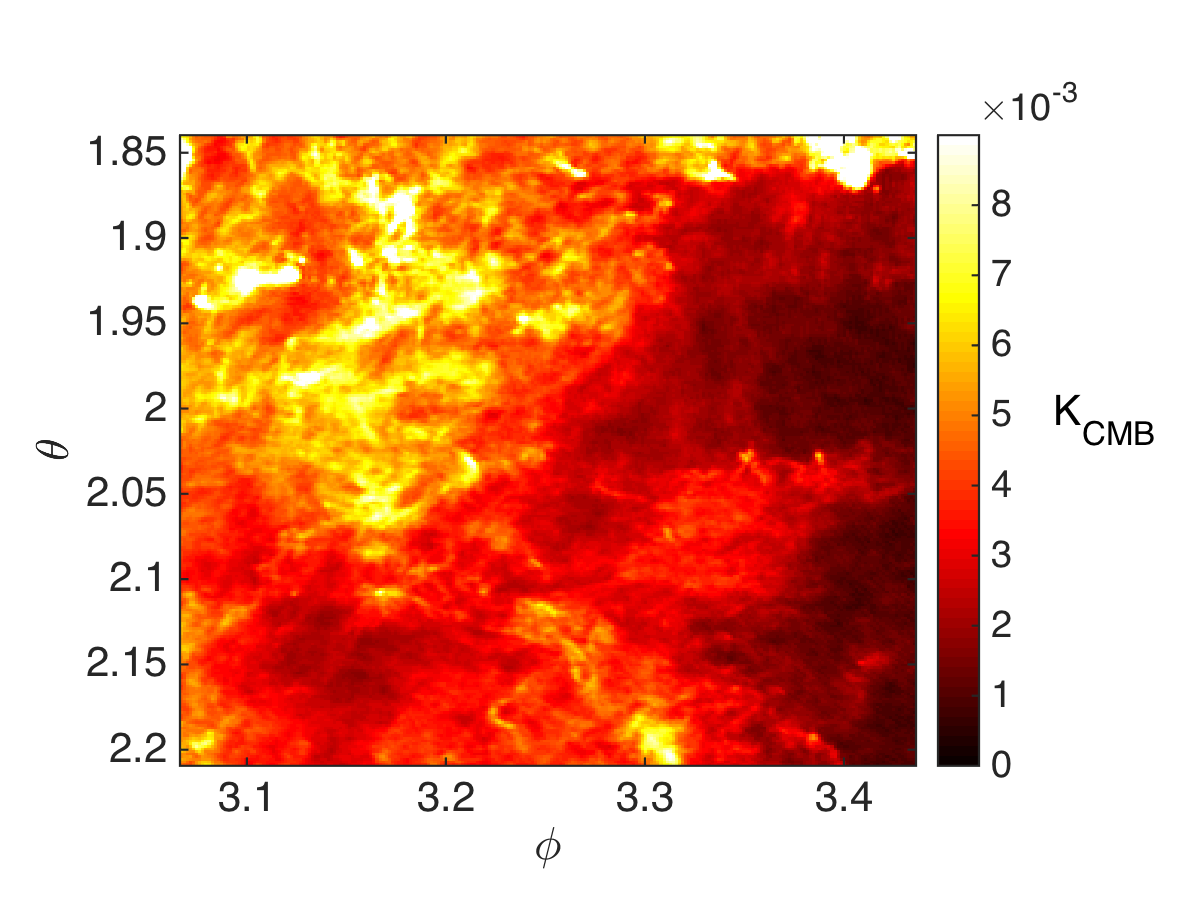}&
\includegraphics[width=0.249\linewidth,  trim=3cm 0.5cm 4.6cm 2cm, clip=true]{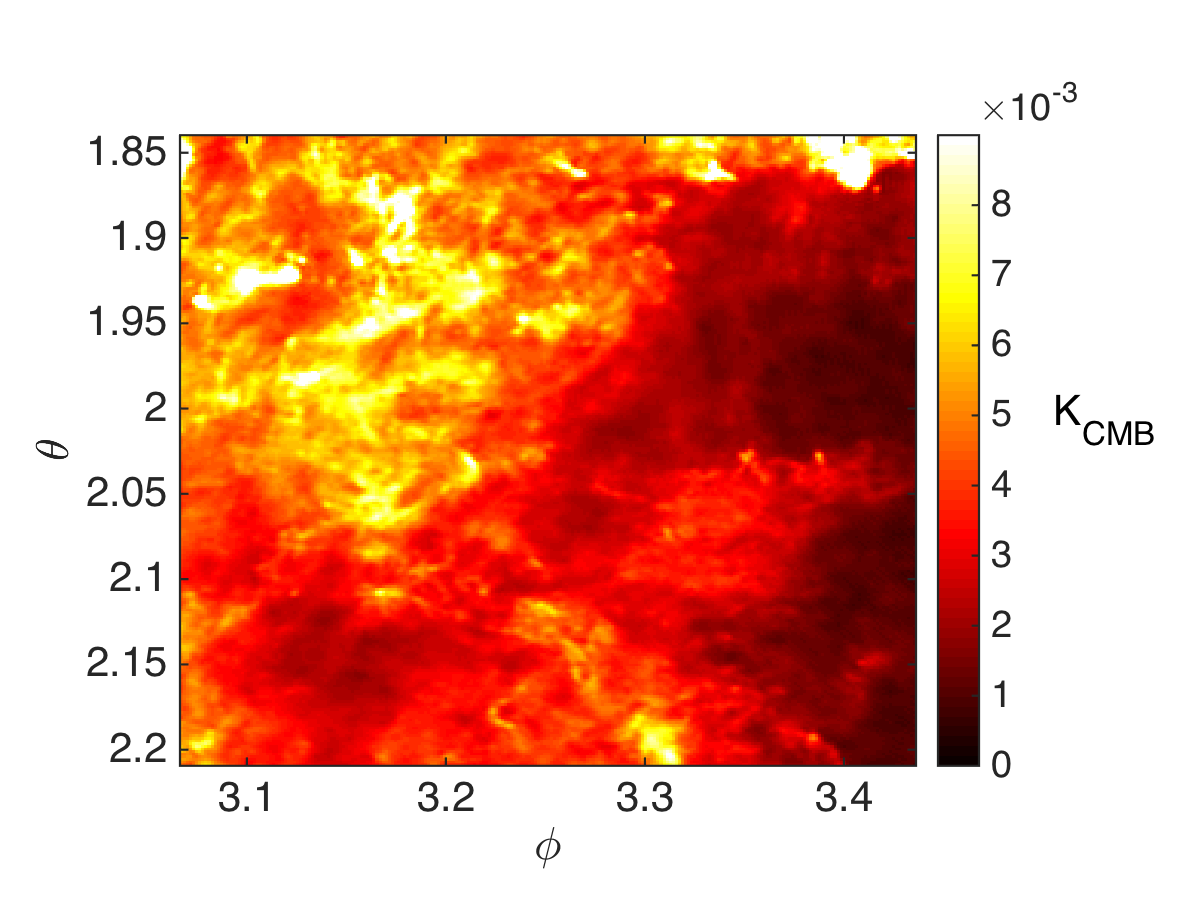}&
\includegraphics[width=0.33\linewidth,  trim=3cm 0.5cm 0.5cm 1.4cm, clip=true]{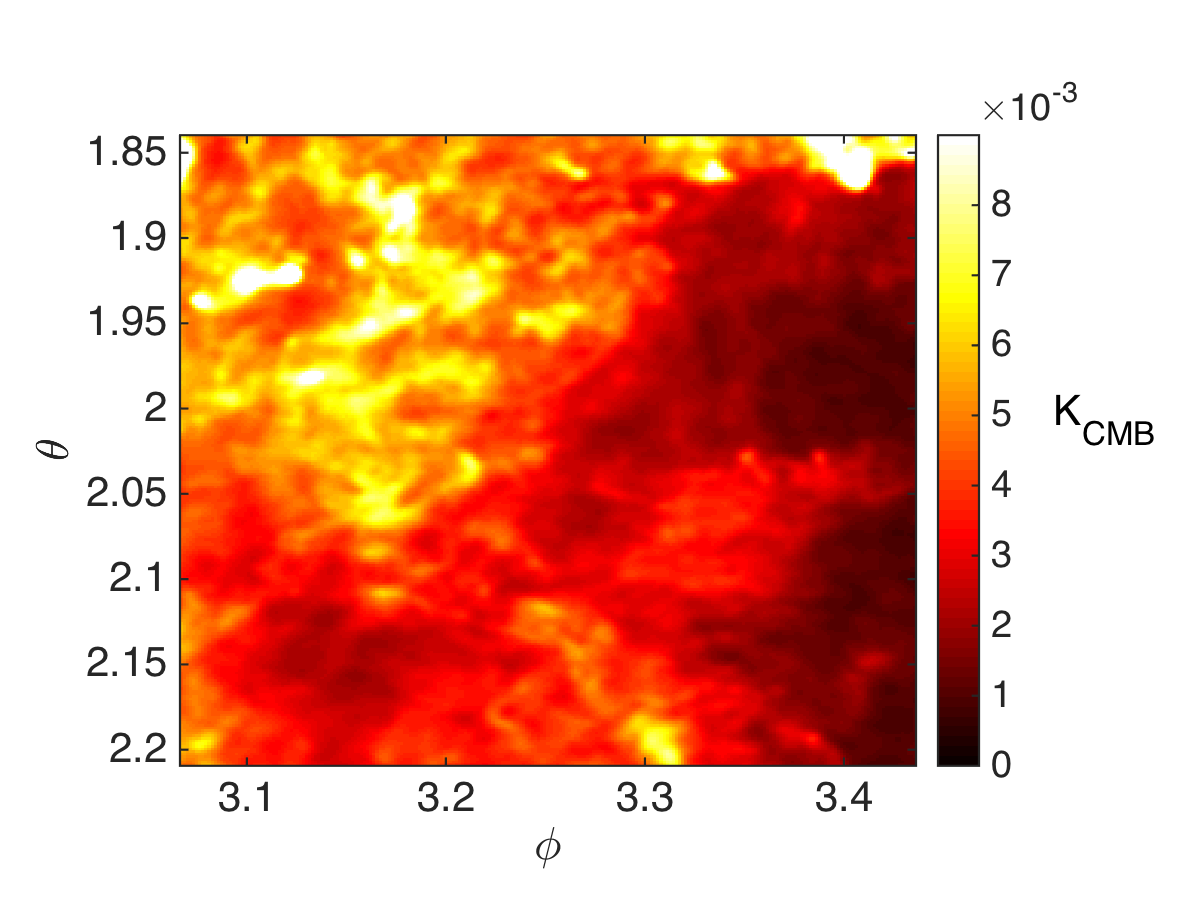}\\

\end{tabular}
\caption{Results from denoising the {\it Planck} 353 GHz total intensity map. The left column shows the acquired data, the middle column shows the denoised data and the right column shows the data denoised by smoothing with a 5 arcmin Gaussian kernel (the standard approach \cite{Planck:2016:orientation_mag_feild}). Each plot shows a zoomed region of the sphere. \referee{The acquired data shown here is a zoom in of that shown in \fig\ref{fig:planck_original_data}, the top figure is the blue region and the bottom figure is the green region.}}
\label{fig:planck_denoising}
\end{center}
\end{figure*}

We solve the denoising problem in the synthesis setting with measurement operator set to the identity and \referee{using the axisymmetric wavelets $(N=1)$}. We set $\eta = 3.0$ but have found the specific value have little effect on the reconstruction. \fig{\ref{fig:planck_denoising}} shows the original map, the result from denoising and a smoothed map. The smoothed map is the original map convolved with a 5 arcmin kernel to replicate the current denoising techniques adopted. It is clear the noise is reduced by our sparse denoising approach, while preserving small scale structure.

\section{Conclusions} \label{sec:conclusions}

We develop a general framework to solve image reconstruction problems on the
sphere by sparse regularisation, minimising the $\el_1$ norm of wavelet
coefficient representations of spherical images.  By developing fast adjoint
operators, we recover convergence guarantees for the resulting convex
optimisation problems.  As examples, we have demonstrated that using our
framework one can solve denoising, inpainting, and deconvolution problems
effectively, and combinations thereof.

\analysissynthesis{We study and compare the analysis and synthesis settings
for solving inverse problems on the sphere for the first time.  The
analysis problem has shown promising results in Euclidean space, hypothesised
to be due to its more restrictive nature. However, the more restrictive nature of the
analysis framework in Euclidean space does not carry over to the spherical
setting.  
The most restrictive solution-space on the sphere depends on the interplay
between the adopted sampling scheme, the selection of the analysis/synthesis
problem, and any weighting of the $\ell_1$ norm.  We examine a variety of
sampling schemes on the sphere, including the DH \cite{driscoll:1994} and MW
\cite{mcewen:fssht} sampling theorems (leading to theoretically exact
spherical harmonic transforms) and the KKM \cite{khalid:optimal_sampling}
sampling scheme (leading to approximate but highly accurate spherical
harmonic transforms). DH, MW and KKM sampling requires $\sim4L^2$, $\sim2L^2$,
and $L^2$ samples, respectively.}

\analysissynthesis{To examine the analysis and synthesis problems and the impact of the various
sampling schemes considered, we study results from a simple inpainting problem
at low resolution.  In the numerical results shown in
\fig{\ref{fig:low_res_reconstruction}} and \fig{\ref{fig:low_res_snr}} we find
that the synthesis setting typically out-performs the analysis setting \referee{for suboptimal sampling schemes}.
Moreover, reconstruction fidelity is enhanced further by adopting more
efficient sampling schemes that require fewer samples to capture the
information content of signals on the sphere.  \referee{These findings are robust to the choice of test signal.}}

\analysissynthesis{\referee{We hypothesise that differences between the analysis and synthesis settings on the sphere are due to restrictions of the solution-space.  Elad {\it et al.} \cite{elad:2007} showed theoretically and with simulations that there are fundamental differences between the analysis and synthesis settings due to the solution-spaces.  As in Euclidean space, we find settings with a more restricted solution-space yield superior
performance. 
In contrast to Euclidean space, it is the synthesis
setting rather than the analysis setting that typically results in a more
restrictive solution-space on the sphere.  This is due to the inefficiency of
spherical sampling schemes and the weighting introduced in the $\ell_1$ norm.
Since the KKM sampling exhibits the optimal number of samples on the sphere there is no appreciable different between the size of the solution-spaces of the analysis and synthesis problems and the fidelity of spherical images recovered by the analysis and synthesis approaches are similar.}
In addition, the sparsity of band-limited signals is further promoted by more
efficient sampling schemes when considering a sparse representation that
captures spatial localisation \cite{mcewen:css2}, such as wavelets.}

We also demonstrate solving inverse problems in a number of high resolution
settings, facilitated by our fast adjoint operators. We solve inpainting, deconvolution, and combined inpainting and deconvolution problems, all in the presence of noise, 
using both axisymmetric and directional wavelets.  For all inverse problems considered our sparse regularisation techniques yield excellent reconstruction fidelity.

Our framework for solving inverse problems on the sphere can be applied to
many real-world problems. We have shown that our framework can be used to effectively denoise 353 GHz channel observations
from the {\it Planck} satellite, which will be useful for studying Galactic magnetic fields.

\ifCLASSOPTIONcaptionsoff
  \newpage
\fi

\bibliographystyle{IEEEtran}
\bibliography{bib_myname,bib_journal_names_long,bib}


\appendix

\section{Fast Wigner Transform Adjoints}\label{app:fast_so3_adjoints}

Standard convex optimisation methods require not only the application
of the operators that appear in the optimisation problem but 
also their adjoints.  Moreover, these methods are typically iterative,
necessitating repeated application of each operator and its adjoint.
Thus, to solve optimisation problems that incorporate Wigner
transform operators fast algorithms to apply both
the operator and its adjoint are required to render high-resolution
problems computationally feasible.  

Here we develop fast algorithms to perform adjoint forward and adjoint
inverse Wigner transforms for the extension of the MW sampling scheme to the rotation group \cite{mcewen:so3}. 
The fast adjoint follows by taking the adjoint of each stage of the
fast standard transforms \cite{mcewen:so3} and applying these in
reverse order.
The forward and inverse transforms can be found in \cite[Sec.~3]{mcewen:so3}. Here we use notation consistent with that work, \referee{where $\dlmnhalfpim=D_{mn}^{\ell}(0,\pi/2,0)$.} The fast adjoint of the
forward transform is as follows:
\begin{align}
  \Gmnm^\dagger 
  &= \img^{-(\m-\n)}
  \sum_{\ell=0}^{L-1} 
  \dlmnhalfpim \:
  \dlmnhalfpin \:
  \wigc{\f}{\el}{\m}{\n}
  \spcend ,\\
    \Fmnmp^\dagger
  &= 
  (2\pi)^2 
  \sum_{\m\p=-(\elmax-1)}^{\elmax-1}
   \Gmnm^\dagger\:
  \weight (\m\p - \m\p{}\p)
  \spcend,\\
  \tilde{G}_{m n}^\dagger(\beta_b) &= \frac{1}{2L-1}
  \sum_{m\p=-(L-1)}^{L-1}\Fmnm^\dagger e^{\img m\p \beta_b}\\
  F_{m n }^\dagger(\beta_b) &= 
   \begin{cases}
   \:  \tilde{G}_{m n}^\dagger(\beta_b) + (-1)^{m+n}\tilde{G}_{m n}^\dagger(-\beta_b) \: ,\\ &\!\!\!\! \!\!\!\! \!\!\!\! \!\!\!\! \!\!\!\! \!\!\!\! \!\!\!\!  \!\!\!\! \!\!\!\!  b \in \{ 0, 1, \dots, L-2\}   \\
   \:  \tilde{G}_{m n}^\dagger(\beta_b)\: , & \!\!\!\! \!\!\!\! \!\!\!\! \!\!\!\! \!\!\!\! \!\!\!\! \!\!\!\!   \!\!\!\! \!\!\!\! b = L-1\\
    \end{cases}\\
    \f^\dagger(\eulaiang, \eulbiang, \eulciang) &= \frac{1}{(2M-1)(2N-1)}\nonumber\\
    &\sum_{n=-(N-1)}^{N-1}  \sum_{m=-(M-1)}^{M-1} F_{m n }^\dagger(\beta_b) e^{\img(m\alpha_a + n\gamma_g)}.
\end{align}
Similarly, the fast adjoint of the inverse transform is as follows:
\ba
\tilde{f}(\eulaiang, \eulbiang, \eulciang) = 
 \begin{cases}
 f(\eulaiang, \eulbiang, \eulciang), & \saai \in \{ 0, 1, \dotsc, \elmax-1 \}\\
 0  \: , & \saai \in \{ \elmax, \dotsc, 2\elmax-2 \}
 \end{cases}
  \spcend ,
\ea
\begin{align}
F^\dagger_{m n m\p} &= \sum_{a=0}^{2M-2} \sum_{b=0}^{L-1} \sum_{g=0}^{2N-2} \tilde{f}(\eulaiang, \eulbiang, \eulciang)e^{-\img(m\eulaiang + n\eulbiang+ m\p\eulciang)}\\
 \wigc{\f}{\el\dagger}{\m}{\n} &= \img^{-(n-m)} \!\!\!\!\!\!\!\sum_{m\p=-(L-1)}^{L-1}\!\!\!\! \frac{2\ell+1}{8\pi^2}   \dlmnhalfpim \: \dlmnhalfpin \: F_{m n m\p}^\dagger.
\end{align}
These fast adjoint algorithms scale as $\order(N L^3)$ (where \mbox{$N \ll L \sim M$}) and are implemented in the {\tt SO3} code.

\end{document}